\documentclass[aps,prx,superscriptaddress,amsmath,amsfonts,amssymb,showpacs,floatfix,reprint,nofootinbib,longbibliography] {revtex4-2}

\usepackage{CJK}
\usepackage{physics}
\usepackage{soul}
\usepackage{color} 

\usepackage{multirow}
\usepackage{lipsum}  

\usepackage[percent]{overpic}
\usepackage{amsmath}
\usepackage{anyfontsize}
\usepackage[table,dvipsnames]{xcolor}
\usepackage{tikz}
\usepackage{bbding}

\usepackage[dvipsnames]{xcolor}

\definecolor{myblue1}{RGB}{50, 120, 160}

\usepackage{xr-hyper}
\usepackage[colorlinks,linktocpage,bookmarks=false]{hyperref}
\makeatletter
\newcommand*{\addFileDependency}[1]{%
  \typeout{(#1)}
  \@addtofilelist{#1}
  \IfFileExists{#1}{}{\typeout{No file #1.}}
}
\makeatother

\newcommand{\circo}[2][]{\tikz[baseline=(char.base)]\node[
    shape=circle,
    fill=#1,
    draw,
    inner sep=2pt,
    text=white
    ] (char) {#2};}

\usepackage[normalem]{ulem} %

\newcommand{\dlgnew}[1]{#1}
\newcommand{\refthreenew}[1]{{#1}}

\usepackage{soul,xcolor}
\setstcolor{red}
\definecolor{lightblue}{rgb}{.90,.95,1}
\definecolor{greeny}{rgb}{0.9,1,0}
\sethlcolor{greeny}

\definecolor{myrulecolor}{RGB}{150,20,0}%
\colorlet{mylinkcolor}{violet}
\colorlet{mycitecolor}{YellowOrange}
\colorlet{myurlcolor}{Aquamarine}

\definecolor{cblue}{RGB}{55,126,184}
\definecolor{ogreen}{RGB}{238,255,204}
\hypersetup{colorlinks=true,linkcolor=cblue,citecolor=cblue,urlcolor=cblue}

\usepackage{bm}

\usepackage{graphicx}
\graphicspath{{FIG/}}
\usepackage{multirow}

\usepackage{makecell}

\usepackage{booktabs} %

\AtBeginDocument{
	\heavyrulewidth=.08em
	\lightrulewidth=.05em
	\cmidrulewidth=.03em
	\belowrulesep=.65ex
	\belowbottomsep=0pt
	\aboverulesep=.4ex
	\abovetopsep=0pt
	\cmidrulesep=\doublerulesep
	\cmidrulekern=.5em
	\defaultaddspace=.5em
}%

\newcommand{\beq}{\begin{equation}}
\newcommand{\eeq}{\end{equation}}

\renewcommand\[{\begin{equation}}
\renewcommand\]{\end{equation}}

\newcommand{\mTOnePL}{\vb*{m}_{\mathsf{T_{1,p}}} }
\newcommand{\mTOneICE}{\vb*{m}_{\mathsf{T_{1,i}}} }

\newcommand{\mTOnePLScalar}{{m}_{\mathsf{T_{1,p}}} }

\newcommand{\mEScalar}{m_{\text{E}}}

\newcommand{\bfm}{{\vb*{m}}}

\newcommand{\sfX}{\mathsf{X}} 
\newcommand{\sfY}{\mathsf{Y}} 
\makeindex

\begin{document} 
	\begin{CJK*}{UTF8}{gbsn} %
		\title{ 
  An Atlas of Classical Pyrochlore Spin Liquids  
  }
  \author{Daniel Lozano-G\'omez} 
        \email{daniel.lozano-gomez@tu-dresden.de}
        \affiliation{Institut f\"ur Theoretische Physik and W\"urzburg-Dresden Cluster of Excellence ctd.qmat, Technische Universit\"at Dresden, 01062 Dresden, Germany}
        \affiliation{Department of Physics and Astronomy, University of Waterloo, Waterloo, Ontario N2L 3G1, Canada}
		\author{Owen Benton}  
  \affiliation{School of Physical and 
  Chemical Sciences, Queen Mary University
  of London, London, E1 4NS, United Kingdom}
        \author{Michel J. P. Gingras}  \affiliation{Department of Physics and Astronomy, University of Waterloo, Waterloo, Ontario N2L 3G1, Canada}
        \affiliation{Waterloo Institute for Nanotechnology, University of Waterloo, Waterloo, Ontario N2L 3G1, Canada}
		\author{Han Yan (闫寒)} 
    \email{hanyan@issp.u-tokyo.ac.jp}
    \affiliation{Institute for Solid State Physics, The University of Tokyo,  Kashiwa, Chiba 277-8581, Japan}
		\date{\today}
  
\begin{abstract}
Frustrated magnetism in the pyrochlore lattice magnet has proven to be a most fruitful setting for the experimental and theoretical search for spin liquids.
Besides the canonical case of spin ice, recent works have identified a variety of new classical and quantum spin liquids engendered by the generic nearest-neighbor anisotropic spin Hamiltonian for that lattice. 
However, a general framework for the thorough and systematic classification and characterization of these exotic states of matter has been lacking, as has an \textit{exhaustive} list of all possible spin liquids that this model can support and, perhaps most interesting, what is the corresponding structure of their emergent field theory description. 
In this work, we develop such a theoretical framework to identify the interaction parameters stabilizing different classical spin liquids and derive their corresponding  effective generalized Gauss's laws at low temperatures. 
Combining this with Monte Carlo simulations, we systematically identify all classical spin liquids for the general nearest-neighbor anisotropic spin Hamiltonian on the pyrochlore lattice. 
In doing so, we uncover new spin liquid models with exotic forms of generalized Gauss's law and multipole conservation laws. 
Our approach allows us to compile an atlas of all spin liquids realized in the phase diagram, providing a global picture of their mutual connections in parameter space and transitions between them.
Our work will inform future theoretical and experimental studies of classical and quantum spin liquids on the pyrochlore lattice and help rationalize the exotic properties of pyrochlore magnets.
\end{abstract}

\maketitle

\end{CJK*}

\tableofcontents

\section{Introduction}
The exploration of quantum and classical spin liquids (CSLs) stable down to the lowest temperatures has been one of the most active pursuits in many-body physics over the past thirty years~\cite{Wen2007,Balents2010,Springer_frust,ZhouRevModPhys.89.025003,savaryQuantumSpinLiquids2016,Gingras_2014,Knolle_Moessner_2019,moessner2021book}. 
These phases, whose spins are disordered yet highly correlated, develop at low temperatures and are described by effective gauge theories emerging from the microscopic spin degrees of freedom (DOFs). 
For example, CSLs often realize the electrostatic sector of gauge theories \cite{Yan2024a,Yan2024b,Fang2024PhysRevB}.
When quantum fluctuations are introduced, these CSLs may become quantum spin liquids, which exhibit topological entanglement, fractionalized excitations, and, in gapped cases, topological order- some of the most fascinating phenomena in quantum matter.
Classical spin liquids typically arise from finely tuned spin-spin interactions on specific lattice geometries, often featuring geometric frustration ~\cite{Bergman_band}.
This frustration stabilizes highly degenerate classical ground states ~\cite{Balents2010,Springer_frust} 
leading to collective paramagnetic behavior
~\cite{villain1979} and, in some cases, providing a potential route to quantum spin liquid (QSL) phases~\cite{Balents2010,savaryQuantumSpinLiquids2016,Gingras_2014,Knolle_Moessner_2019}.

In particular, the pyrochlore lattice, a network of corner-sharing tetrahedra~\cite{Balents2010,Gardner-RMP,Hallas-AnnRevCMP,Rau2019ARCMP,Smith2025-AnnRevCMP}, has provided a fruitful landscape for the search of spin liquids. 
The first and most canonical example is classical spin ice~\cite{Harris1997PhysRevLett,Bramwell-Science,castelnovoSpinIceFractionalization2012,udagawaSpinIce2021},
 experimentally realized  in   $\rm Ho_2 Ti_2 O_7$~\cite{Harris1997PhysRevLett,Bramwell-Ho2Ti2O7, FennelPhysRevBSpin_Ices,
 Fennelscience.1177582} and $\rm Dy_2 Ti_2 O_7$~\cite{RamirezDTO,FennelPhysRevB.Dy2Ti2O7,Morris2009Science}. 
At low temperatures, this classical spin liquid realizes an effective Maxwell U(1) electrostatics theory on the parent (or premedial) diamond lattice~\cite{HenleyARCMP}, 
whose ground states are divergence-free ``electric field'' configurations, and elementary excitations are sources and sinks (i.e. ``charges'') of that field~\cite{castelnovoSpinIceFractionalization2012}~\footnote{Often, the dual language of magnetic field and magnetic monopoles is used in the literature~\cite{Gingras_2014,Castelnovo2008Nature}.}.
In other materials~\cite{Rau2019ARCMP,Smith2025-AnnRevCMP}, in which quantum effects~\cite{Rau_quantum} are expected to be more important than in the (Ho,Dy)$_2$(Ti,Sn,Ge)$_2$O$_7$ 
classical spin ice compounds~\cite{Zhou_spin_ices},  
a quantum counterpart of spin ice realizing full-fledged quantum electrodynamics, namely quantum spin ice~\cite{Hermele2004,Gingras_2014}, has been under intense theoretical~\cite{Molavian2007,Onoda_quantummeling-PRL,leeGenericQuantumSpin2012,Benton2012PhysRevB,shannon2012PRL,Huang2018PhysRevLett,Benton_DO_2020,Pace2021PhysRevLett,hosoiUncoveringFootprintsDipolarOctupolar2022,DesrochersPRL,yan2023experimentally} and experimental investigations~\cite{Kimura2013_Przr,Pr2Zr2O7_disorder,Sibille2018_PrHf,CZO.Gao,CZO.Gaulin,Sibille2020NatPhys,poree2023fractional,Gao2025,Smith_Ce2Zr2O7,YahnePhysRevX.14.011005}.\\

Zooming out to the entire landscape, a much wider spectrum of exotic classical spin liquids have been discovered upon considering the generic nearest-neighbor anisotropic bilinear spin model on the pyrochlore lattice. 
This family of models, parameterized by four symmetry-allowed anisotropic spin-spin 
couplings of exchange origin ($\{J_{zz}, J_\pm, J_{\pm\pm}, J_{z\pm} \}$ \cite{Curnoe2008PhysRevB,ThompsonPRL2011,RossPRX2011,Yan-2017,KTC_2024_phase}), has been successful in characterizing the low-temperature physics of several pyrochlore compounds  $\rm R_2 M_2 O_7$~\cite{Hallas-AnnRevCMP,Rau2019ARCMP,Smith2025-AnnRevCMP}, where $\rm R$ is a trivalent rare earth ion and $\rm M$ is a non-magnetic tetravalent transition metal ion~\cite{Gardner-RMP}.\\

These classical spin liquids are characterized by emergent \textit{generalized} Gauss's laws \cite{Xu2006PhysRevB,Pretko2017PhysRevBa,Pretko2017PhysRevBb,Yan2024a,Yan2024b}. 
Among them, classical spin ice~\cite{Bramwell-Science}, the  Heisenberg antiferromagnet  (HAFM) spin liquid~\cite{Reimers1992,Moessner1998PRB} and the XXZ model~\cite{Taillefumier2017PhysRevX} host one or several copies of emergent Gauss's law~\cite{HenleyARCMP}.
Other cases, such as the so-called pinch-line spin liquid~\cite{Benton2016NatComm} and  the rank-1$-$rank-2 spin liquid~\cite{lozanogomez2023arxiv} are more exotic; these  are  
\textit{higher-rank} spin liquids, whose generalized Gauss's laws are imposed on tensorial (rank-2) electric fields.
Such more complex versions of electrodynamics, which are allowed on the lattice due to the lack of Lorentz or rotational symmetry, would describe new gapless quantum spin liquids, and are closely related to the field of fractonic states of matter \cite{Pretko2020IJMPA,Nandkishore2019ARCMP,you2024arxivReivew}.
The intrinsic multipole conservation and subsystem symmetries showcased by these higher-rank spin liquids are at the root of a rich physics, from immobile fractionalized excitations dubbed fractons \cite{Vijay2015PhysRevB,Pretko2017PhysRevBa,Pretko2017PhysRevBb} to quantum error-correction \cite{Chamon2005PhysRevLett,Haah2011PhysRevA}.

Apart from spin ice, which occupies a three-dimensional parameter space at zero temperature~\cite{Rau2019ARCMP,Yan-2017,Benton_DO_2020,KTC_2024_phase}, the other classical pyrochlore spin liquids that have been theoretically discovered generally live on the phase boundary of two or more magnetically ordered phases~\cite{Benton2016NatComm,lozanogomez2023arxiv}~\footnote{The existence of spin liquids on so-called breathing pyrochlore lattices has also been explored~\cite{Savary_BP,Yan2020PRL_Rank2}}.
While the phase diagram of the possible long-range magnetic orders for the  general  Hamiltonian has been discussed in detail in the literature~\cite{Wong2013,Yan-2017,Rau2019ARCMP,Hallas-AnnRevCMP} (for a recent in-depth analysis of the irrep fields, and the geometry and topology of the phases and phase boundaries, see the recent work by Chung~\cite{KTC_2024_phase}), a theoretical framework to identify systematically \emph{all} CSLs on the pyrochlore lattice is still lacking.
This raises a key question: do additional, previously overlooked spin liquids exist in the phase diagram? 
In short,  an ``atlas'' of classical spin liquids, i.e., a map showing the exact regions in parameter space where CSLs are realized and what their respective dualities are, and how different CSLs transition into each other remains wanting.\\

In this work, we address these shortcomings by developing such a theoretical framework based on irreducible representations of the symmetries of a tetrahedron (tetrahedral point group, $T_d$) and the inter-tetrahedra constraints created by the spin Hamiltonian, which together enable a comprehensive study of various pyrochlore classical spin liquids. 
Specifically, we identify all the classical spin liquids stabilized down to the lowest temperatures for the generic nearest-neighbor Hamiltonian on the pyrochlore lattice.
Our approach also provides a straightforward recipe for constructing the low-energy effective generalized Gauss's law for the different CSLs. 
We have found in total nine classical spin liquids and derived their gauge theories %
including two cases not previously discussed in the literature and which realize a Maxwell and a rank-2 U(1) electrostatics, respectively. 
This allows us to chart an atlas of classical pyrochlore spin liquids, presenting both the exact parameter space that defines different CSLs, identify their dualities and map the phase transitions between them.
  
Our approach transforms the theoretical search for pyrochlore CSLs from smart craftsmanship to a rigorously organized and systematic procedure.
Our work is also highly experimentally relevant: as shown in several previous studies~\cite{RossPRX2011,Guitteny2013PhysRevB,Jaubert2015PhysRevLett,Yan-2017,Hallas-AnnRevCMP,Sarkis_Yb2Ge2O7,Smith_Ce2Zr2O7,Scheie_YbTiO}, it turns out that many pyrochlore compounds find themselves rather close to the phase boundaries of semi-classical long-range magnetic orders where such multi-phase competition is essential for understanding their physics.
Our work, charting out the CSLs that live on these phase boundaries, will be useful to understand the exotic physics at play in past and future experiments. 
In particular, we discuss the exciting opportunities to realize fracton physics in these frustrated magnetic systems.
Our atlas also provides a road-map for the systematic hunt for pyrochlore quantum spin liquids, as it highlights the parameter spaces of most intense competition between the various spin-spin interactions and indicates the potential nature of the prospective quantum spin liquids descending from their classical limits.

The rest of the paper is organized as follows: in Sec.~\ref{sec:model_results}, we review salient aspects of the generic nearest-neighbor anisotropic bilinear model on the pyrochlore lattice and briefly summarize the results that follow.
In Sec.~\ref{Sec:irrep.and.inter.tetra}, we present the irreducible representation analysis of the spin Hamiltonian and show how this can be used to derive the underlying gauge fields and emergent Gauss's laws characterizing the various classical spin liquid phases of the model.
Using this procedure, we identify in Sec.~\ref{sec:CSL_non_kramers} and \ref{sec:CSL_kramers} the regions in parameter space where both already known as well as, to the best of our knowledge, heretofore unidentified classical spin liquid phases are realized. 
In Sec.~\ref{sec:Connectibity}, we show how the spin liquid regions in the $\{J_{zz}, J_\pm, J_{\pm\pm}, J_{z\pm} \}$ phase diagram intersect each other.
In particular, we show that all the classical spin liquids in the generic nearest-neighbor Hamiltonian can be reached by continuously tuning the interaction parameters without an intermediate symmetry-broken phase bisecting them. 
Then, in Sec.~\ref{sec:numerics}, we provide numerical evidence for the $T\to 0$ stability of these unexplored spin liquids identified in Sec.~\ref{sec:CSL_kramers}. 
\dlgnew{
Section VIII briefly discusses those
phase boundaries where magnetic order supersedes a potential CSL.
Sections \ref{sec:roadmap}-
\ref{sec:Application_other_lattices}
provide an outlook for the future, discussing implications for quantum models, real materials, and 
the application of the framework presented in this paper to other lattices, respectively.
Finally, Sec.~\ref{sec:discussion} 
closes with a brief discussion and conclusion.
}

\begin{table*}[th!]
\def\arraystretch{1.5}%
\caption{
    \label{Table_all_csls}
    Table of all classical spin liquids (CSLs) for the pyrochlore spin system with nearest-neighbor anisotropic spin-spin interactions [Eq.\eqref{eq:Hex1}].
    The third column gives the irreps that achieve minimal $a_\mathsf{X}$ couplings [i.e. ``fluctuating irrep'' in Eq.~\eqref{eq:Ht-diagonal}].
     The corresponding irreducible representation order parameters become the electric field constrained by generalized Gauss's laws. 
    The fourth column shows the corresponding $\{J_{uv}\}$ spin-spin interaction parameters in Eq.~\eqref{eq:Hex1} that define the CSL while the fifth column shows the dimension of that parameter space (an overall scaling of all the parameters is not considered).
    The sixth column lists the corresponding generalized Gauss's laws of the CSLs, whose derivation is discussed in detail in this work. The seventh column lists some of the relevant references for previously identified CSLs. 
    }
    \begin{tabular}{|c|c|c|c|c|c|c|}
    \multicolumn{7}{c}{\rule{0pt}{6ex}  NON-KRAMERS CLASSICAL SPIN LIQUIDS \rule{0pt}{4ex}}\\
	\hline
		CSL $\#$&type and name of CSL & fluctuating irreps& $\{J_{uv}\}$ interaction parameters & \begin{tabular}[c]{@{}c@{}}phase \\ dim.
        \end{tabular}& Gauss's laws &  Refs. \\
    \hline
		~~~1~~~& Single R1U1 (Spin ice) 
        &$\mathsf{T_{1,i}}$
        &\begin{tabular}[c]{@{}c@{}}$J_{zz} > 0$\\ $-J_{zz} < -6J_\pm   \text{ and } 2J_\pm\pm4J_{\pm\pm}$
        \end{tabular}  & {3D} 
        &$\nabla \cdot \vb*{E}^\mathsf{ice} = 0$      & \cite{Harris1997PhysRevLett}\\ 
    \hline

		2& Double  R1U1 ($\rm SL_\perp $)  
        &${\mathsf{T_{1,p}}} , \ {\mathsf{T_{2}}} $   
        & 
        \begin{tabular}[c]{@{}c@{}}
            $J_{\pm\pm} = 0, J_{\pm}< 0$\\ 
            $2J_\pm < 3J_{zz} \text{ and } - J_{zz}$
        \end{tabular} 
        & {1D} 
        & 
        \begin{tabular}[c]{@{}c@{}}
            $\nabla \cdot \vb*{E}^\mathsf{A} = 0$\\
            $ \nabla \cdot \vb*{E}^\mathsf{B} = 0 $ 
        \end{tabular}  
        &    \cite{Taillefumier2017PhysRevX}\\ 
    \hline
		3&  Triple  R1U1 (pHAF)  
        &$\mathsf{T_{1,i}},\ {\mathsf{T_{1,p}}} , \ {\mathsf{T_{2}}}$ 
        & \begin{tabular}[c]{@{}c@{}}$J_{zz} > 0$\\ $ J_{zz}: J_{\pm}: J_{\pm\pm}= 2:-1:0 $\end{tabular} & {1D}                                 
        & $\partial_i E_{ij}^{\mathsf{pHAF}} = 0$     & \cite{Moessner1998PhysRevLett,Taillefumier2017PhysRevX}   \\
        \hline 
		4&  R1U1-R2U1 $(\rm R_1$-$\rm R_2^{\mathsf{T}_1})$ 
        &$\mathsf{T_{1,i}}, \ {\mathsf{E}},\ {\mathsf{T_{1,p}}} $ 
        & \begin{tabular}[c]{@{}c@{}}
        $J_{zz} > 0$\\ 
         $   J_{zz}: J_{\pm}: J_{\pm\pm}= 6:1:-2$
        \end{tabular}  & {1D}   
        & 
        \begin{tabular}[c]{@{}c@{}}
           $\nabla \cdot \vb*{E}^\mathsf{ice} = 0$\\
            $\partial_i E_{ij}^\mathsf{T_1+E} = 0$ 
        \end{tabular}       
        &    \cite{lozanogomez2023arxiv} \\ 
    \hline
		5& R1U1-R2U1 $(\rm R_1$-$\rm R_2^{\mathsf{T}_2})$   &$\mathsf{T_{1,i}}, \ {\mathsf{E}},\ {\mathsf{T_{2}}} $      & \begin{tabular}[c]{@{}c@{}}$J_{zz} > 0$\\ $   J_{zz}: J_{\pm}: J_{\pm\pm}= 6:1:2$\end{tabular}                      & {1D}       
        & 
        \begin{tabular}[c]{@{}c@{}}
           $\nabla \cdot \vb*{E}^\mathsf{ice} = 0$\\
            $\partial_i E_{ij}^\mathsf{T_2+E} = 0$ 
        \end{tabular}    
        &      \cite{lozanogomez2023arxiv}  \\ 
    \hline
     \multicolumn{7}{c}{\rule{0pt}{6ex}  KRAMERS CLASSICAL SPIN LIQUIDS\rule{0pt}{4ex}}\\
    \hline  
		6&\begin{tabular}[c]{@{}c@{}}SV-R2U1\\ (SV $ \equiv$ Scalar-vector)\end{tabular}   
        &$\mathsf{T_{1,-}},\ {\mathsf{E}} , \ {\mathsf{T_{2}}}$ 
        & $J_{z\pm} >0$, Eq.~\eqref{eq:CL6_constraint} 
        & {2D}  
        & 
        \begin{tabular}[c]{@{}c@{}}
           $|\epsilon^{ijk}| \partial_i E_{jk}^{\mathsf{SV}}  = 0$\\
           $\partial_i E_{ij}^{\mathsf{SV}} = 0$
        \end{tabular}  
        &     \\\hline
		7&             \begin{tabular}[c]{@{}c@{}}SV-R2U1$^\ast$\\ (Pinch-line spin liquid)\end{tabular}  
        &$\mathsf{T_{1,-}},\ {\mathsf{E}} , \ {\mathsf{T_{2}}}$ 
        & $J_{z\pm} < 0$,  Eq.~\eqref{eq:CL6_constraint} 
        & {2D}  
        & 
        \begin{tabular}[c]{@{}c@{}}
           $|\epsilon^{ijk}| \partial_i E_{jk}^{\mathsf{SV}}  = 0$\\
           $\partial_i E_{ij}^{\mathsf{SV}} = 0$
        \end{tabular}  
        &   \cite{Benton2016NatComm} \\\hline
        8& Triple  R1U1 (HAFM)              
        &$\mathsf{T_{1,-}},\ {\mathsf{E}} , \ {\mathsf{T_{2}}} , \ {\mathsf{A_{2}}}$ 
        & 
        \begin{tabular}[c]{@{}c@{}}
           $J_{zz}: J_\pm :J_{\pm\pm}:J_{z\pm}$ \\ 
           $= -2   :1: 2: 2\sqrt{2}$
        \end{tabular} 
        & {1D}  
        &  $\partial_i E_{ij}^{\mathsf{HAF}} = 0$         & 
\cite{Moessner1998PhysRevLett} \\\hline
        9& Triple  R1U1 (HAFM$^*$)             
        &$\mathsf{T_{1,-}},\ {\mathsf{E}} , \ {\mathsf{T_{2}}} , \ {\mathsf{A_{2}}}$ 
        & 
        \begin{tabular}[c]{@{}c@{}}
           $J_{zz}: J_\pm :J_{\pm\pm}:J_{z\pm}$ \\ 
           $= -2   :1: 2: -2\sqrt{2}$
        \end{tabular} 
        & {1D} 
        &  $\partial_i E_{ij}^{\mathsf{HAF}} = 0$ &   
        \\\hline
	\end{tabular}
\end{table*}
\section{The Model and summary of main results}
\label{sec:model_results}
\subsection{The nearest-neighbor pyrochlore model}
This section summarizes the pyrochlore anisotropic spin model, the location of CSLs in its ground-state phase diagram, and the corresponding emergent Gauss's laws for readers seeking a quick overview of our paper.
For those interested in a deeper dive into the technicalities, the subsequent sections provide a comprehensive discussion of the construction of the field theory and the details of the classical Monte Carlo (CMC) simulations.
\begin{figure}[h!]
    \centering
    \begin{overpic}[width=0.8\columnwidth]{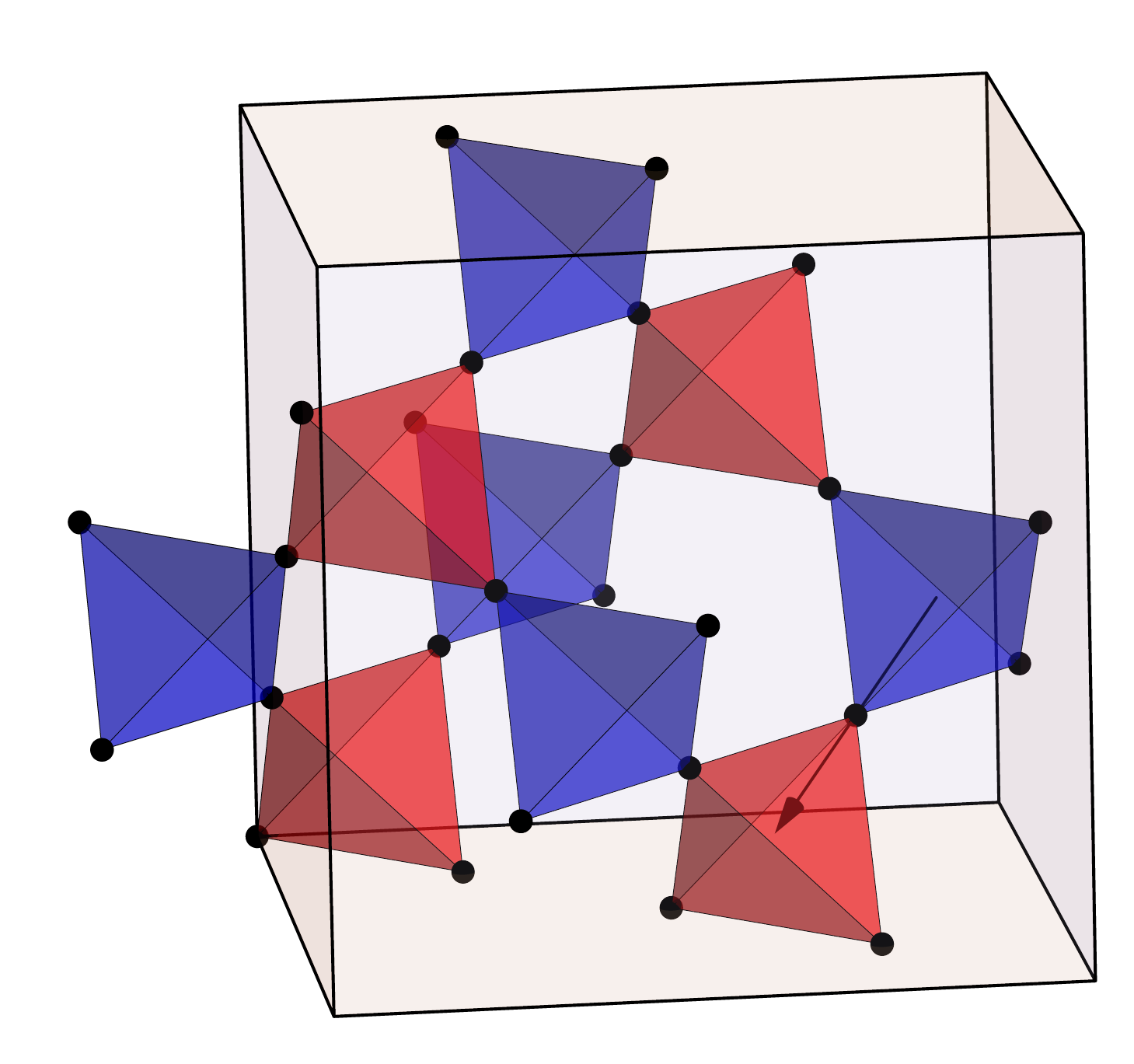}
    \put(55,10){0}
    \put(80,10){3}
    \put(63,29){2}
    \put(78,27){1}
    \end{overpic}
    \caption{Pyrochlore lattice where the spin degrees of freedom  associated to the rare-earth R$^{3+}$ ions in R$_2$M$_2$O$_7$ are located at the corners (black dots) of the tetrahedra.
    The M$^{4+}$ ions reside on another regular pyrochlore lattice (not shown), that interpenetrates the one shown, and whose sites are located at the center of the hexagonal plaquettes defined by the R$^{3+}$ ions~\cite{Gardner-RMP}.
    We color the $A$ and $B$ tetrahedra in red and blue, respectively. 
    Additionally, we illustrate the sublattice labels for the four sites of a single $A$ tetrahedron along with the local $z$ direction of the spin in sublattice $1$ connecting the center of the adjacent tetrahedra sharing this lattice site. 
    }\label{fig:pyrochlore_lattice_unit_cell}
\end{figure}

A prominent family of magnetic pyrochlore materials is the rare-earth pyrochlore oxides R$_2$M$_2$O$_7$ (see Fig.~\ref{fig:pyrochlore_lattice_unit_cell}).
In these materials, a strong crystal electric field (CEF) often produces a well-separated ground-state doublet, with an energy gap significantly larger than the interaction scale~\cite{Bertin_2012,Gaudet2017EffectOC,Rau2019ARCMP}~\footnote{The Tb-based pyrochlores~\cite{Molavian2007,Rau2019ARCMP,Roll_TbTiO} and, to a lesser extent, the Er-based ones~\cite{Rau2019ARCMP,McClarty-Curnoe,Petit_ErTiO,Rau-Petit} are examples where this crystal-field energy gap is quite a bit smaller than in the other 
R$_2$M$_2$O$_7$ compounds, with the consequences of such not yet fully understood.}.
In such a scenario, it is convenient to describe these doublets as pseudospin-$1/2$ degrees of freedom (DOFs),  $S_i^{\alpha}$, and to model the ion-ion superexchange interactions as interactions between these pseudospins~\cite{Onoda-PRB,Onoda_quantummeling-PRL,Molavian2007,leeGenericQuantumSpin2012, savaryCoulombicQuantumLiquids2012,RossPRX2011,  Rau2019ARCMP,Huang2014PhysRevLett,Chibotaru,Rau_quantum}.

The allowed form of nearest-neighbor interactions between pseudospin components can then be determined by symmetry. When doing this, one must remember that the transformation of the pseudospins under lattice symmetries and time-reversal depends on the symmetry of the wave functions making up the CEF doublet~\cite{Rau2019ARCMP}. 
Through this analysis, three cases can be distinguished: `dipolar' Kramers in which the pseudospin transforms like an ordinary spin-$1/2$; `dipolar-octupolar' Kramers in which one component of the pseudospin transforms like a magnetic octupole; and the non-Kramers case where one of the $z$ component of $\vb*{S}_i$ transforms like a magnetic dipole while the two other components transform like electric quadrupole moments (see Table~1 of Ref.~\cite{Rau2019ARCMP}).

In the most general case, the nearest-neighbor Hamiltonian may be written as follows~\cite{RossPRX2011}
\begin{align}
    &&H_{\sf ex}=\sum_{\langle ij \rangle}
    \Big[ 
    J_{zz} S^z_i S^z_j - J_{\pm} (S^+_i S^-_j + S^-_i S^+_j)  \nonumber \\
    && + J_{\pm \pm} ( \gamma_{ij} S^+_i S^+_j + \gamma^{\ast}_{ij} S^-_i S^-_j)
    \nonumber \\
    && - J_{z \pm} ( \gamma^{\ast}_{ij} S^z_i S^+_j + \gamma_{ij} S^z_i S^-_j +  i \leftrightarrow j)
    \Big],
    \label{eq:Hex1}
\end{align}
where $S^z_i$ is the component of the pseudospin directed along the local $C_3$ symmetry axis at each site (aligned with the local [111] direction connecting adjacent  $A$ and $B$ tetrahedra), and $S_i^\pm$ are raising and lowering operators defined relative to that axis. 
\dlgnew{Further details regarding the definition of that local basis are given in   Appendix~\ref{appendix:local_coordinates}}.
The $\gamma_{ij}$ variables are bond-dependent phase factors which depend on the nature of the low energy CEF doublet~\cite{Onoda_quantummeling-PRL, Rau2019ARCMP,leeGenericQuantumSpin2012,RossPRX2011,Huang2014PhysRevLett}.
For the dipolar Kramers~\cite{Rau2019ARCMP,RossPRX2011}
and non-Kramers cases~\cite{Rau2019ARCMP,leeGenericQuantumSpin2012,Onoda_quantummeling-PRL,Onoda-PRB}, the $\gamma_{ij}$ phase factors take on values 
\[
\vb*{\gamma}
=\left(
\begin{array}{cccc}
    0 & 1 & c & c^* \\
    1 & 0 & c^* & c \\
    c & c^* & 0 & 1 \\
    c^* & c & 1 & 0 \\
\end{array}
\right),
\label{eq:gamma-matrix}
\]
where $c \equiv e^{2\pi i/3}$. 
For Kramers ions, all components of the pseudospin are odd under time reversal, while for non-Kramers ions the transverse $S_i^\pm$ components are time-reversal even. 
This has the important consequence that  $J_{z\pm}=0$ for non-Kramers ions~\cite{Onoda_quantummeling-PRL, Onoda-PRB,leeGenericQuantumSpin2012}.
Additionally, for the so-called dipolar-octupolar (DO) doublet Kramers systems~\cite{Huang2014PhysRevLett,Rau2019ARCMP,Benton_DO_2020,hosoiUncoveringFootprintsDipolarOctupolar2022}, such as the pyrochlores with Ce$^{3+}$, Sm$^{3+}$ and Nd$^{3+}$ rare-earth ions~\footnote{The classical Dy$_2$(Ti,Sn,Ge)$_2$O$_7$ spin ices~\cite{Zhou_spin_ices} with magnetic Dy$^{3+}$ are also DO  materials~\cite{Rau2019ARCMP,Huang2014PhysRevLett}, albeit with rather small $J_\pm$, $J_{\pm\pm}$ and $J_{z\pm}$ transverse couplings in Eq.~(\ref{eq:Hex1})~\cite{Rau_quantum}.}, one has $\gamma_{ij}=1$ on all nearest-neighbor bonds~\cite{Huang2014PhysRevLett}.

In the classical description of Eq.~\eqref{eq:Hex1}
we replace the $S^{\pm}$ operators with $S^x \pm i S^y$ and treat $\vb*{S}=(S^x, S^y, S^z)$ as a three-component vector of fixed length $S=1$.
Moreover, within a classical description, or within the context of materials with a larger spin value~\cite{FeF3,Plumb_NaCaNi2F7},  it is meaningful to consider a single-ion anisotropy term~\cite{FeF3} whose simplest uniaxial form can be added to $H_{\sf ex}$ to give:
\begin{equation}
\label{eqn:single.ion}
    H=H_{\sf ex} + H_{\sf si}, \ \ H_{\sf si}=D \sum_i (S^z_i)^2 .
\end{equation} 
 
In this work, we henceforth set $D=0$
and focus on the $H_{\sf ex}$ terms only. Although we consider this simplified case, our field theory analysis could be straightforwardly generalized to the case with a non-zero $H_{\sf si}$ terms
We note that the results presented in this work focus on the classical Kramers and non-Kramers Hamiltonians, as the classical and quantum phase diagrams of the DO Hamiltonian [$\gamma_{ij}=1$ in Eq.~\eqref{eq:Hex1}]~\cite{Huang2014PhysRevLett} have already been studied in detail in previous works~\cite{Benton_DO_2020,hosoiUncoveringFootprintsDipolarOctupolar2022}.

\subsection{Main results}
We comprehensively analyze the parameter space $\{J_{zz}, J_\pm, J_{\pm\pm}, J_{z\pm}\}$ for the classical dipolar spin system to map out \textit{all} classical spin liquids in the phase diagram by combining analytical theory with CMC simulations. 
Analytically, we construct a long-wavelength field theory based on inter-tetrahedra constraints, unveiling various generalized Gauss's laws that characterize the degenerate ground states of different CSLs. 
This theoretical framework proves exact in the limit of a large number of spin components,
$\mathcal{N}$. 
Through CMC simulations, we verify that the three-component classical version of the spin models (henceforth referred to as `${\mathcal N}=3$  model' for short) remain disordered down to the lowest temperatures examined, and then further investigate the system's thermodynamic properties and spin-spin correlations. 
We note that the existence of an emergent Gauss's law from the large-$\mathcal{N}$ field theory is a necessary but not sufficient condition for the manifestation of a CSL -- there are cases in which the large-$\mathcal{N}$ theory predicts a CSL but the ${\mathcal N}=3$  model version eventually orders at low temperatures~\cite{Francini2024nematicR2}. 
Such cases are usually associated with a thermal order-by-disorder selection~\cite{villain1980,ZhitomirskyPRL2012} that is typically induced by  higher-order (fluctuation) terms in the free energy that are not included in the large-$\mathcal{N}$ analysis~\cite{lozanogomez2023arxiv}.

We have identified a total of \emph{nine} classical spin liquids (CSLs). 
The parameter spaces that define them and their corresponding emergent Gauss's laws are summarized in Table~\ref{Table_all_csls}. 
The generalized Gauss's laws include the conventional Maxwell Gauss's law, i.e., a charge as the divergence of a vector field, as well as more exotic forms such as coexisting two or three copies of Maxwell Gauss's law, charges defined by rank-2 matrix electric field as well as a mixture of them. 
In that table, we use a naming convention for the spin liquids based on the structure of the emergent gauge fields describing the low-temperature physics (R1 $\equiv $rank-1 or R2 $\equiv $rank-2), the gauge structure constraining the field (this being U1$ \equiv \rm U (1)$ for all the CSLs described in the present work), and then provide in parentheses the name these CSLs were given (if there was any) when first identified~\cite{Taillefumier2017PhysRevX,lozanogomez2023arxiv,Benton2016NatComm,Harris1997PhysRevLett,Moessner1998PhysRevLett}. 
For example, for the spin ice CSL, the low-energy theory is described by a single rank-1 (R1) field with a $\rm U(1)$ (U1) structure characterized by the emergent Gauss's law acting on a rank-1 field. 
In our notation, this then reads as $\mathrm{R1U1}$ (spin ice) spin liquid. \dlgnew{We note that, although we have classified the CSL in Table~\ref{Table_all_csls} into non-Kramers or Kramers CSLs, all the non-Kramers CSLs can also be regarded as Kramers CSLs in the  fine-tuned case where $J_{z\pm}=0$ in the Kramers Hamiltonian.}~\footnote{
{
In such Kramers case, all the spin components transform as dipoles. This difference would be exposed when studying neutron structure factors for which only the $S_i^z-S_j^z$ correlations are experimentally measurable for a non-Kramers system while for a Kramers system, the transverse spin-spin correlations are experimentally accessible~\cite{Kadowaki_2015,Rau2019ARCMP}. This difference is reflected in the local $g$-factor where its $xy$ component $g_{\perp}$ vanishes for non-Kramers systems, while it  \dlgnew{is generically nonzero for the Kramers cases, with the exception of Kramers ions with dipolar-octupolar crystal field ground states which have $g_\perp=0$~\cite{Rau2019ARCMP}.
}
}
}

\begin{figure}[t!]
    \centering \includegraphics[width=\columnwidth]{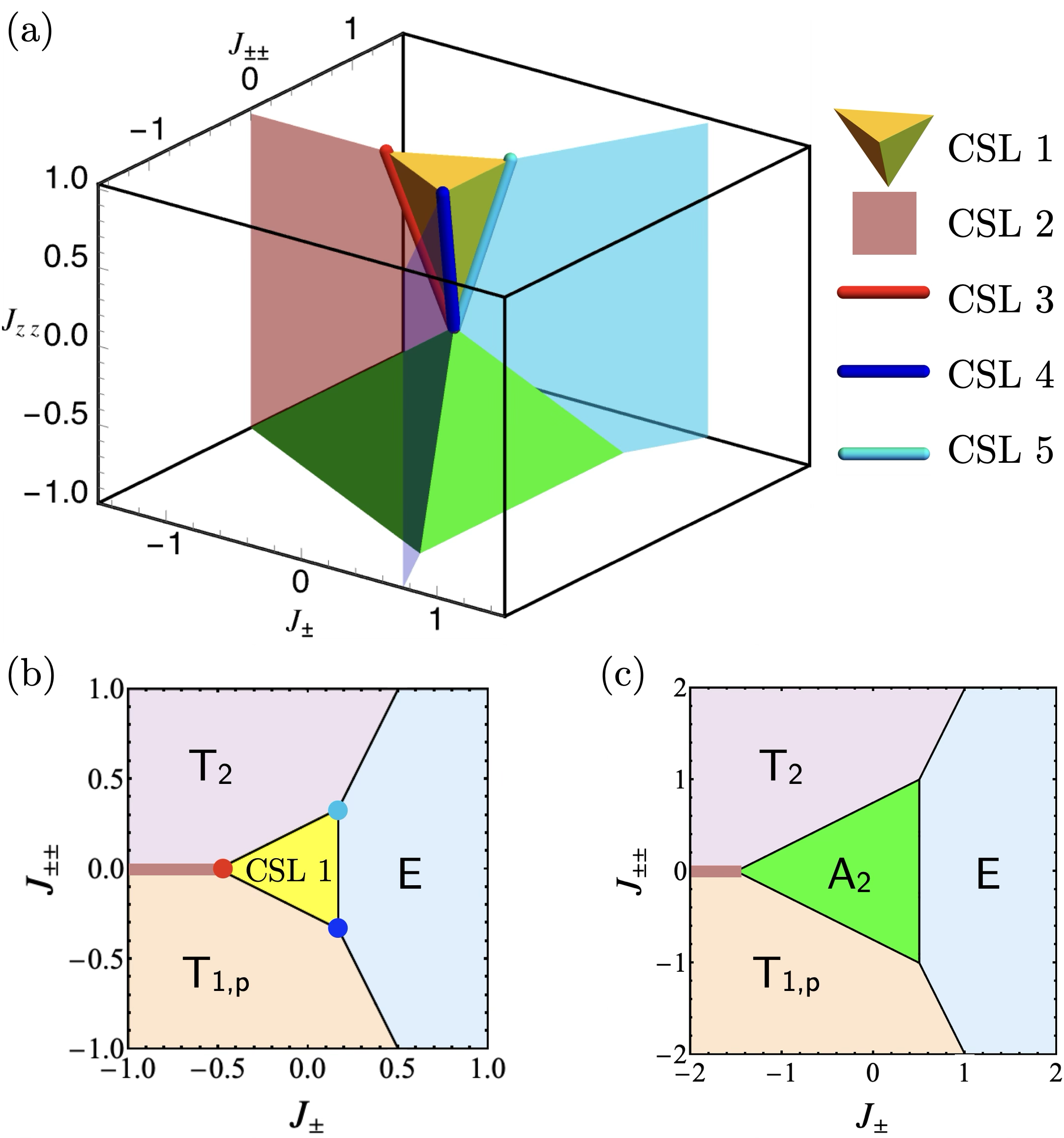} 
    \caption{ (a) Phase diagram of classical spin liquids (CSLs) for the non-Kramers Hamiltonian [Sec.~\ref{sec:CSL_non_kramers}]. 
    There are five CSLs in the phase diagram.
    CSL 1 (spin ice) occupies a $3$D volume, CSL 2 occupies a surface, and the other three CSLs reside on different phase boundary lines.  
    (b) Cross-section of the phase diagram at $J_{zz} = 1$, showing all five CSLs. 
    The irreducible representations of the magnetic ordered phases are also indicated. (c) Cross-section at $J_{zz} = -1$. 
    Only CSL 2 lives in this part of the phase diagram.  }
    \label{fig:nonkramer.phase}
\end{figure}

\begin{figure}[t!]
\centering 
\includegraphics[width=\columnwidth]{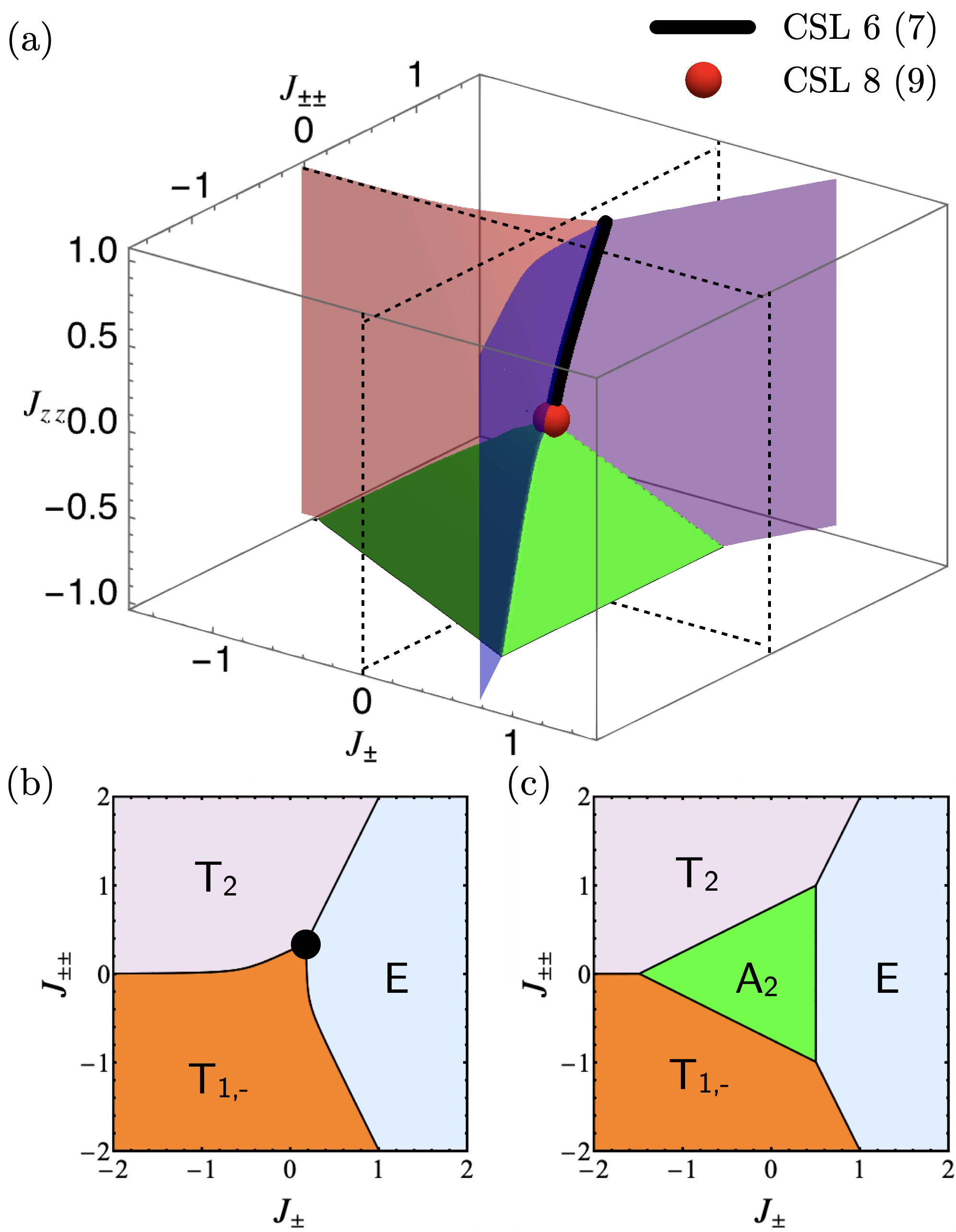}  
    \caption{ (a) Phase diagram of classical spin liquids (CSLs) for the  Kramers Hamiltonian [Sec.~\ref{sec:CSL_kramers}]. 
    CSL 6 and 8 are realized in this phase diagram 
     \dlgnew{(here, $J_{z\pm}$ was taken to be $0.1$, but see comment below in the caption).
   (b) and (c) illustrate cross-sections of the phase diagram at $J_{zz} =1$ and $J_{zz} =-1$, respectively.  }
    The CSL 6 is the scalar-vector rank-2 U(1) spin liquid, and lives on the phase boundary marked by a black line. 
    CSL 8 is the Heisenberg antiferromagnet, corresponding to the red dot.  
    By flipping the sign of the $J_{z\pm}$ coupling, i.e. $J_{z\pm} = -0.1$, the shape of the phase diagram remains invariant due to the duality in Eq.~\eqref{eqn_kramers_duality}. Upon this $J_{z\pm} \rightarrow - J_{z\pm}$,  CSL 6 becomes CSL 7 and  CSL 8 becomes CSL 9.}
    \label{fig:kramer.phase}
\end{figure}

The four-dimensional phase diagram can be qualitatively divided into two regions based on the value of $J_{z\pm}$: one where $J_{z\pm} = 0$ (appropriate to non-Kramers materials), illustrated in Fig.~\ref{fig:nonkramer.phase}, and another where $J_{z\pm} \neq 0$, illustrated in Fig.~\ref{fig:kramer.phase}.
The different CSLs are found within these distinct phase diagrams; the CSLs labeled 1 to 5 live in the $J_{z\pm} = 0$  phase diagram while the CSLs 6 to  9 live in the phase diagram of $J_{z\pm} \neq 0$.
Except for spin ice (CSL 1), all other CSLs reside on the phase boundaries of magnetic orders. 
However, not all phase boundaries separating classical 
pseudospin long-range orders host CSLs. 

Of the nine CSLs identified, several, such as spin ice and the Heisenberg antiferromagnetic (HAFM) model, have been individually explored in prior research. 
However, this study is the first to make a comprehensive study of their effective field theory, phase diagram, and transitions among them. 
Notably, the  CSLs 6 and 9 have not been identified in previous literature, and we discuss below the properties of these two CSLs in some detail.

\section{Irreducible representations and inter-tetrahedra constraints}
\label{Sec:irrep.and.inter.tetra}

\subsection{Diagonalization of the single tetrahedron Hamiltonian}
\label{subsec:single.tetra.ham}

Every nearest-neighbor bond in the pyrochlore lattice belongs uniquely to one tetrahedron ($A$ or $B$ in Fig~\ref{fig:pyrochlore_lattice_unit_cell}). 
The nearest-neighbor Hamiltonian  $H$ in Eq.~\eqref{eqn:single.ion} can therefore be expressed as a sum of terms over individual tetrahedra~\cite{Yan-2017} which include a spin-spin exchange interaction term, $H_{\sf ex}(t)$ and, for generality's sake, a single-ion term, $H_{\sf si}(t)$.
We thus write $H$ as
\beq
H=\sum_t H_t, \ \ H_t=H_{\sf ex}(t) + \frac{1}{2} H_{\sf si} (t),
\eeq
where $H_{\sf ex}(t)$ and $H_{\sf si}(t)$ are the exchange and single-ion Hamiltonians acting on a single tetrahedron, $t$, 
Here, the factor of $1/2$ arises to avoid double-counting of the single ion term, since sites are shared by two tetrahedra. 

\begin{figure}[ht!]
    \centering
    \begin{overpic}[width=0.9\columnwidth]{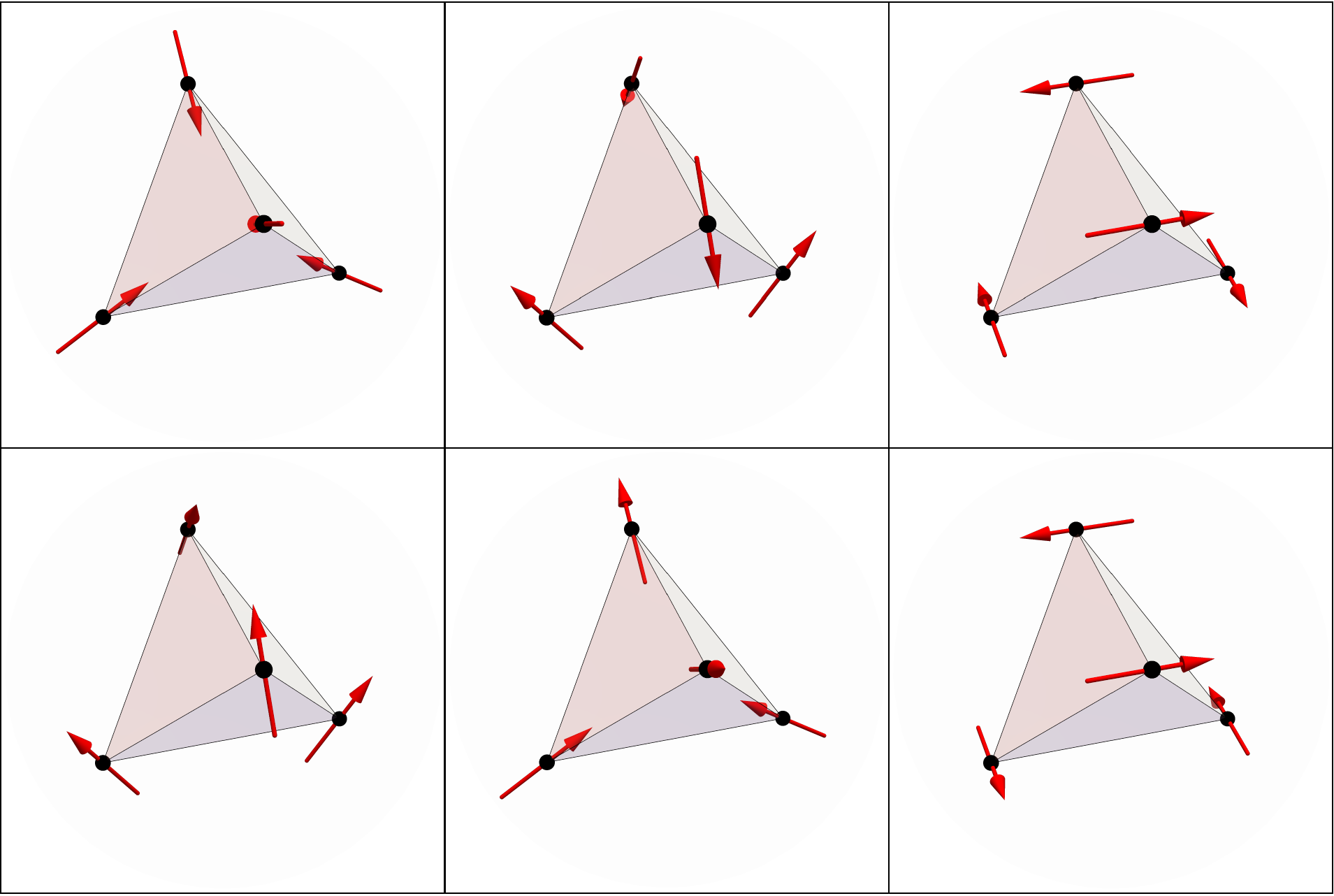}
    \put(2,63){\fontsize{8}{38}$\sf{A_2}$}
    \put(35,63){\fontsize{8}{38}$\sf{E}(\psi_2)$}
    \put(2,29){\fontsize{8}{38} $\sf T_{1,p}$}
    \put(35,29){\fontsize{8}{38} $\sf T_{1,i}$}
    \put(67,63){\fontsize{8}{38} $\sf{E}(\psi_3)$}
    \put(67,29){\fontsize{8}{38} $\sf{T_2}$}
    \end{overpic}
    \caption{Spin configuration of the irreducible representation configurations. 
    \dlgnew{Here, each panel shows a single component of the labeled irrep, i.e. $m^{\alpha}_\mathsf{x}$. 
    For the three dimensional ($\sf T_{1,p}$, $\sf T_{1,i}$ and 
    $\sf T_{2}$)   irreps, this corresponds to one of the components of the corresponding order parameter given in Table~\ref{Table_irrep_local}.
    For the $\mathsf{E}$ irrep, we display both components of the $\vb*{m}_{\mathsf{E}}$ field, $\sf{E}(\psi_2)$ and $\sf{E}(\psi_3)$.}
    }
    \label{fig:irreps}
\end{figure}

The single-tetrahedron Hamiltonian $H_t$ can then be re-expressed as a sum of quadratic terms:
\begin{equation}
\label{eqn:tetra.hamiltonian}
\begin{split}
        H_t= \frac{1}{2} \bigg[& a_{\sf A_2} m_{\sf A_2}^2 + a_{\sf E} {\vb*{m}}_{\sf E}^2 + a_\mathsf{T_2} {\vb*{m}}_\mathsf{T_2}^2 
   +  \bigl (   a_{\sf T_{1,i}} {\vb*{m}}_{\sf T_{1, i}}^2
    \\
  & 
  +  a_{\sf T_{1,p}} {\vb*{m}}_{\sf T_{1, p}}^2
  +  a_{\sf T_{1,ip}} {\vb*{m}}_{\sf T_{1, i}} \cdot  {\vb*{m}}_{\sf T_{1, p}} \bigr ) \bigg],
\end{split}
\end{equation} 
where each ${\vb*{m}}_\sfX$ is a local order parameter for different kinds of magnetic configuration corresponding to a particular irreducible representation (irrep) of $T_d$, the symmetry group of a tetrahedron, see Fig.~\ref{fig:irreps}. \\

We use the ${\vb*{m}}_\sfX$ notation to represent both the vector ${\vb*{m}}_\sfX$ irreps and scalar  ${m}$ irreps for simplicity of  notation. 
Each ${\vb*{m}}_\sfX$ is defined by a specific linear combination of the spin components, which are given in Table~\ref{Table_irrep_local}. The labels ${\sf A_2}, {\sf E}, {\sf T_1}, \mathsf{T_2}$ correspond to different irreps of the point group of a tetrahedron according to which the different fields transform, \dlgnew{where we have omitted the subscript $g$ for short hand notation. For completeness, we also provide an alternative commonly used naming convention for these spin modes, i.e. the labels $\Gamma_\delta$ with $\delta\in\{3,5,7,9\}$, which corresponds to the irreps obtained when considering the $Fd\bar{3}m$ space group instead of the point group $T_d$~\cite{Rau2019ARCMP}.}
The coefficients $a_\mathsf{X}$ are linear combinations of the exchange interaction parameters $\{J_{zz}, J_{\pm}, J_{\pm\pm}, J_{z\pm}\}$ and the single-ion anisotropy $D$, as listed in Table~\ref{Table_irrep_para_local}. \dlgnew{Henceforth, we refer to the set of interaction parameters $\{J_{zz}, J_{\pm}, J_{\pm\pm}, J_{z\pm}\}$ as the $\{J_{uv}\}$ parameters.}
\begin{table*}[ht!]
\def\arraystretch{1.7}%
\caption{\label{Table_irrep_local}
    Local order parameters $\vb*{m}_\mathsf{x}$ as irreducible representation (irrep) of $T_d$, the point group symmetry of a tetrahedron.
   \dlgnew{The first column labels the corresponding irrep naming convention according to the point group (space group) $T_d$ ($Fd\bar{3}m$).}
    The second column gives the corresponding magnetic phases.
    The third column lists the order parameter defined in terms of spins. Here the subscripts $\mathsf{i}$ and $\mathsf{p}$ refer to the ice and planar $\mathsf{T_1}$ ferromagnetic single tetrahedron configurations, respectively.
    }
    \vspace{10mm}
    \setlength{\tabcolsep}{10pt} %
\begin{tabular}{|c|c|c|}
\hline Symmetry & Phase &  Order parameter \\
\hline$A_{2 g}\left(\Gamma_3\right)  $& all-in/all-out & $m_\mathsf{A_{2}}  =  S_1^z+S_2^z+S_3^z+S_4^z$ \\\hline
$E_g\left(\Gamma_5\right)$& $\psi$ phase & $\vb*{m}_\mathsf{E}  =  \begin{pmatrix} S_1^x+S_2^x+S_3^x+S_4^x \\
S_1^y+S_2^y+S_3^y+S_4^y\end{pmatrix} $ \\
\hline  
$T_{2 g}\left(\Gamma_7\right)$& $\!\!\!\!$ Palmer-Chalker~\cite{Palmer-Chalker} $\!\!\!\!$  &
 $\! \!\!\! \! \! \vb*{m}_\mathsf{T_2} =   \!\begin{pmatrix} S_1^y+S_2^y-S_3^y-S_4^y \\ \left(-\frac{\sqrt{3}}{2} S_1^x-\frac{1}{2} S_1^y\right)-\left(-\frac{\sqrt{3}}{2} S_2^x-\frac{1}{2} S_2^y\right)+\left(-\frac{\sqrt{3}}{2} S_3^x-\frac{1}{2} S_3^y\right)-\left(-\frac{\sqrt{3}}{2} S_4^x-\frac{1}{2} S_4^y\right) \\ \left(\frac{\sqrt{3}}{2} S_1^x-\frac{1}{2} S_1^y\right)-\left(\frac{\sqrt{3}}{2} S_2^x-\frac{1}{2} S_2^y\right)-\left(\frac{\sqrt{3}}{2} S_3^x-\frac{1}{2} S_3^y\right)+\left(\frac{\sqrt{3}}{2} S_4^x-\frac{1}{2} S_4^y\right)\end{pmatrix} \!\! \!$
\\
\hline  
$T_{1 g}\left(\Gamma_9\right)$&  $\!\!\!\!$ spin ice $\!\!\!\!$  &
$\vb*{m}_\mathsf{T_{1,i}} =  \begin{pmatrix} S_1^z+S_2^z-S_3^z-S_4^z \\ S_1^z-S_2^z+S_3^z-S_4^z \\ S_1^z-S_2^z-S_3^z+S_4^z\end{pmatrix} $
\\\hline 
$T_{1 g}^{\prime}\left(\Gamma_9\right)$ & 
$\!\!\!\!$ \makecell{splayed \\ ferromagnet} $\!\!\!\!$  &  
$\! \!\!\!  \vb*{m}_\mathsf{T_{1,p}}  =   \!\begin{pmatrix} S_1^x+S_2^x-S_3^x-S_4^x \\ \left(-\frac{1}{2} S_1^x+\frac{\sqrt{3}}{2} S_1^y\right)-\left(-\frac{1}{2} S_2^x+\frac{\sqrt{3}}{2} S_2^y\right)+\left(-\frac{1}{2} S_3^x+\frac{\sqrt{3}}{2} S_3^y\right)-\left(-\frac{1}{2} S_4^x+\frac{\sqrt{3}}{2} S_4^y\right) \\ \left.-\frac{1}{2} S_1^x-\frac{\sqrt{3}}{2} S_1^y\right)-\left(-\frac{1}{2} S_2^x-\frac{\sqrt{3}}{2} S_2^y\right)-\left(-\frac{1}{2} S_3^x-\frac{\sqrt{3}}{2} S_3^y\right)+\left(-\frac{1}{2} S_4^x-\frac{\sqrt{3}}{2} S_4^y\right)\end{pmatrix}
\!\!\!  $\\
\hline
\end{tabular} 
\end{table*}
\begin{table}[th!]
\def\arraystretch{1.5}%
\caption{\label{Table_irrep_para_local}
    Coefficients of the irreducible representations in Eq.~\eqref{eqn:tetra.hamiltonian} in terms of exchange parameters and single-ion anisotropy in Eq.~\eqref{eq:Hex1}.
    }
    \vspace*{5mm}
\begin{tabular}{|c|c|}
\hline Coefficient & \begin{tabular}{c} 
Definition in terms of exchange \\
parameters $\left\{J_{z z}, J_{ \pm}, J_{ \pm \pm}, J_{z \pm}, D \right\}$
\end{tabular} \\
\hline$a_{\mathsf{A_2}}$ & $3 J_{z z} +D$ \\
\hline$a_{\mathsf{E}}$  & $-6 J_{ \pm}$ \\
\hline$a_{\mathsf{T_2}}$  & $2 J_{ \pm}-4 J_{ \pm \pm}$ \\
\hline$a_{\mathsf{T_{1, i }}}$ & $-J_{z z}+D$ \\
\hline$a_{\mathsf{T_{1, p }}}$ & $2 J_{ \pm}+4 J_{ \pm \pm}$ \\
\hline$a_{\mathsf{T_{1, ip}}}$ & $-8 J_{z \pm}$ \\
\hline
\end{tabular}
\end{table}

The coupling between ${\vb*{m}}_{\sf T_{1,i}}$ and ${\vb*{m}}_{\sf T_{1,p}}$, which both transforms according to $\sf T_1$, can be removed by a parameter-dependent redefinition of the fields.
Specifically, we define:
\begin{align}
& {\vb*{m}}_\mathsf{T_{1,-}} = \cos\!{\phi} \; {\vb*{m}}_{\sf T_{1,i}} +\sin\!\phi \; {\vb*{m}}_{\sf T_{1,p}} \label{eq:T1-_def},\\
& {\vb*{m}}_\mathsf{T_{1,+}} = -\sin\!\phi \; {\vb*{m}}_{\sf T_{1,i}} +\cos\!\phi \; {\vb*{m}}_{\sf T_{1,p}} \label{eq:T1+_def},
\end{align}
where 
\begin{align}
&{\bf v}_{-} = \begin{pmatrix} v_{i-} \\ 
v_{p -} \end{pmatrix} \equiv \begin{pmatrix} \cos\phi \\ 
\sin \phi \end{pmatrix}   , \\
&{\bf v}_{+} = \begin{pmatrix}  v_{i+} \\ 
v_{p +} \end{pmatrix} \equiv \begin{pmatrix} -\sin\phi \\ 
\cos \phi \end{pmatrix} , 
\end{align}
are the normalized eigenvectors of the $2 \times 2$
matrix
$$
\begin{pmatrix} a_{\sf T_1, i} & \frac{1}{2} a_{\sf T_1, ip} \\
\frac{1}{2} a_{\sf T_1, ip} & a_{\sf T_1, p} 
\end{pmatrix},
$$
with eigenvalues
\begin{align}
& a_{\sf T_1, -}  = 
\frac{1}{2} 
\left ( a_{\sf T_1, i}+a_{\sf T_1, p}- \sqrt{(a_{\sf T_1, i}-a_{\sf T_1, p})^2 + a_{\sf T_1, ip}^2} \; \right), \nonumber \\
\\
& a_{\sf T_1, +} = \frac{1}{2} 
\left ( a_{\sf T_1, i}+a_{\sf T_1, p} + \sqrt{(a_{\sf T_1, i}-a_{\sf T_1, p})^2 + a_{\sf T_1, ip}^2} \; \right ), \nonumber \\
\end{align}
respectively.
Here, $\phi$ corresponds to the mixing angle between $\bfm_\mathsf{T_{1,p}}$ and $\bfm_\mathsf{T_{1,i}}$ due to non-zero $J_{z\pm}$,
\[
\phi = \frac{1}{2}\arctan \left( \frac{a_{\sf T_{1,ip}}}{ a_\mathsf{T_{1,i}} - a_\mathsf{T_{1,p}} }\right) .
\label{eq:phi_def}
\] 
With this redefinition of the $\sf T_1$ fields, the diagonalized Hamiltonian for a single tetrahedron becomes:
\begin{equation}
\begin{split}
    H_t=\frac{1}{2}&\bigg[ a_{\sf A_2} m_{\sf A_2}^2 + a_{\sf E} {\vb*{m}}_{\sf E}^2 + a_\mathsf{T_2} {\vb*{m}}_\mathsf{T_2}^2 \\
    &+  a_{\sf T_{1,-}} {\vb*{m}}_{\sf T_{1, -}}^2
  +  a_{\sf T_{1,+}} {\vb*{m}}_{\sf T_{1, +}}^2\bigg], 
\end{split} 
\label{eq:Ht-diagonal}
\end{equation}
the sum of the squares of the ${\vb*{m}}_\sfX$ fields is
constrained as a result of the spin normalization:
\begin{equation}
    m_{\sf A_2}^2 + {\vb*{m}}_{\sf E}^2 + {\vb*{m}}_\mathsf{T_2}^2 
   +   {\vb*{m}}_{\sf T_{1, -}}^2
  + {\vb*{m}}_{\sf T_{1, +}}^2 = 16 S^2
\label{eq:totalm2}
\end{equation}
where $S\equiv |\vb*{S} |$ is the spin length.\\

One begins by identifying which term in Eq.~\eqref{eq:Ht-diagonal} has the lowest coefficient $a_\mathsf{X}$'s and use it as a baseline measure by subtracting it from all the other $a_\mathsf{Y}$'s coefficients.
Actually, there can be one or more $a_\mathsf{X}$ being \emph{simultaneously} minimal while also satisfying the spin normalization constrained in Eq.~\eqref{eq:totalm2}. 
Defining the minimal irrep coefficient as 
\begin{align}
    \mathcal{E}_0 \equiv {\rm min} \left( a_{\sf A_2}, a_{\sf E}, a_\mathsf{T_2}, a_{\sf T_{1-}}, a_{\sf T_{1+}} \right),\label{eq:minimum_energy_E0}
\end{align}
we then add and subtract $16\mathcal{E}_0S^2$ from Eq.~\eqref{eq:Ht-diagonal} and use Eq.~\eqref{eq:totalm2} to obtain:
\begin{align}
    H_t= &\; \frac{1}{2}\bigg[ 
    (a_{\sf A_2} - \mathcal{E}_0) m_{\sf A_2}^2 + (a_{\sf E} - \mathcal{E}_0) {\vb*{m}}_{\sf E}^2 \nonumber \\
    &\quad+ %
     (a_\mathsf{T_2}- \mathcal{E}_0) {\vb*{m}}_\mathsf{T_2}^2 
   +  (a_{\sf T_{1,-}}- \mathcal{E}_0) {\vb*{m}}_{\sf T_{1, -}}^2 \nonumber \\
   &\quad+ %
   (a_{\sf T_{1,+}}- \mathcal{E}_0) {\vb*{m}}_{\sf T_{1, +}}^2 + 16 S^2 \mathcal{E}_0 \bigg].
  \label{eq:Ht-diagonal-def}
\end{align}
Each multiplicative prefactor in Eq.~\eqref{eq:Ht-diagonal-def} is either vanishing (when $a_\mathsf{X}=\mathcal{E}_0$) or strictly positive.
From that equation, we can directly read off the ground state constraints for any  $\left\{J_{z z}, J_{ \pm}, J_{ \pm \pm}, J_{z \pm}, D \right\}$ parameter set:
\begin{align}
& \sum_\sfX {\vb*{m}}_\mathsf{X}^2 = 16S^2 \  \forall\ \mathsf{X} \ \ 
    \text{such that} \ a_\mathsf{X}=\mathcal{E}_0;
    \label{eq:constraint_general_X}
    \\
   \nonumber \\
    & {\vb*{m}}_\mathsf{Y}=0 \ \forall \ \mathsf{Y} \ \ 
    \text{such that} \ a_\mathsf{Y}>\mathcal{E}_0.
    \label{eq:constraint_general}
\end{align}
These constraints are applied to every tetrahedra $t$. 
In the zero temperature ($T\to 0$) limit, the components of the $\mathsf{X}$ irreps are free to fluctuate and these therefore correspond to the  
fluctuating DOFs that remain at low $T$.
On the other hand, the $\mathsf{Y}$ irreps become ultimately frozen and introduce constraints on the ground state spin configurations. 
If there are enough DOFs among the  ${\vb*{m}}_\mathsf{X}$ fields, then the first condition [Eq.~\eqref{eq:constraint_general_X}] does not sufficiently pin down the value of each ${\vb*{m}}_\mathsf{X}$ to induce magnetic order. 
In such a case, we may treat ${\vb*{m}}_\mathsf{X}$'s as almost freely fluctuating at the level of a single tetrahedron.
As we discuss in Subsection \ref{subsec:inter} below, 
Eq.~\eqref{eq:constraint_general} will be used in conjunction with the inter-tetrahedra constraints defined by Eq.~\eqref{eq:constraint_general_X} to derive the Gauss's laws that emerge in the system and control its ground state properties.

\subsection{Inter-tetrahedra constraints and coarse-graining}
\label{subsec:inter} 

Equation~\eqref{eq:constraint_general} establishes a generalized set of constraints which, in a ground state, must be fulfilled on every tetrahedra of the lattice.
The problem of fulfilling these constraints is rendered nontrivial by the fact that these are not independent: neighboring tetrahedra share a spin.
So the ground state condition on one tetrahedron, say of type $A$, must also be respected by its four neighboring tetrahedra, say of type $B$.

To understand the consequences of this, it is useful to divide the pyrochlore tetrahedra into two sets of tetrahedra, labeled $A$ and $B$, such that the neighbors of an $A$ tetrahedron are all $B$ tetrahedra and vice versa, see Fig. \ref{fig:pyrochlore_lattice_unit_cell}. 
The centers of the $A$ and $B$ tetrahedra each form a pair of face-centered cubic (FCC) lattices, the two types together forming a bipartite diamond lattice. 

One can then place the $A$ tetrahedra into ground state configurations independently since they do not overlap with each other.
Specifying a ground state configuration on the $A$ tetrahedra amounts to setting the values of all the ${\vb*{m}}_\mathsf{Y} = 0$ according to the energetic constraint given by Eq.~\eqref{eq:constraint_general} (i.e. those for which $a_\mathsf{Y} > \mathcal{E}_0$), while freely choosing a certain combination of the other ${\vb*{m}}_\sfX$ fields, up to spin normalization constraints.
Since every spin is included in exactly one $A$ tetrahedron, and since the map between the ${\vb*{m}}$ fields and the spins is invertible, this construction completely determines the allowed spin configurations in the $A$ tetrahedra when these are taken independently.

However, the ${\vb*{m}}_\sfX$'s on $A$ tetrahedra have an additional constraint. 
In order to be a ground state, the spin configuration, or equivalently the ${\vb*{m}}_\sfX$'s fields, must also satisfy the constraints on the $B$ tetrahedra.
In other words, the constraint requirements on a $B$ tetrahedron constrains the ${\vb*{m}}_\sfX$ fields on the four $A$ tetrahedra surrounding it~\footnote{ 
The mutual constraint between the $A$ and $B$ tetrahedra can be seen more clearly by writing the four spins belonging to the B tetrahedron in terms of the values of the  ${\vb*{m}}_\sfX$ fields on the four surrounding $A$ tetrahedra. }. 
Using the relationship between the spins in the $B$ tetrahedron and the ${\vb*{m}}_\sfX$ fields on the $A$ tetrahedra, the constraint on the ${\vb*{m}}_\sfX$'s on the $B$ tetrahedra can be expressed in terms of the fields on the surrounding $A$ tetrahedra. 
On each $B$ tetrahedron, one obtains a set of equations of the following general form:
\begin{eqnarray}
     &\vb*{m}_\sfY ({\vb*{R} }_B)& = 0\nonumber\\ 
        & \implies& \sum_{i=1}^{4} \sum_{\sfX  \in {\rm gs}} 
    \sum_{\alpha} c_{i \alpha \sfX } \,
    {m}^{\alpha}_\sfX (\vb*{R}_B+\vb*{u}_i)=0,
     \label{eq:Bconstraint}
\end{eqnarray}
where, on the second line, the $\vb*{u}_i$'s  are the FCC lattice vectors connecting the center of each $B$ tetrahedron to the centers of the four neighboring $A$ tetrahedra. In this exact expression, the sum taken over the $\sfX$ irreps is performed over those $\sfX$  in the $A$ tetrahedra that are not constrained to vanish according to Eq.~\eqref{eq:constraint_general},
with sum over $\alpha$ running over all components of the $\vb*{m}_\sfX$ fields.
The $c_{i \alpha \sfX}$'s are constant (transformation)  coefficients obtained by expressing the $B$ tetrahedra ${\vb*{m}}_\sfX$ fields in terms of those on the $A$ tetrahedra. 
The specific value of these $c_{i \alpha \sfX}$ %
can be worked out using the linear relations between the fields ${\vb*{m}}_\sfX$ and the spins $\vb*{S}$ using Table~\ref{Table_irrep_local}~\footnote{A spin $\vb*{S}_i$ shared by two adjacent $A$ and $B$ tetrahedra is written in terms of the $\vb*{m}(\vb*{R}_B)$ of the $B$ tetrahedra, i.e.~$\vb*{S}_i [\vb*{m}(\vb*{R}_B)]$. 
Then, using the definition of the $\vb*{m}(\vb*{R}_A)$ fields for the $A$ tetrahedron provided in Table~\ref{Table_irrep_local}, a relationship between the fields on the $A$ and $B$ tetrahedra is obtained.}.

The  constraints on the $\vb*{m}^{\alpha}_\sfX$  prescribed by Eq.~\eqref{eq:Bconstraint} can be expressed into a form reminiscent of Gauss's law by applying a coarse-graining procedure, i.e. 
\begin{equation}
 \begin{split}
  & \sum_{i=1}^{4} \sum_{\sfX  \in {\rm gs}} 
    \sum_{\alpha} c_{i \alpha \sfX }
    {m}^{\alpha}_\sfX (\vb*{R}_B+\vb*{u}_i)=0\\
   & \xrightarrow{\text{coarse-grained}}
      \sum_{\sfX  \in {\rm gs}} 
    \sum_{\alpha} D_{ \alpha \sfX }
    {m}^{\alpha}_\sfX=0,
\end{split}
    \label{eq:Bconstraint_1}   
\end{equation}
where $D_{ \alpha \sfX }$ are a generalized set of derivatives labeled by the irrep component upon which these are applied. 

To derive such a continuum description, we assume, for each ${\vb*{m}}_\sfX$, the existence of some underlying smoothly varying field ${\vb*{m}}_\sfX(\bf{r})$ defined at all points in space (here we slightly abuse the notation to use ${\vb*{m}}_\sfX$ to refer to both the irreps on the discrete lattice and the coarse-grained fields).
We further suppose that if a smoothly varying field 
${\vb*{m}}_\sfX(\bf{r})$ is evaluated at the center of an $A$ tetrahedron, it takes the value of ${\vb*{m}}_\sfX$ on that tetrahedron center.
Then, assuming slow variations of the fields, we approximate $ {m}^{\alpha}_{\sfX} ({\vb*{R}}_B-{\vb*{u}}_i)$
 in Eq.~\eqref{eq:Bconstraint} by a low-order gradient expansion of $ {\vb*{m}}_{\sfX}(\bf{r})$ about  ${\bf r}={\vb*{R}}_B$. Keeping the lowest order terms in the resulting expansion of Eq.~\eqref{eq:Bconstraint} establishes the effective Gauss's law(s) governing the long-wavelength fluctuations of the   ${\vb*{m}}_\sfX(\bf{r})$ fields within the ground state.

Working through this procedure yields the following set of correspondences between the microscopic constraints on the $B$ tetrahedra and the constraints on the derivatives of the coarse-grained fields:

\begin{widetext}
\begin{align}
m_{\sf A_2}=0 \implies & m_{\sf A_2}=  -\frac{a_d}{ \sqrt{3}}\sin\phi
\vb*{\nabla} \cdot {{\vb*{m}}}_\mathsf{T_{1,+}}+\frac{a_d}{ \sqrt{3}}\cos\phi
\vb*{\nabla} \cdot {{\vb*{m}}}_\mathsf{T_{1,-}}=0 , \label{eq:A2constraint}   \\
\nonumber\\
{\vb*{m}}_{\sf E}=0 \implies &
{\vb*{m}}_{\sf E}=  \frac{a_d}{ 2\sqrt{3}} \cos\phi
\begin{pmatrix}
  2 \partial_x {m}^x_\mathsf{T_{1,+}}
-\partial_y {m}^y_\mathsf{T_{1,+}} 
-\partial_z {m}^z_\mathsf{T_{1,+}}   \\
 \sqrt{3}(\partial_y {m}_\mathsf{T_{1,+}}^y
-\partial_z {m}_\mathsf{T_{1,+}}^z)
\end{pmatrix} 
+  \frac{a_d}{ 2\sqrt{3}}  \sin\phi\begin{pmatrix}
 2 \partial_x {m}^x_\mathsf{T_{1,-}}
-\partial_y {m}^y_\mathsf{T_{1,-}} 
-\partial_z {m}^z_\mathsf{T_{1,-}}   \\
\sqrt{3} (\partial_y {m}_\mathsf{T_{1,-}}^y
-\partial_z {m}_\mathsf{T_{1,-}}^z)
\end{pmatrix} 
\nonumber \\
&   \hphantom{{\vb*{m}}_{\sf E}=}
+ \frac{a_d}{ 2\sqrt{3}} 
\begin{pmatrix}
\sqrt{3} (
-\partial_y {m}_\mathsf{T_2}^y+\partial_z {m}_\mathsf{T_2}^z) \\
 2 \partial_x {m}^x_\mathsf{T_2}
-\partial_y {m}^y_\mathsf{T_2} 
-\partial_z {m}^z_\mathsf{T_2} 
\end{pmatrix}  
=0 ,
\label{eq:Econstraint}  \\
\nonumber\\
{\vb*{m}}_\mathsf{T_2}=0 \implies &
 {\vb*{m}}_\mathsf{T_2}=   \frac{a_d}{ 2\sqrt{3}} \begin{pmatrix}
2\partial_x {m}_{\sf E}^2 \\
 -  \sqrt{3}  \partial_y {m}_{\sf E}^1  - \partial_y {m}_{\sf E}^2\\
 \sqrt{3}  \partial_z {m}_{\sf E}^1  - \partial_z {m}_{\sf E}^2   
\end{pmatrix}
+ \frac{a_d}{ 2 }   \cos\phi \vb*{\nabla} \times \vb*{m}_\mathsf{T_{1,+}} +
 \frac{a_d}{ 2 } \sin\phi \vb*{\nabla} \times \vb*{m}_\mathsf{T_{1,-}}=0,
\label{eq:T2constraint}  \\
\nonumber\\
{\vb*{m}}_\mathsf{T_{1,-}}=0 \implies&
{\vb*{m}}_\mathsf{T_{1,-}}=  \frac{a_d}{  \sqrt{3}}   \cos\phi \vb*{\nabla} {m}_{\sf A_2}
+ \frac{\sqrt{3} a_d}{ 2 } \sin\phi\cos\phi 
\begin{pmatrix}
\partial_y {m}_\mathsf{T_{1,+}}^z + \partial_z {m}_\mathsf{T_{1,+}}^y \\
\partial_z {m}_\mathsf{T_{1,+}}^x + \partial_x {m}_\mathsf{T_{1,+}}^z \\
\partial_x {m}_\mathsf{T_{1,+}}^y + \partial_y {m}_\mathsf{T_{1,+}}^x 
\end{pmatrix}  \nonumber \\
& \hphantom{{\vb*{m}}_\mathsf{T_{1,-}}=}  + \frac{a_d}{ 2 \sqrt{3}} \sin\phi  \begin{pmatrix}
2 \partial_x {m}_{\sf E}^1 \\
- \partial_y {m}_{\sf E}^1 +  \sqrt{3} \partial_y {m}_{\sf E}^2 \\
- \partial_z {m}_{\sf E}^1 -  \sqrt{3}  \partial_z
{m}_{\sf E}^2
\end{pmatrix}
+ \frac{a_d}{ 2  }    \sin\phi\vb*{\nabla} \times \vb*{m}_\mathsf{T_2} 
=0,
\label{eq:constraint-}  \\
\nonumber\\
{\vb*{m}}_\mathsf{T_{1,+}}=0  \implies &
{\vb*{m}}_\mathsf{T_{1,+}}=    -\frac{a_d}{   \sqrt{3} } \sin\phi  \vb*{\nabla} {m}_{\sf A_2}
 -\frac{ \sqrt{3} a_d}{ 2}\sin\phi\cos\phi 
\begin{pmatrix}
\partial_y {m}_\mathsf{T_{1,-}}^z + \partial_z {m}_\mathsf{T_{1,-}}^y \\
\partial_z {m}_\mathsf{T_{1,-}}^x + \partial_x {m}_\mathsf{T_{1,-}}^z \\
\partial_x {m}_\mathsf{T_{1,-}}^y + \partial_y {m}_\mathsf{T_{1,-}}^x 
\end{pmatrix} \nonumber \\
&  \hphantom{{\vb*{m}}_\mathsf{T_{1,+}}=}
+ \frac{a_d}{ 2 \sqrt{3} } \cos\phi   \begin{pmatrix}
2 \partial_x {m}_{\sf E}^1 \\
- \partial_y {m}_{\sf E}^1 +  \sqrt{3}   \partial_y {m}_{\sf E}^2 \\
- \partial_z {m}_{\sf E}^1 -  \sqrt{3}  \partial_z
{m}_{\sf E}^2
\end{pmatrix} 
 + \frac{a_d}{ 2   }   \cos\phi \vb*{\nabla} \times \vb*{m}_\mathsf{T_2} 
=0.
\label{eq:constraint+}  
\end{align}
\end{widetext}
Here, $a_d$ is the distance between two neighboring tetrahedron centers. 
We first give here the \emph{full} irrep-dependence of the lowest-order gradient expansion of a given field $\vb*{m}_\sfY$ in terms of all the remaining $\vb*{m}_\sfX$ fields, \emph{and not just} those that are frozen in the ground-state manifold.

In practice, Gauss's laws for a particular system are obtained by only considering the differential equations resulting from the frozen $\vb*{m}_\mathsf{Y}$ fields on the left-hand sides of  Eq.~\eqref{eq:A2constraint}-\eqref{eq:constraint+}  while only keeping terms involving the fluctuating $\vb*{m}_\mathsf{X}$ fields on the right-hand sides.
In the next subsection, we provide an example of this procedure for the classical spin ice spin liquid.

\subsection{The non-Kramers case}
\label{subsec:non.kramer.inter}

\dlgnew{To set the stage for the disucssion of CSLs in  Sec.~\ref{sec:CSL_non_kramers} and Sec.~\ref{sec:CSL_kramers}, we discuss here the case of non-Kramers ions.
Thanks to their effective spin Hamiltonian that has $J_{z\pm}=0$, the irrep analysis and inter-tetrahedra constraints take simpler forms than for Kramers ions.
Considering non-Kramers ions first will allow us to illustrate the  methodology  by referring step-by-step to the simplest case, that  of the Gauss's law arising in spin ice.}

For the non-Kramers ions, and as discussed in Sec.~\ref{sec:model_results}, $J_{z\pm} = 0$, so that $a_\mathsf{T_{1,ip}} =0$ (see Table \ref{Table_irrep_para_local}).
As a consequence, the irreps $\vb*{m}_\mathsf{T_{1,i}}$ and $\vb*{m}_\mathsf{T_{1,p}}$ do not mix, hence $\phi=0$ [see Eq.~\eqref{eq:phi_def}].
The inter-tetrahedra constraints then take a rather compact form 
\begin{widetext}
\begin{align}
\label{eq:nonK.constraintA2}
    m_\mathsf{A_2}= 0 & \implies  
    m_\mathsf{A_2}= \frac{a_d}{\sqrt{3}} \vb*{\nabla} \cdot \vb*{m}_\mathsf{T_{1,i}} = 0,\\
 \nonumber\\
 	\bfm_\mathsf{E}  = 0 &   \implies 
	\bfm_\mathsf{E} =
	\frac{a_d}{2\sqrt{3}}\begin{pmatrix}
		    2\partial_x m_\mathsf{T_{1,p}}^x  -  \partial_y m_\mathsf{T_{1,p}}^y  -  \partial_z m_\mathsf{T_{1,p}}^z    \\
		  \sqrt{3}  ( \partial_y m_\mathsf{T_{1,p}}^y  -  \partial_z m_\mathsf{T_{1,p}}^z  )
	\end{pmatrix}	+ \frac{a_d}{2\sqrt{3}}
	\begin{pmatrix} 
   \sqrt{3}   ( -  \partial_y m_\mathsf{T_2}^y  + \partial_z m_\mathsf{T_2}^z ) 
  \\
     2 \partial_x m_\mathsf{T_2}^x  -  \partial_y m_\mathsf{T_2}^y  -  \partial_z m_\mathsf{T_2}^z   
	\end{pmatrix}
=0,  \label{eq:double_U1}\\
 \nonumber\\
 \bfm_\mathsf{T_2}  = 0 & \implies
	\bfm_\mathsf{T_2}  = \frac{a_d}{2\sqrt{3}}
     \begin{pmatrix}
		2 \partial_x  m_\mathsf{E}^2  \\
		-  \sqrt{3}  \partial_y m_\mathsf{E}^1   -  \partial_y m_\mathsf{E}^2 \\
		\sqrt{3}  \partial_z m_\mathsf{E}^1 -  \partial_z m_\mathsf{E}^2   
	\end{pmatrix}
	+ \frac{a_d}{2 }    \vb*{\nabla}\times \bfm_\mathsf{T_{1,p}} = 0, \\
 \nonumber\\
    \bfm_\mathsf{T_{1,i}}  = 0   & \implies 
    \bfm_\mathsf{T_{1,i}}  = 
	  \frac{a_d}{\sqrt{3}}  \vb*{\nabla} m_\mathsf{A_2}  = 0,  \\
 \nonumber\\
 \label{eq:nonK.constraint+}
 \bfm_\mathsf{T_{1,p}}  = 0  & \implies 
 \bfm_\mathsf{T_{1,p}}  =
	  \frac{a_d}{2\sqrt{3}}
	\begin{pmatrix}
		2 \partial_x  m_\mathsf{E}^1  \\
		-  \partial_y m_\mathsf{E}^1 +  \sqrt{3}  \partial_y m_\mathsf{E}^2  \\
		-  \partial_z m_\mathsf{E}^1 -  \sqrt{3}  \partial_z m_\mathsf{E}^2
	\end{pmatrix}
	+  \frac{a_d}{2 }  \vb*{\nabla}\times \bfm_\mathsf{T_2} = 0. 
\end{align}
\end{widetext}
 
Before proceeding, we note that for the non-Kramers Hamiltonian, the local $z$ spin degrees of freedom transform as (time-odd) magnetic dipoles while the local $xy$ spin degrees of freedom transform as (time-even) magnetic quadrupoles~\cite{leeGenericQuantumSpin2012}.
This forbids a bilinear mixing of these degrees of freedom, imposing $J_{z\pm}=0$. 
This restriction is further reflected in Eq.~\eqref{eq:nonK.constraintA2}-\eqref{eq:nonK.constraint+} where each constraint results in differential equations involving fields constructed by the local $z$ spin degrees of freedom (the $\mathsf{A_2}$ and the $\mathsf{T_{1,i}}$ fields) \emph{or}  by the local $xy$ spin degrees of freedom (the $\mathsf{E}$, the $\mathsf{T_{1,p}}$, and the $\mathsf{T_{2}}$ fields), with those two sets being decoupled.
This distinction allows us to further classify the constraints in Eq.~\eqref{eq:nonK.constraintA2}-\eqref{eq:nonK.constraint+} into two independent sets of constraints.

For illustration purposes, let us take Eq.~\eqref{eq:nonK.constraintA2} as an example to show how these field-theory constraints are derived.
We label the four sublattice sites on the $B$ tetrahedron to be  $0,1,2,3,$, and find
\[
m_\mathsf{A_2} = S_0^z+ S_1^z + S_2^z  +S_3^z.  
\]
The $B$ tetrahedron  is joined with an $A$ tetrahedron along the $[111]$ cubic direction by the spin $\vb*{S}_0$. 
We see that on the $A$ tetrahedron, $S_0^z$ can  be expressed as linear combinations of $\bfm$ fields on that $A$ tetrahedron as
\[\label{eqn.}
S_0^z = \frac{1}{4}\left(m_\mathsf{A_2} + [111]\cdot\bfm_\mathsf{T_{1,i}}  \right) .
\]
Repeating a similar manipulation for the other spins $\vb*{S}_{1,2,3}$ results in the follow expression relating $m_\mathsf{A_2}$ on a $B$ tetrahedron, $m_\mathsf{A_2}(\vb*{R}_B)$, to the neighboring four $A$ tetrahedra, namely
\[
\begin{split}
  &m_\mathsf{A_2}(\vb*{R}_B ) = \\  
  &\frac{1}{4} \sum_{i}\left(m_\mathsf{A_2}(\vb*{R}_B+\vb*{u}_i)  +\frac{a_d}{\sqrt{3}}  \vb*{\hat{u}}_i \cdot \bfm_\mathsf{T_{1,i}}  (\vb*{R}_B+\vb*{u}_i)\right), 
\end{split}
\]
Then, applying the constraint $m_\mathsf{A_2} =0$ and taking the ${\vb*{m}}_\sfX$ fields to the continuous limit, we obtain
\[
m_\mathsf{A_2}  = 0 \implies   \vb*{\nabla} \cdot \bfm_\mathsf{T_{1,i}} = 0,
\]
which gives Eq.~\eqref{eq:nonK.constraintA2}.
The other equations in Eqs.~(\ref{eq:A2constraint}-\ref{eq:constraint+}) and  Eqs.~(\ref{eq:nonK.constraintA2}-\ref{eq:nonK.constraint+}) can be derived in the same fashion by considering the other ${\vb*{m}}_\sfX$ fields on the left side of those equations one at a time, although the algebra is considerably more lengthy.

\subsection{Why coarse-grained field theory works}
\refthreenew{The expansion about the $\Gamma$ point is justified by two microscopic properties of the nearest-neighbor Hamiltonian. First, every nearest-neighbor bond belongs to a single tetrahedron, and hence the Hamiltonian is an exact sum of single-tetrahedron Hamiltonians. Any spin configuration that minimizes one tetrahedron can be repeated with the same four-sublattice pattern throughout the pyrochlore lattice. This places every tetrahedron in its local ground-state sector and saturates the global lower bound. An exact $\vb*{q}=\vb*{0}$ ground state therefore exists throughout the nearest-neighbor parameter space, independently of any continuum or mean-field approximation.

Second, the full-lattice interaction matrix, evaluated throughout the Brillouin zone, shows that for every CSL the bottom flat bands touch the relevant dispersive bands at $\Gamma$ [Figs.~\ref{fig:E_T2_T1planar_Jzpm_pos}(a), \ref{fig:E_T2_T1planar_Jzpm_neg}(a), and~\ref{fig:dual_HAFM}(a)]. This touching controls the leading long-wavelength correlations described by the gauge theory. The $\vb*{q}=\vb*{0}$ construction does not require modes at other momenta to be absent, nor does it assume that all spectral weight is concentrated at $\Gamma$; it identifies the soft sector controlling the $\vb*{q}\to\vb*{0}$ singularity. 
Moreover, the sublattice and spin projection entering a particular neutron-scattering structure factor can suppress the visibility of this singularity even when it remains present in the full correlation matrix. The absence of a visible pinch point in such a channel therefore does not imply the absence of the underlying $\Gamma$-point soft mode.

The coarse-grained constraints in Eqs.~\eqref{eq:A2constraint}--\eqref{eq:nonK.constraint+} are not introduced as independent phenomenological assumptions. Their starting point is the exact lattice compatibility relation in Eq.~\eqref{eq:Bconstraint} between the irrep fields on neighboring tetrahedra. In momentum space, this exact relation can be expanded systematically about the $\Gamma$ point in powers of $qa$, where $a$ is the lattice length scale. After imposing the locally frozen irrep components, Eqs.~\eqref{eq:A2constraint}--\eqref{eq:nonK.constraint+} are obtained by retaining the first nonvanishing terms in this expansion. The omitted contributions enter at higher orders in $qa$ and therefore become asymptotically small relative to the retained terms as $\vb*{q}\to\vb*{0}$.

The microscopic full-lattice spectrum shows that, whenever a CSL exists, the bottom flat bands touch the relevant dispersive bands at $\Gamma$. The gradient expansion is therefore applied precisely to the soft sector controlling the long-wavelength singular correlations. The resulting Gauss's laws and pinch-point structures are controlled statements in the regime $qa\ll1$; we do not assume that the truncated continuum theory describes the full Brillouin zone. Higher-gradient terms can modify nonuniversal finite-$q$ details, but not the leading long-wavelength constraint structure within the CSL regime. The comparisons with the full-lattice large-$\mathcal{N}$ calculation and CMC then independently test how accurately this asymptotic description captures the microscopic correlations at accessible length scales.}

In Appendix Sec.~\ref{app.theory.largeN}, we present a quantitative comparison between correlation functions computed from the field theory and the large-$\mathcal{N}$ theory, finding excellent agreement between the two descriptions.

\subsection{Large-$\bm{\mathcal{N}}$ approximation}

\label{appendix:large-N}

\dlgnew{The rewriting of the spin Hamiltonian  [Eq.~\eqref{eq:Hex1}] in terms of irreps and associated ${\vb*{m}}_\sfX$ fields, along with the latter's long-wavelength description and the Gauss's laws that emerge from the local constraints on the ${\vb*{m}}_\sfX$'s, will be shown to result, as illustrated  in Sec.~\ref{subsec:non.kramer.inter},  in a classical field theory description of all CSLs on the pyrochlore lattice at zero temperature (see Sec.~\ref{sec:CSL_non_kramers} and Sec.~\ref{sec:CSL_kramers} below).
For completeness, it is desirable to also consider a semi-analytical framework that (i) is defined at the lattice scale, (ii) makes possible calculations of the spin-spin correlation functions at nonzero temperature, thus providing a setting to compare with classical Monte Carlo results at nonzero temperature, and (iii) recovers the coarse-grained field theory description at zero temperature. 
To this effect, we employ the so-called large-$\mathcal{N}$ approximation.
}

\dlgnew{
This approximation, also known as the self-consistent Gaussian approximation (SCGA), is a classical approximation where the hard constraint on the spin length, $|\vb*{S}_i|^2=S^2$, is replaced by a soft, averaged spin-length constraint, $\langle |\vb*{S}_i|^2 \rangle =S^2$, which is energetically imposed by introducing a Lagrange multiplier $\lambda 
\sum_i\vb*{S}_i^2 $. 
This approximation results in a Gaussian theory that can be exactly solved~\cite{ConlonAbsentPhysRevB.81.224413,SCGA_Canals_kagome,SCGA_Canals_checker,SCGA_Canals_pyrochlore} and has the aforementioned properties, in particular recovering the key features of the effective long-wavelength theory a number of CSLs in the pyrochlore lattice~\cite{Benton2016NatComm,Yan2020PRL_Rank2,lozanogomez2023arxiv,Chung_flatband}. 
}

\dlgnew{We consider a generic bilinear anisotropic spin Hamiltonian, such as the one in Eq.~\eqref{eq:Hex1}, which can be written as
}
\begin{equation}
\mathcal{H}=\frac{1}{2}\sum_{i\mu,j\nu}\sum_{\alpha,\gamma}S_{i\mu}^\alpha\, \vb*{J}_{i\mu,j\nu}^{\alpha,\gamma}\,  S_{j\nu}^\gamma, \label{eq:general_Hamiltonian}
\end{equation}
\dlgnew{
where the sub-indices $i$ and $j$ label the primitive FCC vectors $\vb*{R}_i$, while $\mu$ and $\nu$ denote the sublattice basis. 
The $\vb*{q}$ space spin-spin correlation function, for $\alpha$ and $\gamma$ Cartesian spin components and for sublattices $\mu$ and $\nu$,
is given by }
\begin{equation}
   {\mathcal {S}}^{\alpha\gamma}_{\mu\nu}=\langle S_{\mu}^\alpha(\vb*{q}) S_{\nu}^\gamma(-\vb*{q})\rangle =
   (\beta \vb*{J}_{\mu\nu}^{\alpha\gamma}(\vb*{q})+\lambda )^{-1} .\label{eq:SCGA-chi}
\end{equation}
\dlgnew{Here, the  Lagrange multiplier coefficient  $\lambda$  must
be determined self-consistently at each temperature $T=1/\beta$ by solving the soft spin-length constraint which,
using Eq.~\eqref{eq:SCGA-chi}, reads in 
${\vb*{q}}$ space as
}
\begin{equation}
S^2=\frac{1}{N}\sum_{m,\vb*{q} }(\beta \varepsilon_m(\vb*{q})+\lambda)^{-1},
\label{eq:lagrange_multiplier_constraint}
\end{equation}
\dlgnew{
where $\varepsilon_m(\vb*{q})$ is the $m$-th eigenvalue of the interaction matrix
$\vb*{J}_{i\mu,j\nu}^{\alpha,\gamma}$ 
in Eq.~\eqref{eq:general_Hamiltonian}
expressed in momentum space (i.e., $\vb*{J}_{\mu\nu}^{\alpha\gamma}(\vb*{q})$).
}
\

\subsection{Step-by-step procedure for deriving Gauss's law}
\label{subsec:recipe}

\refthreenew{Our classification is complete for the general nearest-neighbor anisotropic pyrochlore Hamiltonian in Eqs.~\eqref{eq:Hex1} and~\eqref{eqn:single.ion}. It follows from the exact single-tetrahedron decomposition in Eq.~\eqref{eq:Ht-diagonal}. The coefficients $a_\mathsf{X}(\{J_{uv}\},D)$ multiply the complete set of tetrahedral irreps. For any microscopic parameter set, the locally allowed ground-state sectors are therefore exactly those for which $a_\mathsf{X}=\mathcal{E}_0\equiv\min_\mathsf{Y}a_\mathsf{Y}$, while every sector with $a_\mathsf{Y}>\mathcal{E}_0$ is frozen out of the ground-state manifold [Eqs.~\eqref{eq:constraint_general_X} and~\eqref{eq:constraint_general}].

We enumerate every realizable pattern of minimum coefficients: all single-irrep regions, including the finite-volume spin-ice phase, every pairwise-degeneracy hypersurface, and every lower-dimensional intersection involving three or more degenerate irreps. This is a finite algebraic enumeration performed before any $\vb*{q}=\vb*{0}$ expansion or large-$\mathcal{N}$ calculation. The corresponding phase diagrams are displayed in Figs.~\ref{fig:nonkramer.phase} and~\ref{fig:kramer.phase}.}

\refthreenew{The coarse-grained field theory, the full-lattice large-$\mathcal{N}$ calculation, and  CMC are distinct methods with different assumptions and complementary roles. The coarse-grained theory derives the long-wavelength gauge constraints from the microscopic inter-tetrahedra compatibility relations; the large-$\mathcal{N}$ calculation tests the spectrum and correlations on the full lattice while relaxing the local spin-length constraint; and CMC directly samples the original microscopic Hamiltonian with fixed-length three-component spins. Their agreement therefore provides nontrivial cross-checks rather than repeated applications of a single approximation. We retain a candidate in the atlas as a CSL only when it survives all three checks. Candidates that satisfy the coarse-grained and large-$\mathcal{N}$ criteria but develop symmetry-breaking order in CMC are excluded from the atlas and discussed in Sec.~\ref{sec:Unsuccessfuk_CSL}.}

We now use the irreps and inter-tetrahedra constraints to identify the CSLs that the model supports for non-Kramers ions.
We consider all combinations of coefficients $a_\mathsf{X}$'s that can be consistently minimal with the $a_\mathsf{Y}$'s simultaneously not minimal in the single-tetrahedron Hamiltonian. 
The procedure for constructing the field theory is then:
\begin{itemize}
    \item \emph{Step $\#1$}: Compute the $a_{\mathsf{X}}$ coefficients for all irrep fields and identify all the minimum energy irreps for which $a_{\mathsf{X}}=\mathcal{E}_0$ according to Eq.~\eqref{eq:minimum_energy_E0}. This will ultimately allow us to determine the fields fluctuating in the ground state.
    \item \emph{Step $\#2$}: Then, identify all $\bfm_\mathsf{Y} =0$ as constrained by Eq.~\eqref{eq:constraint_general} which become frozen and are thermally depopulated in the limit $T\to 0$.
    \item \emph{Step $\#3$}: Then, identify all $\bfm_\mathsf{X}$'s that are free to fluctuate.
    \item \emph{Step $\#4$}: Next, identify the constraints on the $\bfm_\mathsf{X}$'s  imposed by $\bfm_\mathsf{Y} =0$ frozen fields through the inter-tetrahedra relations  Eqs.~(\ref{eq:A2constraint}-\ref{eq:constraint+}) for the Kramers ions or  Eqs.~(\ref{eq:nonK.constraintA2}-\ref{eq:nonK.constraint+}), for the non-Kramers ions.
    \item \emph{Step $\#5$}: Finally, check that in the large-$\mathcal{N}$ theory, the $a_{\mathsf{X}}=\mathcal{E}_0$ parameters identified above yield low-energy flat bands reflecting a possible extensive ground-state degeneracy characteristic of a classical spin liquid phase.
    The number of flat band corresponds to the number of DOFs in the emergent electric fields minus the number of constraints.
\end{itemize}

In this procedure, steps $\#0$ to $\#3$ allow one to identify a set of constraints (Gauss's laws) for the allowed fluctuating fields, $\bfm_\mathsf{X}$, of a given model. Step $\#4$ then verifies that these constraints
\emph{may result} in a CSL by studying the degeneracy of the ground-state manifold in the large-$\mathcal{N}$ limit.
Subsequently, we performed CMC simulations  (reported in Sec.~\ref{sec:numerics}) for the lattice Hamiltonian in Eq.~\eqref{eq:Hex1} to find out  which of the sets of parameters identified above as candidate CSLs are indeed such (for the ${\mathcal N}=3$  model) down to the lowest temperature.

To illustrate further, let us discuss now a simple and concrete example to demonstrate how to execute the above protocol, considering specifically the classical spin ice spin liquid. 
This spin liquid is a non-Kramers case so $a_\mathsf{T_{1,ip}} =0$, and the irreps $\mTOnePL$ and $\mTOneICE$ do not mix. 
Following the procedure, we thus have:
\begin{itemize}
    \item \emph{Step $\#1$}: For this choice of $a_{\mathsf{X}}$ parameters, 
   we have that $\mathcal{E}_0=a_{\mathsf{T_{1,i}}}$ such that $a_{\mathsf{A_2}}$, $a_{\mathsf{E}}$,  $a_{\mathsf{T_{1,p}}}$ and $a_{\mathsf{T_2}}
   > \mathcal{E}_0$.
    \item  \emph{Step $\#2$}: $a_\mathsf{T_{1,i}}$ is minimal and the others $a_{\mathsf{X}}$ are not (ref. Table.~\ref{Table_irrep_para_local}) and, therefore, all other $\bfm_\mathsf{Y} =0$ for $\mathsf{Y}= 
    {\mathsf{A_2}}, {\mathsf{E}}, \mathsf{T_{1,p}}$ and
    $\mathsf{T_{2}}$. 
    \item \emph{Step $\#3$}: As stated just above, the only field able to thermally fluctuate in the present case is $\bfm_\mathsf{T_{1,i}}$. 
    \item \emph{Step $\#4$}: 
    While $\mTOneICE$ remains fluctuating, and despite a set of fields being frozen, the  allowed spin patterns associated with the latter constrain 
    $\mTOneICE$.
    Indeed, according to Eq.~\eqref{eq:nonK.constraintA2}, the field $\bfm_\mathsf{T_{1,i}}$ is constrained by $m_\mathsf{A_2} = 0 \implies \vb*{\nabla} \cdot \vb*{m}_\mathsf{T_{1,i}} = 0$. 
    This is exactly the well-known emergent Maxwell's Gauss's law for classical spin ice ~\cite{HenleyARCMP,castelnovoSpinIceFractionalization2012}, with the charge-free condition imposed for the ground state configurations.
    \item \emph{Step $\#5$}: Lastly, the band structure for this model yields two low-energy flat bands reflecting the extensive degeneracy associated with this spin liquid.
        This is because the emergent electric field, $\vb*{m}_\mathsf{T_{1,i}}$, 
        has three DOFs, and there is one constraint, namely ${m}_{\mathsf{A_2}}=0$ (recalling that $A_2$ is a one-dimensional irrep; see Table \ref{Table_irrep_local}).
\end{itemize}

\section{CSLs for Non-Kramers Spins}
\label{sec:CSL_non_kramers}

\subsection{Identification of CSLS for non-Kramers spins}

We are now ready to employ the methodology developed in Sec.~\ref{Sec:irrep.and.inter.tetra} to systematically scan  the whole parameter space $\{J_{zz}, J_{\pm}, J_{\pm\pm}, J_{z\pm}\}$ of the nearest-neighbor Hamiltonian in Eq.~\eqref{eq:Hex1} without the single-ion anisotropy term in Eq.~\eqref{eqn:single.ion}. 
In this section, we present all the CSLs identified in non-Kramers spin models.
As discussed above, in this case $a_{\mathsf{T_{1,ip}}}$ vanishes, so $\mTOnePL$ and $\mTOneICE$ are the correct linear combinations of the $T_d$ irreps in the diagonalized form of the single-tetrahedron Hamiltonian.
Applying Step \#4, we then use  Eqs.~(\ref{eq:nonK.constraintA2}-\ref{eq:nonK.constraint+}) to derive the Gauss's laws. 
Following this procedure, we identified five distinct types of CSLs which we now proceed to discuss.
The non-Kramers spin liquids are listed in Table~\ref{Table_all_csls} as CSL 1-5, and their location in the phase diagram is indicated in Fig.~\ref{fig:nonkramer.phase}. 

\vspace{1em}  \textbf{CSL 1. Single  R1U1 (Spin ice):}
The first case is the well-known spin ice. While our model considers three-component spins rather than strictly Ising-like (one-component) spins, the ground state properties remain unchanged.
In this case, the coefficient $a_{\mathsf{T_{1,i}}}$ is minimal, and all the other coefficients are greater than $a_{\mathsf{T_{1,i}}}$.
The effective physics of this spin liquid was already analyzed as the simplest example of the inter-tetrahedra coarse-graining rule  
at the end of Sec.~\ref{subsec:non.kramer.inter}, Eqs.~\eqref{eq:nonK.constraintA2}.

As shown there, a Gauss's law is obtained, given by
\begin{equation}
\label{EQN_spin_ice}
m_\mathsf{A_2}  = 0 \implies \vb*{\nabla} \cdot \vb*{m}_\mathsf{T_{1,i}}\equiv \vb*{\nabla} \cdot \vb*{E}^\mathsf{ice}  = 0,
\end{equation} 
which is equivalent to the Maxwell U(1) Gauss's law.
As is well-known for this regime, the charge (in some literature referred to as magnetic monopole) is measured by $m_\mathsf{A_2} = S_1^z +S_2^z +S_3^z +S_4^z$, and imposes the microscopic 2-in$-$2-out ``ice-rule'' for the ground states~\cite{Bramwell-Science,Castelnovo2008Nature}.

\vspace{1em} 
\textbf{CSL 2. Double  R1U1 (SL$_\perp$)}: 
The second CSL that we consider in the phase diagram arises when $a_{\mathsf{T_{1,p}}} $ and $a_{\mathsf{T_{2}}}$ are simultaneously minimal, while the other $a_\mathsf{Y}$'s  are not. 
In this case, the vanishing (i.e.~frozen out) field $\bfm_{\sf E} = 0$  imposes constraints on the $\bfm_{\mathsf{T_{1,p}}}$ and $\bfm_{\mathsf{T_{2}}}$ fluctuating fields, with the constraints given by Eq.~\eqref{eq:double_U1}. 
This can be written as  two copies of a Maxwell U(1) Gauss's law:
\begin{equation}
\bfm_\mathsf{E}  = 0 \implies     \vb*{\nabla} \cdot  \vb*{E}^\mathsf{A} = 0 \ \text{and}\ 
  \vb*{\nabla} \cdot \vb*{E}^\mathsf{B} = 0,
\end{equation}
 where
\begin{align} 
  &\vb*{E}^\mathsf{A} = (2  m_\mathsf{T_{1,p}}^x , -    m_\mathsf{T_{1,p}}^y -\sqrt{3}m_\mathsf{T_{2}}^y, \sqrt{3}m_\mathsf{T_{2}}^z -   m_\mathsf{T_{1,p}}^z  ),\\ 
  &\vb*{E}^\mathsf{B} = (2 m_\mathsf{T_{2}}^x ,  \sqrt{3}    m_\mathsf{T_{1,p}}^y - m_\mathsf{T_{2}}^y, -   \sqrt{3}   m_{T_\mathsf{1,p}}^z -   m_\mathsf{T_{2}}^z).
 \end{align}
This CSL was previously identified and studied in Ref.~\cite{Taillefumier2017PhysRevX}. 
There, it was found that the system remains in a spin liquid state down to a low temperature and that the dipolar spin-spin correlations in that regime are indeed described by a double U(1) Gauss's law.
At very low temperatures, a hidden order quadrupolar (spin nematic) develops in which the spins spontaneously select a particular axis in the local $xy$-plane along which they to co-align. 
Nevertheless, the spin-spin correlations themselves remain algebraic.

\vspace{1em} 
\textbf{CSL 3. Triple  R1U1 (pHAF)}: 
In this case, $a_{\mathsf{T_{1,p}}} $, $a_{\mathsf{T_{1,i}}} $ and $a_{\mathsf{T_{2}}} $ are simultaneously minimal while the other ones are not. 
For these parameters,   CSL 1 and CSL 2 become degenerate. 
As a consequence, this CSL is described by an emergent spin ice $\times$ Double U(1) field.
This gives rise to three copies of Maxwell U(1) Gauss's laws which can be written compactly as 
\begin{equation}
    \partial_i E_{ij}^{\mathsf{pHAF}} = 0,
    \label{eq:pHAF_gauss}
\end{equation}
where $\vb*{E}^{\mathsf{pHAF}}$ is a 3$\times 3$  matrix,
\begin{equation}
\vb*{E}^{\mathsf{pHAF}} = \left[ (\vb*{E}^\mathsf{ice})^\mathrm{T} ,(\vb*{E}^\mathsf{A})^\mathrm{T} ,(\vb*{E}^\mathsf{B})^\mathrm{T}   \right].
\end{equation}

We note that this CSL 3 state corresponds to the so-called pseudo-Heisenberg AFM point in parameter space~\cite{Taillefumier2017PhysRevX}. The set of $\{J_{uv}\}$ parameters for which this CSL is realized corresponds to a \emph{local} Heisenberg antiferromagnetic model where $J_{\pm}/J_{zz}=-1/2$. Indeed, the above description of this CLS is consistent with the understanding that its thermodynamic properties are identical to that of the HAFM model~\cite{Moessner1998PhysRevLett,Moessner1998PRB}, but now for spins described in their local basis, hosting three copies of U(1) Maxwell Gauss's law, one for each component of $\vb*{S}$. Consequently, the Gauss' law in Eq.~\eqref{eq:pHAF_gauss} simply describes a vanishing moment in the local basis, where $\vb*{E}^\mathsf{ice}$ corresponds to the local $z$-moment, while the remaining fields to the local $xy$-moment.

\vspace{1em}{\textbf{CSL 4. R1U1-R2U1}  $(\textbf{R}_1$-$\textbf{R}_2^{\mathsf{T}_1})$:}
 In this case,   $a_{\mathsf{E}} $,   $a_{\mathsf{T_{1,p}}} $ and $a_{\mathsf{T_{1,i}}} $ are minimal, and the other $a_\mathsf{Y}$'s are  not. 
The constraints give two co-existing Gauss's laws,
 \begin{align}
    m_\mathsf{A_2} & =  0   \implies   \vb*{\nabla} \cdot \vb*{E}^\mathsf{ice}  = 0,\\ 
    \bfm_\mathsf{T_{2}} & = 0     \implies  \nonumber \\
	&\sqrt{3}\vb*{\nabla}\times \bfm_\mathsf{T_{1,p}}   + 
     \begin{pmatrix}
		2 \partial_x  m_\mathsf{E}^2  \\
		-  \sqrt{3}  \partial_y m_\mathsf{E}^1   -  \partial_y m_\mathsf{E}^2 \\
		\sqrt{3}  \partial_z m_\mathsf{E}^1 -  \partial_z m_\mathsf{E}^2   
	\end{pmatrix} = 0 .
    \label{eq:Mt2_CSL4}
\end{align} 

The first equation is the same Maxwell Gauss's law as for spin ice, whereas the second equation  can be rewritten in a  more compact form by introducing a rank-2 ``electric field'' $E_{ij}^\mathsf{T_1+E}$ with off-diagonal anti-symmetric and diagonal traceless components, defined as 
\begin{equation}
\label{eqn_Et1e}
     \vb*{E}^\mathsf{T_1+E} = 
    \begin{pmatrix}
	2 \mEScalar^2 &  \sqrt{3}\mTOnePLScalar^z &  - \sqrt{3}\mTOnePLScalar^y \\
	- \sqrt{3}\mTOnePLScalar^z & -\sqrt{3} \mEScalar^1 - \mEScalar^2 &   \sqrt{3}\mTOnePLScalar^x \\
	\sqrt{3}\mTOnePLScalar^y &  - \sqrt{3} \mTOnePLScalar^x & \sqrt{3} \mEScalar^1 - \mEScalar^2
	\end{pmatrix} .
\end{equation}

Using this  $\vb*{E}^\mathsf{T_1+E}$ rank-2 tensor, the following vector charge Gauss's law, i.e.~a higher-rank Gauss's law, is obtained from Eq.~\eqref{eq:Mt2_CSL4}:
\begin{equation}
\label{EQN_T1_r2u1}
	\partial_i E_{ij}^\mathsf{T_1+E}  = 0.
 \end{equation}
We note that the higher-rank Gauss's law in Eq.~\eqref{EQN_T1_r2u1} resembles that of a rank-2 U(1) gauge theory, but it is not the symmetric tensor version studied in Refs.~\cite{Pretko2017PhysRevBa,Pretko_fractonsPhysRevB.98.115134}. 
In particular, this CSL 4  is characterized by \emph{infinitely many} multipole-charge conservation laws. 
Defining $\rho_j = \partial_i E_{ij}^\mathsf{T_1+E}$, we obtain the following conserved quantities:
\begin{align}
&    f(x+y+z)(\rho_1 + \rho_2 + \rho_3),\\
&   g(x-y-z)(\rho_1- \rho_2 - \rho_3),\\
   & k(-x+y-z)(-\rho_1 + \rho_2 - \rho_3),  \\
& l(-x-y+z)(-\rho_1 - \rho_2 + \rho_3),  
\end{align}
where $f(w)$, $g(w)$, $k(w)$, $l(w)$ are arbitrary polynomials of the argument $w$.  
As an example, if we define $w= x+y+z$, the dipole $w (\rho_1 +\rho_2 + \rho_3    )$, quadrupole $w^2 (\rho_1 +\rho_2 + \rho_3    )$, etc., are all conserved along the $[111]$ direction. 
Such multipole-charge conservation laws apply to the four $[111]$ cubic crystalline directions. 

This CSL 4, and also the next one  discussed just below, CSL 5,  was very recently identified and studied in Ref.~\cite{lozanogomez2023arxiv}. 
This CSL is interesting in that it displays two distinct spin-liquidity regimes as a function of temperature.
At intermediate temperature, the energy-degenerate $\mathsf{E}$ and both $\mathsf{T_{1}}$ irreps are thermally populated, resulting in the so-called rank-1$-$rank-2 ($\rm R_1$-$\rm R_2$) spin liquid in which two-fold and four-fold pinch points are observed in the spin-spin correlations in reciprocal space.
At low temperatures, the model exhibits a thermal depopulation of the $\mathsf{E}$ and the $\mathsf{T_{1,p}}$ irreps associated with an entropic selection of the $\mathsf{T_{1,i}}$, thus leading to a spin-ice-like CSL at the lowest temperatures. 
The phenomenology observed in this model is loosely reminiscent of the liquid-to-liquid transition in some molecular liquids such as phosphorous~\cite{P-L2L}, sulfur~\cite{S-L2L}, and silicon~\cite{Si-L2L}. 
We note that the additional $\mathsf{T_1}$ superscript label for CSL 4 indicates that the rank-2 tensor field  $\vb*{E}^\mathsf{T_1+E}$ 
has a $\bfm_\mathsf{T_{1,p}}$ field dependence whereas its dual, CSL 5, has a dependence upon the 
$\bfm_\mathsf{T_{2}}$
field, as is further discussed next.

\vspace{1em} {\textbf{CSL 5. R1U1-R2U1 }$(\textbf{R}_1$-$\textbf{R}_2^{\mathsf{T}_2})$ }
  In this case, $a_{\mathsf{E}} $,   $a_{\mathsf{T_{2}}} $ and   $a_{\mathsf{T_{1,i}}} $   are at their minimal value while the other ones are not. 
This spin liquid is related to the CSL 4 above via a duality that corresponds to local $\pi/2$ rotation defined as 
    \begin{equation}
  \begin{aligned} 
   \label{eqn_non_kramer_duality}
    J_{\pm\pm}  & \to \;  -J_{\pm\pm}, \\
     (S^x, S^y)  & \to \; (- S^y,S^x),
  \end{aligned}
\end{equation}
or, alternatively, $S_j^\pm \to \pm  iS_j^\pm $. In terms of irreps, this duality implies $m_\mathsf{E}^1 \leftrightarrow m_\mathsf{E}^2 $ and ${\vb*{m}}_\mathsf{T_2} \leftrightarrow {\vb*{m}}_\mathsf{T_{1,p}}$. 
Implementing these dualities on the rank-2 tensor $\vb*{E}^\mathsf{T_1+E}$, we obtain the dual rank-2 tensor $ \vb*{E}^\mathsf{T_2+E}$. 
Altogether, the Gauss's laws describing this CSL 5 are 
\begin{equation}
\label{EQN_T2_r2u1}
\begin{split}
    m_\mathsf{A_2}  =  0 \implies   &\vb*{\nabla} \cdot \vb*{E}^\mathsf{ice}  = 0, \\
\bfm_\mathsf{T_{1,p}}  = 0    \implies  & \partial_i E_{ij}^\mathsf{T_2+E}  = 0,
\end{split} 
\end{equation} 
where $\vb*{E}^\mathsf{T_2+E}$ is defined similarly as $\vb*{E}^\mathsf{T_1+E}$, but now, by making replacement  ${\vb*{m}}_\mathsf{T_{1,p}} \rightarrow {\vb*{m}}_\mathsf{T_2}  $ and $(m_E^1,m_E^2)\rightarrow (-m_E^2,m_E^1)$.
Thanks to the duality~\eqref{eqn_non_kramer_duality}, the thermodynamical properties of this CSL 5 are identical to those of CSL 4. 
However, the spin-spin correlations themselves are different~\cite{lozanogomez2023arxiv}. 

\dlgnew{Lastly, we note that, given the current parametrization of their spin-spin couplings, the non-Kramers compound $\rm Tb_2Ti_2O_7$~\cite{Rau2019ARCMP,Gardner-RMP} and its off-stoichiometric variant $\rm Tb_{2+x}Ti_{2-x} O_{7+y}$~\cite{Takatsu_TbTiO} have been proposed to lie close to the parameter regime of this CSL~\cite{lozanogomez2023arxiv}.}
We return to this matter in Sec.~\ref{sec:proxy-liquid-materials-TTO}.

\subsection{Summary of the non-Kramers spins Hamiltonian}

To summarize, the non-Kramers spin model hosts five CSLs in total, listed in Table~\ref{Table_all_csls}. 
In the of $J_\pm - J_{\pm\pm} - J_{zz}$ phase diagram illustrated in Fig.~\ref{fig:nonkramer.phase}.
Only the spin ice  (CSL 1) exists as a phase occupying a 3D volume in parameter space; all other CSLs sit on phase transition planes or lines separating long-range ordered phases.
The spin ice phase takes the shape of a three-edged pyramid. 
The double U(1) CSL  (CSL 2) occupies a plane attached to an edge of the pyramid. 
The  CSLs 3, 4, 5 are located on the three edges of the pyramid defined by the spin ice phase, a feature further manifested by the fact that their respective Gauss's laws incorporate the Maxwell U(1) Gauss's law of spin ice.

Various interesting points on the phase diagram were previously studied in the literature. 
Obviously, the most intensively studied one is spin ice.
Besides, the  CSLs 2 and 3 and their double and triple U(1) Gauss's laws, respectively, have also been investigated in Ref.~\cite{Taillefumier2017PhysRevX}. On the other hand, the $\rm R_1$-$\rm R_2^{\mathsf{T}_1}$ CSL 4 was very recently  studied in Ref.~\cite{lozanogomez2023arxiv}. 
There, the Hamiltonian of that model was coined the name dipolar-quadrupolar-quadrupolar (DQQ) model because it is realized at the triple point where two long-range ordered quadrupolar phases meet the disordered spin ice (magnetic dipolar) phase discussed Refs.~\cite{Onoda-PRB,leeGenericQuantumSpin2012}. 
The CSL 5 ($\rm R_1$-$\rm R_2^{\mathsf{T}_2} $) is realized at the dual point of the DQQ model, referred to as DQQ$^*$ in Ref.~\cite{lozanogomez2023arxiv}, so these the CSL 4 and CSL 5 share identical thermodynamics. 

We note that, although there are also three triple-phase-boundary points for the case where $J_{zz}<0$ (See Fig ~\ref{fig:nonkramer.phase}), which have three degenerate $a_\mathsf{X}$'s, \emph{all of these} exhibit a finite-temperature phase transition into an ordered state as found from classical Monte Carlo simulations. 
We direct the reader to Sec.~\ref{sec:Unsuccessfuk_CSL} for more details on these other specific models that we refer to as ``failed CSL candidates''.

\section{CSLs for Kramers Spins}
\label{sec:CSL_kramers}

\subsection{Identification of CSLs for Kramers spins}

In this section, we discuss the case of the spin model for
Kramers ions. 
In terms of the bilinear Hamiltonian in Eq.~\eqref{eq:Hex1}, Kramers spins allow for a nonzero  $J_{z\pm}$ interaction. 
When rewriting the Hamiltonian in terms of the irreps as discussed in Sec.~\ref{subsec:single.tetra.ham}, an additional term coupling the two  $\mathsf{T_1}$ irreps, $\bfm_\mathsf{T_{1,p}}$ and $\bfm_\mathsf{T_{1,i}}$, is now allowed in Eq.~\eqref{eqn:tetra.hamiltonian}. 
Again, this originates from the fact that both $S_i^z$ and $S_i^\pm$ are time-odd, transforming as real spin-1/2 operators, and are coupled in Eq.~\eqref{eq:Hex1}.
As previously discussed, the interaction between $\bfm_\mathsf{T_{1,p}}$ and $\bfm_\mathsf{T_{1,i}}$  via  $a_{\mathsf{T_1,ip}}$ in Eq.~\eqref{eqn:tetra.hamiltonian}, can be eliminated by a change of basis within the $\mathsf{T_{1,i}}$ and $\mathsf{T_{1,p}}$ sector, giving a new basis defined by the fields $\bfm_\mathsf{T_{1,+}}$ and $\bfm_\mathsf{T_{1,-}}$ which diagonalizes the Hamiltonian $H_t$. 
The inter-tetrahedra constraints are then modified to be expressed in terms of  these new $\bfm_\mathsf{T_{1,\pm}}$ fields. 
This is discussed in detail from Eq.~\eqref{eq:T1-_def} to Eq.~\eqref{eq:Ht-diagonal}. 
We now proceed to discuss the CSLs that arise when $J_{z\pm}$, or equivalently, when $a_\mathsf{T_{1,ip}}$, is non-zero. As done above for the non-Kramers case, we apply the procedure discussed in Sec.~\ref{subsec:recipe} with the inter-tetrahedra constraints of Eqs.~(\ref{eq:A2constraint}-\ref{eq:constraint+}) to systematically search for the CSLs that Kramers ions can exhibit on the pyrochlore lattice.  
For the Kramers Hamiltonian, we have found in total four different types of CSLs, which are discussed in detail below.
These CSLs are listed in Table~\ref{Table_all_csls} as CSL 6-9, and their phase diagram is shown in Fig.~\ref{fig:kramer.phase}.
There are actually two phase diagrams, for $J_{z\pm} >0$ and $J_{z\pm} <0$ respectively, but the geometry of the two phase diagrams are identical as a consequence of an exact duality of the Kramers Hamiltonian associated to a change of sign in the $J_{z\pm}$ coupling~\cite{Rau2019ARCMP} which we discuss in further detail below.

\vspace{1em} 
\textbf{CSL 6. Scalar-vector-charge  (SV) R2U1 CSL: }
This CSL is stabilized when 
the parameters $a_\mathsf{E}$, $a_\mathsf{T_2}$, and $a_\mathsf{T_{1,-}}$ are simultaneously minimal, that is when
\begin{equation} 
\begin{split}
    J_{z\pm}>& 0, \\
    J_{zz} >& -\frac{1}{\sqrt{2}}J_{z\pm},\\
    J_{\pm}=&\frac{1}{12}\left( J_{zz}+\sqrt{J_{zz}^2+24J_{z\pm}^2}\right), \\
    J_{\pm\pm}=&\frac{1}{6}\left( J_{zz}+\sqrt{J_{zz}^2+24J_{z\pm}^2}\right).
\end{split}\label{eq:CL6_constraint}
\end{equation} 
In the 3D phase diagram of fixed $J_{z\pm}$, these equations parametrize a line. 
The effective Gauss's law for this CSL has a scalar charge and a vector charge defined for its rank-2, traceless electric field,
\begin{align}
	m_{\mathsf{A_2}} = 0 \implies & |\epsilon^{ijk}| \partial_i E_{jk}^{\mathsf{SV}}  = 0, \label{eq:CL6_Gauss_Law_1} \\
	\vb*{m}_{\mathsf{T_{1,+}}} = 0 \implies & \partial_i E_{ij}^{\mathsf{SV}} = 0\label{eq:CL6_Gauss_Law_2},
\end{align}
where  
\begin{equation}
\begin{split}
      \vb*{E}^{\mathsf{SV}} =& \begin{pmatrix}
	2 \mEScalar^1 & \sqrt{3} m_{T_2}^z    &    - \sqrt{3} m_{T_2}^y   \\
	- \sqrt{3} m_{T_2}^z  & -\mEScalar^1 + \sqrt{3} \mEScalar^2 &  \sqrt{3} m_{T_2}^x \\
	 \sqrt{3} m_{T_2}^y  &-\sqrt{3} m_{T_2}^x& -\mEScalar^1 - \sqrt{3} \mEScalar^2
\end{pmatrix}\\& - 
3\sin\phi
\begin{pmatrix}
	0 & m_{T_{1,-}}^z &     m_{T_{1,-}}^y \\
	 m_{T_{1,-}}^z &0&   m_{T_{1,-}}^x \\
	 m_{T_{1,-}}^y &   m_{T_{1,-}}^x & 0
\end{pmatrix} .
\end{split}
\label{eq:rank2_CSL6}
\end{equation}    
Here,  $|\epsilon^{ijk}|$ is the absolute value of the Levi-Civita symbol.
The Gauss's law does not exhibit continuous rotational symmetry due to the contraction with   $|\epsilon^{ijk}|$, which is not a covariant tensor.
The correlation $\langle {E}^{\mathsf{SV}}_{ij}(\vb*{q}) {E}^{\mathsf{SV}}_{kl}(-\vb*{q})\rangle $  exhibits both pinch-lines and four-fold pinch points, similar features to those observed in the pinch-line spin liquid~\cite{Benton2016NatComm}. 

Defining $\rho_0 \equiv  |\epsilon^{ijk}| \partial_i E_{jk}^{\mathsf{SV}} $ and $\rho_j \equiv  \partial_i E_{ij}^\mathsf{SV}$, 
we identify the following infinitely many conservation laws:
\begin{align}
    &f(x+y+z)(-\rho_0 + \rho_1 + \rho_2 + \rho_3) ,\\
    &g(x-y-z)(-\rho_0 + \rho_1 - \rho_2 - \rho_3), \\
    &k(-x+y-z)(-\rho_0 - \rho_1   +\rho_2 - \rho_3), \\
    &l(-x-y+z)(-\rho_0 - \rho_1 -\rho_2 + \rho_3).  
\end{align}
As before, $f(w)$, $g(w)$, $k(w)$, $l(w)$ are arbitrary polynomials of the argument $w$. We note that this particular CSL has not been previously studied, therefore we identify this as a novel higher-rank CSL. However, it is crucial to note that this CSL is dual to the pinch-line spin liquid~\cite{Benton2016NatComm}. This duality implies that the Gauss's laws and gauge fields of these CSLs have the same form resulting in the observation of both pinch-lines and multifold pinch points in their correlation functions. Lastly, we note that, although a similar gauge field as that in Eq.~\eqref{eq:rank2_CSL6} describing the low-temperature physics of the pinch-line spin liquid was identified in Ref.~\cite{Benton2016NatComm}, the precise details of the Gauss's laws, the conserved quantities, and the gauge charges associated with it were not discussed therein.

\vspace{1em} 
\textbf{CSL 7. SV R2U1* (Pinch-line spin liquid): }
This CSL has the same degeneracy of minimal parameters as CSL 6: $a_\mathsf{E}$, $a_\mathsf{T_2}$, and $a_\mathsf{T_{1,-}}$ which, in terms of spin exchange parameters, amount to the conditions 
\begin{equation}
\begin{split}
    J_{z\pm}<& 0,\quad J_{zz} > \frac{1}{\sqrt{2}}J_{z\pm},\\ 
    J_{\pm}=&\frac{1}{12}\left( J_{zz}+\sqrt{J_{zz}^2+24J_{z\pm}^2}\right), \\
    J_{\pm\pm}=&\frac{1}{6}\left( J_{zz}+\sqrt{J_{zz}^2+24J_{z\pm}^2}\right).
\end{split}\label{eq:CL7_constraint}
\end{equation}

The corresponding emergent Gauss's laws have the same functional form as those of CSL 6. 
The low-temperature physics of both spin liquids are described by a rank-2 gauge field constraint by the Gauss's laws in Eq.~\eqref{eq:CL6_Gauss_Law_1} and Eq.~\eqref{eq:CL6_Gauss_Law_2}. 
However, the precise definition of the $\vb*{m}_\mathsf{T_{1,-}}$ and therefore the gauge fields themselves are distinct. 
The CSLs 6 and 7 should therefore be considered as different.
First, the microscopic Hamiltonian parameters defining the two CSLs are different.
Second, in the phase diagram, the two CSLs do not directly connect to each other as they are separated by the limit of $J_{z\pm} = 0$, which corresponds to the distinct CSL 5. 
Nonetheless, CSL 6 and 7 are related by the duality~\cite{Rau2019ARCMP,KTC_2024_phase}  
\[
\label{eqn_kramers_duality}
\begin{split}
   J_{z\pm} &\rightarrow -J_{z\pm}, \\
    (S^x, S^y, S^z) &\rightarrow (-S^x, -S^y, S^z).
\end{split}
\]
More generally, this duality implies that the phase diagram boundaries between magnetic orders are identical after reversing the sign of $J_{z\pm}$, but the magnetic orders themselves are not identical on the two sides. 
This duality and the one of Eq.~\eqref{eqn_non_kramer_duality} are two different dualities for two different models. 
The duality in Eq.~\eqref{eqn_non_kramer_duality} only applies to the non-Kramers Hamiltonian and is not (on its own) a duality of the Kramers Hamiltonian. 
The former one for $J_{z\pm}$ in Eq.~\eqref{eqn_kramers_duality} 
only applies to the Kramers Hamiltonian while in the non-Kramers case, the consequential transformation on the $xy$ components of the pseudospins in Eq.~\eqref{eqn_non_kramer_duality} amounts to a trivial reflection symmetry.
This CSL  7 was studied in  Ref.~\cite{Benton2016NatComm}. 
Here, our observation of an infinite number of conserved charges, the duality of Eq.~\eqref{eqn_kramers_duality}, and also the overall structure of the phase diagram complement that of a  previous study~\cite{Benton2016NatComm}.  
We provide in Sec.~\ref{sec:numerics} below numerical evidence for the persistence of the spin liquid behavior for this phase for distinct values of $J_{z\pm}$ down to the lowest temperatures.
\\

\vspace{1em} 
\textbf{CSL 8. Triple  R1U1 (HAFM)}:  
The second to last CSL that we identify is the well-known Heisenberg antiferromagnet (HAFM)~\cite{Moessner1998PhysRevLett,Moessner1998PRB,Reimers1992,villain1979}. 
This CSL corresponds to the point where the CSL 6 meets the AIAO phase with the $a_\mathsf{E}$, $a_\mathsf{T_2}$, $a_\mathsf{T_{1,-}}$, and $a_\mathsf{A_2}$ parameters all being minimal. 
The spin-spin $\{J_{uv}\}$ parameters for this CSL are parametrized as 
\[
\label{EQN_HAF_parameter}
\begin{split}
& J_{z\pm}> 0, \quad J_{zz} = -\frac{1}{\sqrt{2}}|J_{z\pm}|,\\
\quad 
& J_\pm = \frac{1}{2\sqrt{2}}|J_{z\pm}|, \quad J_{\pm\pm} = \frac{1}{\sqrt{2}}|J_{z\pm}| .
\end{split}
\] 
This point corresponds to the isotropic $O(3)$ Heisenberg antiferromagnetic model in the global spin basis. 
The Gauss's law for this CSL are three copies of U(1), written in a compact form as 
\[ \label{eqn.HAFM.gauss}
m_{\mathsf{T_{1,+}}} = 0 \implies  \partial_i E_{ij}^{\mathsf{HAF}} = 0,
\]
where
\[
 E_{ij}^{\mathsf{HAF}} =  E_{ij}^{\mathsf{SV}} -2 \delta_{ij} (\tan\phi) m_\mathsf{A_2} .
\]
We see here that the scalar $m_\mathsf{A_2}$ field adds exactly to the trace component to the otherwise traceless $E_{ij}^{\mathsf{SV}}$ rank-2 field such that  
$E_{ij}^{\mathsf{HAF}}$ acquires the full nine degrees-of-freedom associated to a $3\times 3$ matrix. 
These DOFs correspond to the nine components of the degenerate $m_\mathsf{X}$ fields, whose trace is the $m_\mathsf{A_2}$ field, two diagonal traceless elements are the $\vb*{m}_\mathsf{E}$ field,   three off-diagonal antisymmetric elements are the $\vb*{m}_\mathsf{T_{2}}$ field, and three off-diagonal symmetric elements are the $\vb*{m}_\mathsf{T_{1,-}}$ field. 
Altogether, these nine degrees of freedom and the three Gauss's laws in Eq.~\eqref{eqn.HAFM.gauss} result in six degenerate low-energy flat bands describing the ground state manifold of this CSL~\cite{Moessner1998PhysRevLett}.

\vspace{1em}  
\textbf{CSL 9. Triple  R1U1 (HAFM*)}: 
The last CSL for the Kramers Hamiltonian that we identify is dual to the HAFM CSL 8, with $J_{z\pm} < 0$ and parametrized  as 
\[
\label{EQN_HAF_star_parameter}
\begin{split}
& J_{z\pm}< 0, \quad J_{zz} = -\frac{1}{\sqrt{2}}|J_{z\pm}|,\\
\quad 
& J_\pm = \frac{1}{2\sqrt{2}}|J_{z\pm}|, \quad J_{\pm\pm} = \frac{1}{\sqrt{2}}|J_{z\pm}| .
\end{split}
\]
Just as the CSL 6, this CSL has not been previously studied and we therefore identify it as a novel CSL.
The emergent Gauss's law of this CSL is identical to that of the HAFM CSL.
Nevertheless, this CSL should be considered a different CSL, as the dual of HAFM via Eq.~\eqref{eqn_kramers_duality}. 
Indeed, the Hamiltonian for this CSL in the global spin basis is different from the HAFM as we further discuss in Appendix~\ref{appendix:global_basis_hamiltonian}. Moreover, for the HAFM$^*$ the $\mathsf{T_{1,-}}$ irrep corresponds to the colinear ferromagnet meaning that a single tetrahedron in the ground-state configuration of this spin liquid can have a finite magnetization, as opposed to those in the regular HAFM.

\subsection{Summary of the Kramers spin Hamiltonian}

To summarize, the  Hamiltonian for Kramers ions hosts a total of four CSLs,  as listed in Table~\ref{Table_all_csls}. 
In this case, the CSLs 6 and 7 are dual to each other, and so are CSLs 8 and 9, with the two dual partners in each pair yielding identical thermodynamic properties (see Sec.~\ref{sec:numerics}). 
Similar to the non-Kramers, the CSLs identified in the Kramers Hamiltonian separate distinct long-range ordered phases where the CSL 8 (9) are realized at the phase boundary between four phases, namely the $\mathsf{A_2}$, the $\mathsf{E}$, the $\mathsf{T}_{1,-}$ and the $\mathsf{T}_2$ phases, localized at the tip of a pyramid corresponding to an $\mathsf{A_2}$ phase as shown in Fig.~\ref{fig:kramer.phase}.
On the other hand, the CSL 6 (7) is realized upon leaving %
these maximally-degenerated points by gaping out the $\mathsf{A_2}$ phase, as shown in Fig.~\ref{fig:kramer.phase}. 
 
The classical spin liquids that we identified above constitute the complete list of classical spin liquids that can be realized in Kramers and non-Kramers pyrochlores with only nearest-neighbor bilinear interactions. 
We reiterate that the present section and Sec.~\ref{sec:CSL_non_kramers} only discuss the non-Kramers and Kramers nearest-neighbor bilinear spin Hamiltonians, respectively, but do not consider the special case of DO Kramers systems as this particular spin model has been exhaustively discussed in  previous works~\cite{Benton_DO_2020,hosoiUncoveringFootprintsDipolarOctupolar2022}. 
In the next section, we discuss how and in what circumstances these spin liquids are connected and transition from one to another and what the consequences for their respective gauge theory are as the exchange parameters $\{J_{zz},J_\pm, J_{\pm\pm},J_{z\pm}\}$ are tuned.

\section{Connectivity of the classical spin liquids and evolution of their effective Gauge theories}
\label{sec:Connectibity}
\begin{figure}[ht!]
    \centering
    \begin{overpic}[width=1.1\columnwidth]{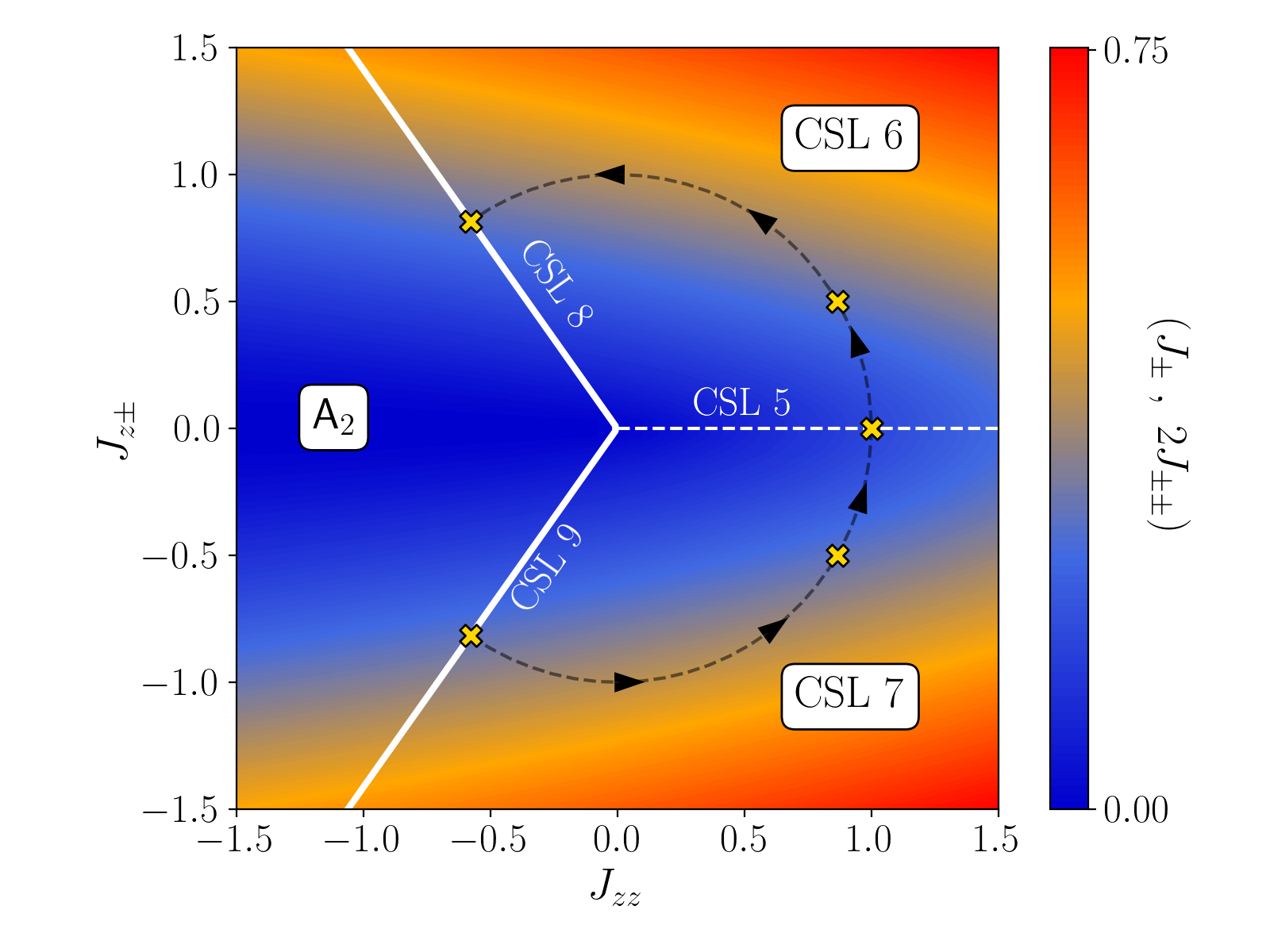}
    \end{overpic}
    \caption{Phase diagram illustrating how the classical spin liquids (CSLs) transition into/connect with each other. 
    The diagram is drawn with respect to $J_{zz}$ and $J_{z\pm}$, while $J_{\pm}$ and $J_{\pm\pm}$ take values \dlgnew{(indicated by the color bar)} based on Eqs.~(\ref{eq:CL6_constraint},\ref{eq:CL7_constraint}), \dlgnew{with the constraint $J_{\pm\pm}=2J_\pm$.}
    The upper right wedge is CSL 6, and lower right edge is CSL 7.
    The solid white lines, corresponding to   CSL 8  and   CSL 9, set the boundary between the spin liquid regime and the symmetry-breaking $\mathsf{A_2}$ (AIAO) phase. 
    The dashed white line separating   CSL 6 and CSL 7 is the CSL 5. 
    The reflection symmetry of the phase diagram about the $x-$axis is a consequence of the duality of Eq.~\eqref{eqn_kramers_duality}.
    Lastly, the black dashed line indicates a way of tuning between the different spin liquids where the yellow crosses represent models for which the structure factors are shown in Fig.~\ref{fig:gauge_tunning}. }\label{fig:phase_diagram_connectivity}
\end{figure}

In this section, we discuss the overall landscape of the CSLs that the anisotropic bilinear spin Hamiltonian of Eq.~\eqref{eq:Hex1} on the pyrochlore lattice harbors.
A hierarchy structure exists for these CSLs:
generally, the CSLs with more degrees-of-freedom  (DOFs) of low-energy fluctuations (more components in the emergent electric field and fewer constraints) require more $a_\mathsf{X}$'s [see Eq.~\eqref{eqn:tetra.hamiltonian}] to be degenerate. 
As a consequence, such CSLs live in a subspace of low dimension in $\{J_{uv}\}$ parameter space.
Furthermore, the CSLs with more DOFs are often delimited by other CSLs with more constrained Gauss's laws -- the latter typically arising from lifting up some of the $a_\mathsf{X}$'s above that of the lowest degeneracy value.
\dlgnew{
An interesting context for such a consideration arises when the $\{J_{uv}\}$ parameters are varied, driving  the system from one CSL to another. 
A concrete example, discussed just below, is the $\{J_{uv}\}$-driven evolution of the Hamiltonian that keeps the system in CSL 6 before ultimately inducing a transition to CSL 8 -- as highlighted in Fig.~\ref{fig:phase_diagram_connectivity} -- a phenomenology to which we shall return in Sections \ref{sec:roadmap} and \ref{sec:proxy-liquid-model}.
}

For example, the CSL 3 (pHAF) is a CSL with 6 DOFs of low energy fluctuations. These DOFs stem from the degeneracy of the 3-dimensional irreps, yielding 9 components of the emergent gauge field, constrained by 3 Gauss's laws, resulting in the 6 DOFs.
In particular, starting from the CSL 3 with the degenerate $\mathsf{T_{1,i}}$, $\mathsf{T_{1,p}}$, and $\mathsf{T_{2}}$ irreps, by lifting the degeneracy of one of the irrep modes, 
\begin{equation*}
(a_\mathsf{T_{1,i}}, \  a_{\mathsf{T_{1,p}}} , \ a_{\mathsf{T_{2}}} )
\rightarrow (a_\mathsf{T_{1,p}},  \ a_{\mathsf{T_{2}}}), 
    \end{equation*}
we obtain the CSL 2 (Double U(1)). 
On the other hand, by lifting the degeneracy 
$(a_\mathsf{T_{1,i}}, \ a_{\mathsf{T_{1,p}}} , \ a_{\mathsf{T_{2}}})$
$\rightarrow$  $a_{\mathsf{T_{1,i}}}$, we obtain the CSL 1 (i.e. spin ice).

Another example is the transformation from CSL 8 to CSL 6 (and similarly CSL 9 to CSL 7), which is effectuated by the following degeneracy lifting
\begin{equation*}
    ( a_{\mathsf{A_{2}}},\ a_{\mathsf{E}} ,   \ a_\mathsf{T_{1,-}} ,  \
a_{\mathsf{T_{2}}}  )
\rightarrow 
(a_{\mathsf{E}} ,   \ a_\mathsf{T_{1,-}} ,  \
a_{\mathsf{T_{2}}}).
\end{equation*} 
In these transitions, the resulting Gauss's laws for the CSL 6 (CSL 7) are obtained from the previous Gauss' laws for CSL 8 (CSL 9) by reducing the number of free components of the corresponding gauge field and/or introducing new constraints.

Let us now discuss the transitions between different CSLs in more detail. 
First, let us elaborate on an important fact about  CSLs 6 and 7.
As discussed in Sec.~\ref{sec:CSL_kramers}, the parameter space identified in Eq.~\eqref{eq:CL6_constraint} for each one of these CSLs corresponds to a 2D surface, with no symmetry-breaking phase transition detected in Monte Carlo simulations down to the lowest temperatures, as we shall discuss in Sec.~\ref{sec:numerics} below.
At low temperatures, effective Gauss's laws are imposed on an emergent rank-2 field constructed in terms of the $\mathsf{E}$, $\mathsf{T}_2$, and $\mathsf{T}_{1,-}$ irreps. 
This is the case for any point in the parameter space characterizing these two CSLs. 
However, it is crucial to realize that, although the form of the effective gauge theory remains the same, the precise constituents  of the rank-2 field expressed in terms of spins do differ.
Such variation is captured by the dependence of the rank-2 tensor on the  parameters $\{v_{i\pm},v_{p\pm}\}$ that define the $\mathsf{T_{1,-}}$ irrep [Eqs.~(\ref{eq:T1-_def},\ref{eq:T1+_def})]. 
Indeed, by tuning the constituents of the rank-2 fields, we are also effectively tuning the definition of the charge in these spin liquids. 
Consequently, one may understand the 2D spin liquid manifolds as a continuous set of models where the $\mathsf{T_{1,\pm}}$ of the low-energy emergent gauge fields are smoothly evolving as a function of $J_{z\pm}$, while the full gauge fields are constrained to fulfill a set of Gauss's laws.

Having discussed how the effective gauge theories describing the CSL 6 and the CSL 7 smoothly evolve as the microscopic Hamiltonian parameters are modified while keeping the system in one of these two CSLs, we now turn to consider how these two CSLs are connected to each other. Figure~\ref{fig:phase_diagram_connectivity} illustrates a phase diagram where the region of stability of multiple CSLs is shown.
There, we take $J_{zz}$ and $J_{z\pm}$ to be free parameters while $(J_{\pm},J_{\pm\pm})$ are fixed by Eq.~\eqref{eq:CL6_constraint}.
We first note that the two parameter spaces defining the CSL 6 and CSL 7 meet on the line defined by $\{J_{zz},J_\pm ,J_{\pm\pm}, J_{z\pm}\}=\{J,J/6,J/3,0\}$ with $J>0$. 
This line, shown as the white dashed line in Fig.~\ref{fig:phase_diagram_connectivity}, corresponds to the CSL 5~\cite{lozanogomez2023arxiv}, where the rank-2 fields decompose into one rank-2 field and one rank-1 field, each fulfilling an independent Gauss's law.
In particular, for the CSL 6, the gauge field $\vb*{E}^\mathsf{SV}$ in Eq.~\eqref{eq:rank2_CSL6} and the Gauss's laws in Eqs.~\eqref{eq:CL6_Gauss_Law_1}-\eqref{eq:CL6_Gauss_Law_2} decompose into  
the rank-1 $\vb*{E}^{\mathsf{ice}}$ and the rank-2 $\vb*{E}^\mathsf{T_1+E}$ electric fields constrained by the Gauss's laws in Eq.~\eqref{EQN_T2_r2u1}.
Importantly, this is the limit of vanishing $J_{z\pm}$ which characterizes non-Kramers ions discussed in Sec.~\ref{sec:CSL_non_kramers}~\footnote{Here we note that the $\{J_{uv}\}$ parameters corresponding to the CLS 5 in Fig.~\ref{fig:phase_diagram_connectivity} are such that the ratio $J_{zz}: J_{\pm}: J_{\pm\pm}= 6:1:2$ is always fulfilled therefore identifying a 1D region as a function of $J_{zz}$ as illustrated in Fig.~\ref{fig:phase_diagram_connectivity}. The evolution of these parameters on that boundary is given by the parametrization of the CSLs 6 and 7, Eqs.~\eqref{eq:CL6_constraint} and ~\eqref{eq:CL7_constraint}, while setting $J_{z\pm}=0$.}.

The other boundaries of the CSL 6 and 7 parameter spaces are located at $J_{zz}=-\left |\frac{1}{\sqrt{2}}J_{z\pm}\right |$ with $J_{z\pm}$ positive for the  CSL 6 while it is negative for the CSL 7. 
These two boundaries correspond to the CSL 8 (for $J_{z\pm}>0$), and CSL 9 (for $J_{z\pm}<0$), respectively. 
These boundaries, marked by solid white lines in Fig.~\ref{fig:phase_diagram_connectivity}, separate the CSL 6 and 7 from the $A_2$ (all-in/all-out, AIAO) magnetically ordered phase.
As discussed at the beginning of this section, the CSLs 8 and 9, ``viewed'' as the boundary of CSLs 6 and 7 with the magnetically ordered $\mathsf{A_2}$ phase, have a higher degeneracy.
The CSL 8 and 9 are thus CSLs with a higher number of low-energy-fluctuating DOFs [Eq.~\eqref{eqn.HAFM.gauss}], and contain all of the fluctuating DOFs of the CSLs 6 and 7 [Eq.~(\ref{eq:CL6_Gauss_Law_1})], respectively.

\begin{figure}[ht!]
    \centering
    \begin{overpic}[width=\columnwidth]{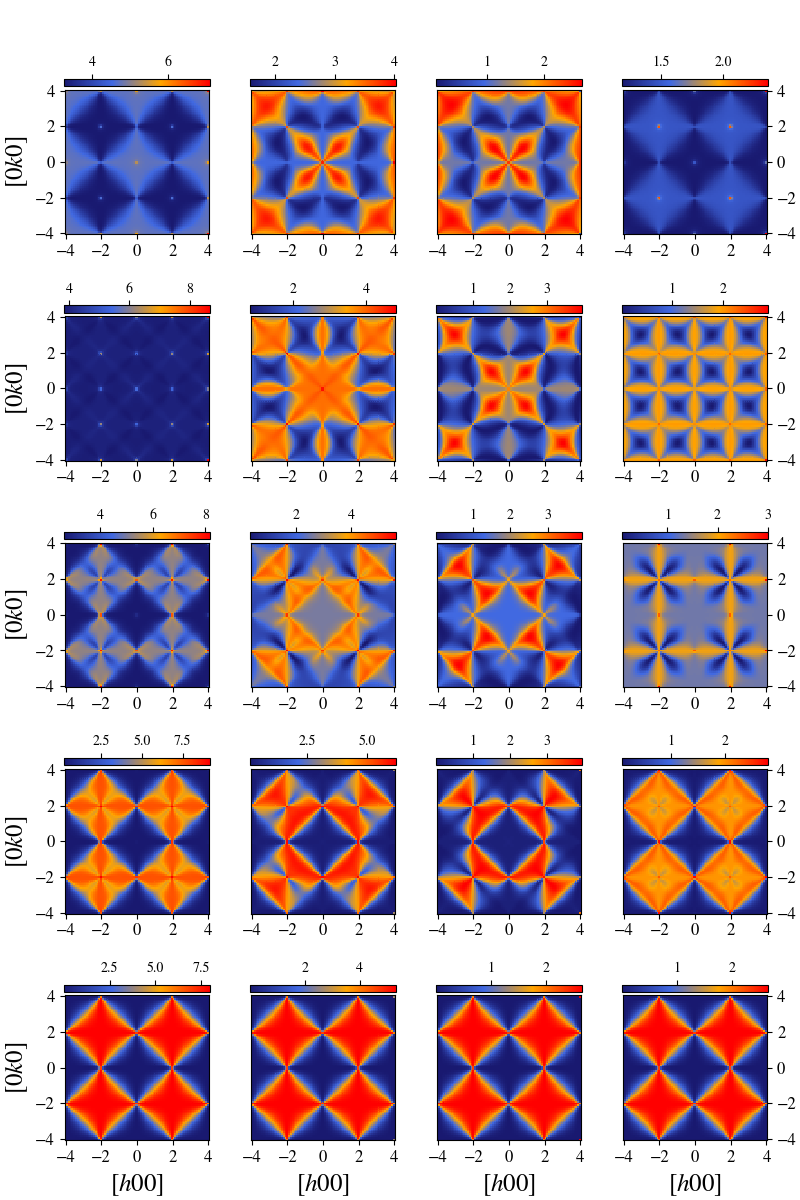}
    \put(8,97) {\fontsize{8}{38} $\mathcal{S}(\vb*{q})$}  
    \put(23,97) {\fontsize{8}{38} $\mathcal{S}_\perp(\vb*{q})$}  
    \put(39,97) {\fontsize{8}{38} $\mathcal{S}_\perp^{\rm SF}(\vb*{q})$}  
    \put(54,97) {\fontsize{8}{38} $\mathcal{S}_\perp^{\rm NSF}(\vb*{q})$} 
    \put(24.7,88.1){\fontsize{21}{38} \textcolor{white}{$\circ$}}  
    \put(24.7,69.3){\fontsize{21}{38} \textcolor{white}{$\circ$}}   
    \put(24.7,31.7){\fontsize{21}{38} \textcolor{white}{$\circ$}}  
    \put(24.7,12.8){\fontsize{21}{38} \textcolor{white}{$\circ$}} 
    \put(58.8,69.5){\fontsize{13}{38} \textcolor{white}{$\square$}}  
    \put(58.8,50.6){\fontsize{12}{38} \textcolor{white}{$\square$}}  
    \put(58.8,31.8){\fontsize{12}{38} \textcolor{white}{$\square$}}   
    \end{overpic}
    \caption{Evolution of the spin structure factor (first column),  neutron structure factor (second column), polarized neutron structure factor in the SF channel (third column), and polarized neutron structure factor in the NSF channel (fourth column) in the $(hk0)$ plane, computed using the large-$\mathcal{N}$ approximation.
    In this figure, each row corresponds to a model parameterized by Eq.~\eqref{eq:CL6_constraint} labeled by the yellow crosses in Fig.~\ref{fig:phase_diagram_connectivity} starting from the upmost row with the CSL 9 (HAFM*) and ending in the bottom row with the CSL 8 (HAFM), as indicated by the black circling arrows in Fig.~\ref{fig:phase_diagram_connectivity}. 
    The white circles (squares) indicate the location of a two-fold (four-fold) pinch point. Definitions of each structure factor are given in Appendix \ref{appendix:correlation_functions}.}
    \label{fig:gauge_tunning}
\end{figure}

We have now discussed how the CSLs 6  and 7 display  a smoothly varying definition of their electric field subject to the emergent Gauss's laws, and how they are connected to other CSLs on the boundary within the parameter space that define them. 
From an experimental context,  such a change in the electric field can be exposed via 
the equal time polarized and unpolarized neutron structure factors which evolve as a function of the $J_{z\pm}$ and $J_{zz}$ coupling parameters. 
Figure~\ref{fig:gauge_tunning} illustrates the spin structure factor, unpolarized and polarized (spin flipping (SF) channel and non-spin flipping (NSF) channel) neutron structure factor in the $(hk0)$ plane for five different sets of parameters marked by yellow crosses in Fig.~\ref{fig:phase_diagram_connectivity}, which correspond to CSLs 9, 7, 5, 6, 8, in this respective order.
The data illustrated here was obtained using a large-$\mathcal{N}$ approximation which yields qualitatively similar results to those obtained with CMC simulations, except for the case with $J_{z\pm}=0$ in the $T\to 0$ limit which displays disorder-by-disorder selection of a spin ice ground state~\cite{lozanogomez2023arxiv}. 
In this figure, and heretofore, we have considered an isotropic $g$ tensor assuming that the fluctuations in all spin components are taken on equal footing in generating the neutron scattering structure factor~\cite{lozanogomez2023arxiv,Kadowaki_2015}.
We refer the reader to Appendix~\ref{appendix:correlation_functions} for the explicit definition of these correlation functions and for a short discussion regarding the choice of an isotropic $g$ tensor used in these calculations. 
For the CSLs 5, 6 and 7, their effective gauge theories are characterized by an emergent Gauss's law of a rank-2 electric field. 
The latter is responsible for the presence of four-fold pinch points in the correlation functions, besides the familiar two-fold pinch points, see Fig.~\ref{fig:gauge_tunning} where white circles (squares) in the unpolarized neutron structure factor (polarized NSF channel) indicate the location of two-fold (four-fold) pinch points. Although these anisotropic features remain observable in the structure factors, the overall intensity distribution significantly changes reflecting the evolving nature of the gauge fields and their functional dependence on the parametrization used in Fig.~\ref{fig:phase_diagram_connectivity}. 

Rather importantly, we note that the CSL 9 (HAFM*) also exhibits four-fold pinch points in its neutron structure factor at the $\Gamma$ point, although its Gauss's law is not associated with a higher-rank field. 
This four-fold pinch point is instead the consequence of the neutron (magnetic moment) scattering transverse projector
~\cite{Castelnovo2019rods}, which is non-analytical at the $\Gamma$ point, and \emph{is not} caused by an emergent gauge structure of the model. 
We therefore note that, although the observation of four-fold pinch points is often a strong indication of an emergent higher-rank gauge theory, when observed at the $\Gamma$ point, it is strictly neither a necessary nor a sufficient condition for the identification of such an underlying gauge theory~\footnote{We note that the CSL 5 exhibits a four-fold pinch point at  $(020)$ and symmetry-related points in its unpolarized neutron structure factors shown in the third row of Fig.~\ref{fig:gauge_tunning}. The spin structure factor of the same model, only shows two-fold pinch points at this $q$-point.
 These four-fold pinch points are produced by the combination of \emph{two overlapping} two-fold pinch points, one coming from the rank-1 field (spin ice) and another from the rank-2 field (the $\vb*{E}^\mathsf{T_2+E}$  field), and cannot be taken on their own as an indication of an emergent rank-2 gauge theory. }. 
This analysis therefore highlights the importance of studying the different accessible scattering channels to expose the anisotropic features associated with the realization of an emergent gauge theory, in addition to the development of an accompanying field theory analysis, as elaborated in Sec.~\ref{Sec:irrep.and.inter.tetra}, to unequivocally identify a spin liquid phase and its underlying gauge theory.

As a final remark, we note that by combining the results summarized by the phase diagrams in Figs.~\ref{fig:nonkramer.phase},~\ref{fig:kramer.phase}, and \ref{fig:phase_diagram_connectivity}, we have established that \emph{all} CSLs in Table.~\ref{Table_all_csls} are \emph{continuously} connected without having to pass through any conventional long-range ordered phase in the $\{J_{zz},J_{\pm},J_{\pm\pm},J_{z\pm}\}$ parameter space.
For example, from Figs.~\ref{fig:kramer.phase} and \ref{fig:phase_diagram_connectivity}, we see that the Kramers CSLs 6, 7, 8 and 9, and the non-Kramers CSL 5 connect to each other.
In addition, Fig.~\ref{fig:nonkramer.phase} shows that all the other non-Kramers CSLs connect to the CSL 5. 
The parameter space of the CSLs is delineated by a ``main body'' (i.e. the extended volume phase in parameter space that corresponds to the spin ice state) while various surfaces and lines define the other CSLs that extend out of the body.

So far, we have presented an analysis of the gauge theories describing the low-temperature behavior of all possible CSLs realized by the generic anisotropic nearest-neighbor bilinear Hamiltonian on the pyrochlore lattice.
Moreover, we have identified the features in reciprocal space correlation functions that can allow to an identification of those CSLs.
In the next section, we present a CMC numerical study of the heretofore unexplored new CSLs identified in this work, namely CSL 6 and CSL 9.
In doing so, we compare the CMC results on these models with the predictions obtained from their respective gauge theory and the large-$\mathcal{N}$ approximation.

\section{Beyond Large-$\bm{\mathcal{N}}$: Thermodynamics and structure factors from classical Monte Carlo}
\label{sec:numerics}

In this section, we present the results from CMC simulations of the two previously unidentified CSL 6 and CSL 9 discussed in Sec.~\ref{sec:CSL_kramers}. Our simulations confirm that these unidentified CSLs
do not exhibit magnetic order down to the lowest temperatures. The complete set of all possible CSLs of the most general nearest-neighbor bilinear Hamiltonian is presented in
Table~\ref{Table_all_csls}. With the exception of CSL 1, all of these CSLs correspond to regions in parameter space where two or more symmetry-breaking ordered phases compete. We note, however, that the identification of a model at such boundaries is not a sufficient condition for a CSL to be realized and stable down to the lowest temperatures. Indeed, the models whose ground state manifold is characterized by a minimal $\mathsf{A_2}$ and two additional irrep fields present a symmetry-breaking transition at low temperatures~\cite{Francini2024nematicR2}. This transition is not predicted by the field theory which does not fully capture the $|\vb*{S}|^2 = 1$ constraint of every spin. We discuss these special cases in Sec.~\ref{sec:Unsuccessfuk_CSL} below. 
 
In the following, we present the thermodynamics of  CSLs 6, 7, and 9 obtained via CMC simulations. The CSL 7 has already been discussed in Ref.~\cite{Benton2016NatComm}, but its dual CSL 6, obtained via  Eq.~\eqref{eqn_kramers_duality}, has, to the best of our knowledge,  not been discussed before.
Although the two CSLs have the same effective theory, their structure factors display drastically different patterns. 
The same applies to CSL 9, which is dual to HAFM via the same equation \eqref{eqn_kramers_duality}.
We show the specific heat and spin structure factor for the three CSLs marked in  Fig.~\ref{fig:phase_diagram_connectivity} by golden crosses,  and whose structure factors are displayed in Fig.~\ref{fig:gauge_tunning} in the first (uppermost row), second, and fourth rows. 

For our simulations, we considered a system defined with an FCC unit cell composed of $4L^3$ spins with $L=10$ and updated the system via a Gaussian spin-flip update~\cite{Alzate-Cardona_2019}, and an over-relaxation update~\cite{ZhitomirskyPRL2012,Creutz}, in addition to averaging up to $10$ independent CMC runs. 
We performed $8\times10^4$ thermalization sweeps and $ 2\times10^5$ measurement sweeps where we measured the energy, specific heat, and spin structure factors of the systems considered. 

\begin{figure}[th!]
    \centering
    \begin{overpic}[width=\columnwidth]{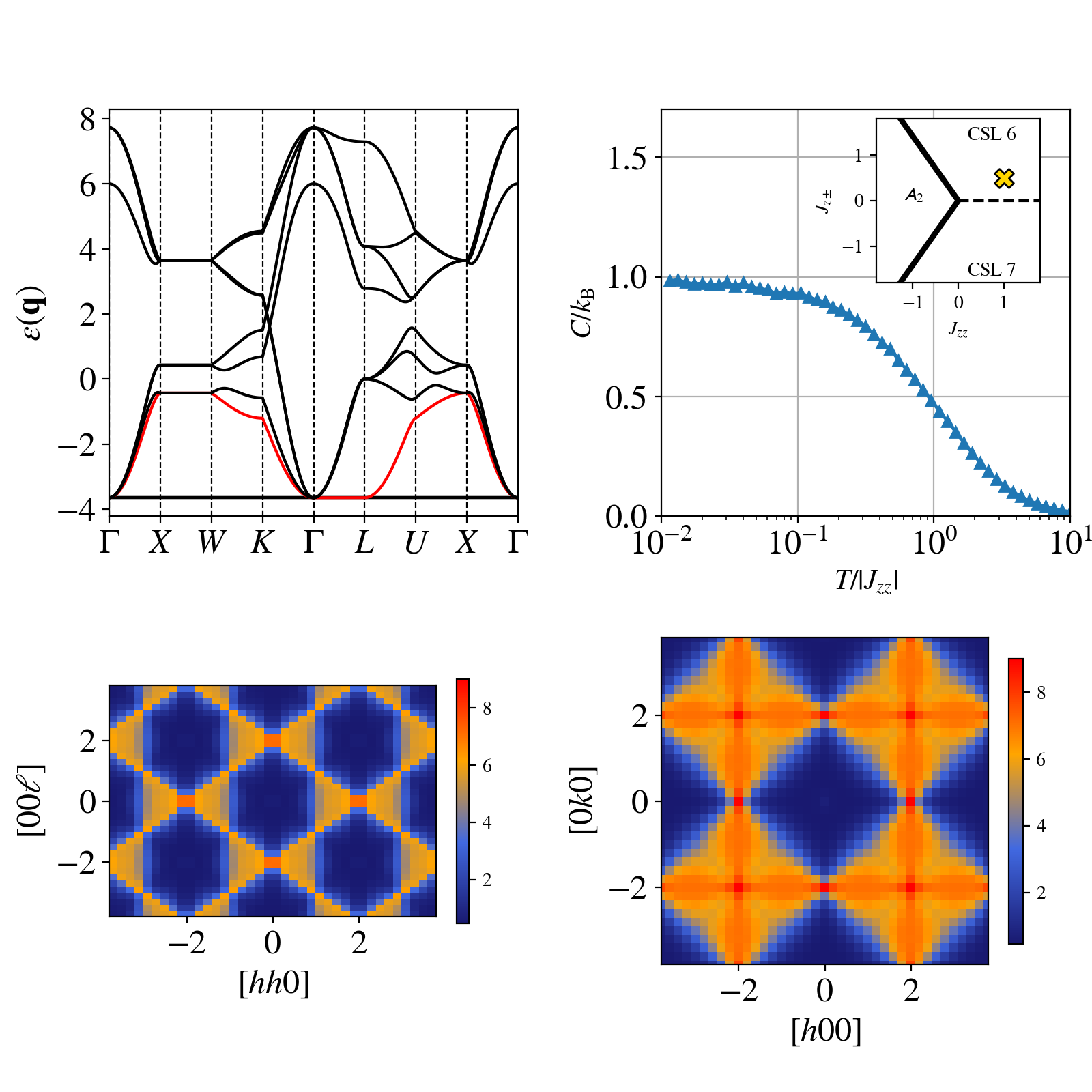}
    \put(0,92) {(a)}  
    \put(50,92) {(b)}  
    \put(0,40) { (c)}  
    \put(50,40) {(d)}  
    \end{overpic}
    \caption{ 
    (a) Diagonalized interaction matrix [Eq.~\eqref{eqn:tetra.hamiltonian}] bands in reciprocal space for CSL 6, using parameters $\{J_{zz}, J_{z\pm} \} =\{1,0.5\}$ [Eq.~\eqref{eq:CL6_constraint}] where the lowest dispersive band is colored in red. 
    There are four flat bands at the bottom in the spectrum. (b) Specific heat of the model, showing no sharp phase transitions from the paramagnetic phase down to the lowest temperature. 
    The inset in this figure shows the location of the model in the phase diagram shown in Fig.~\ref{fig:phase_diagram_connectivity}. 
   (c,d) spin structure factors in the $(hh\ell)$ and $(hk0)$ planes of the model. 
    }
    \label{fig:E_T2_T1planar_Jzpm_pos}
\end{figure}

\begin{figure}[th!]
    \centering
    \begin{overpic}[width=\columnwidth]{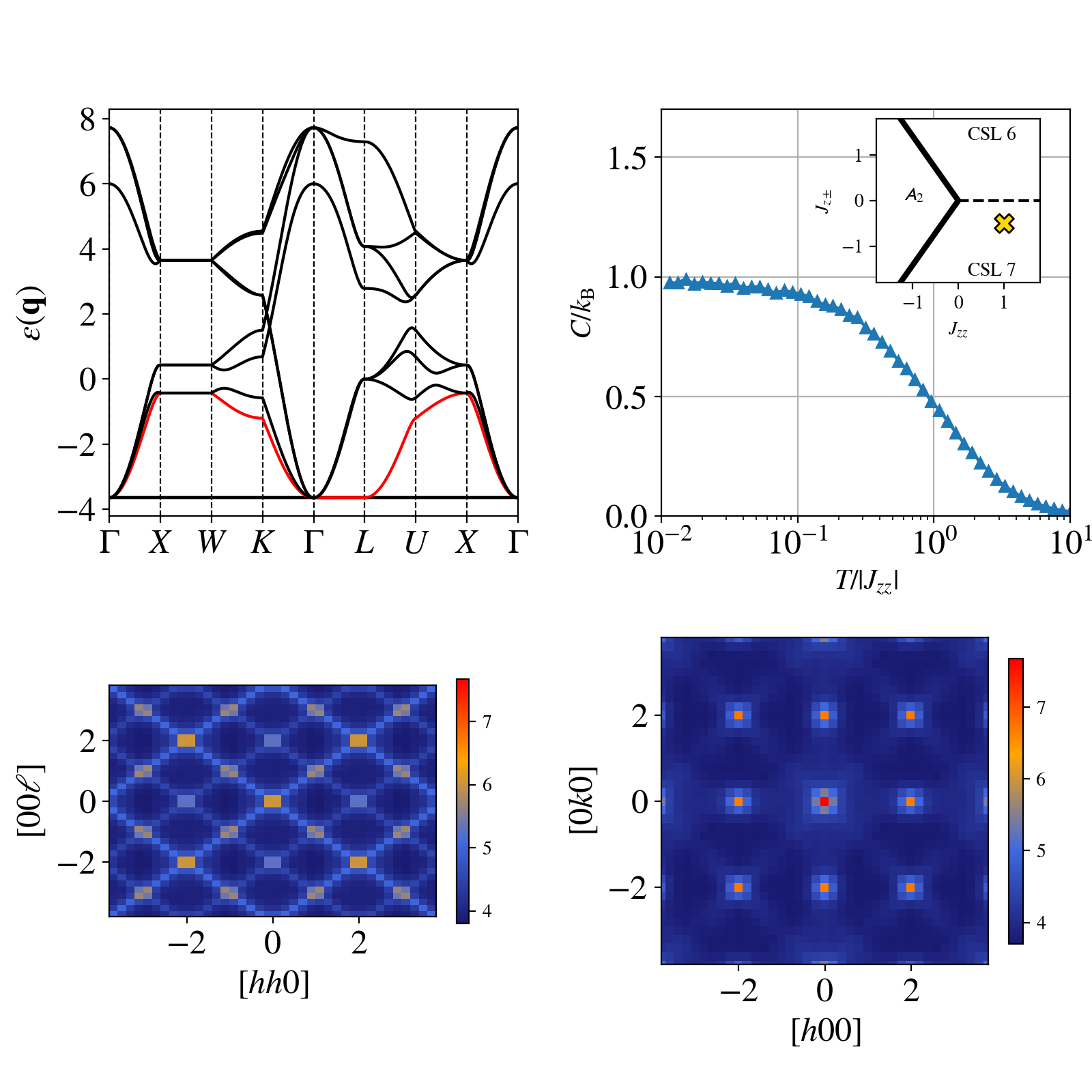}
    \put(0,92) {(a)}  
    \put(50,92) {(b)}  
    \put(0,40) { (c)}  
    \put(50,40) {(d)}  
    \end{overpic}
    \caption{ 
    (a) Diagonalized interaction matrix [Eq.~\eqref{eqn:tetra.hamiltonian}] band in reciprocal space for CSL 7, using parameters $\{J_{zz}, J_{z\pm} \} = \{1, -0.5\}$ [Eq.~\eqref{eq:CL7_constraint}]  where the first dispersive band is colored in red.  
    The spectrum is identical to Fig.~\ref{fig:E_T2_T1planar_Jzpm_pos} due to the duality of Eq.~\eqref{eqn_kramers_duality} between the two models. (b) Specific heat of the model, showing no sharp phase transitions from the paramagnetic phase down to the lowest temperature. 
    The inset in this figure shows the precise location of the model in the phase diagram shown in Fig.~\ref{fig:phase_diagram_connectivity}. 
   (c,d) spin structure factors 
   in the $(hh\ell)$ and $(hk0)$ planes of the model. 
    Although CSL 6 and CSL 7 are dual to each other, their spin structure factor is different. We note that the spin structure factor shown in panel (d) resembles the one predicted by SCGA in Fig.~\ref{fig:gauge_tunning}, where high-intensity signals are observed at the $\Gamma$, $[220]$, and $[200]$ points. }
    \label{fig:E_T2_T1planar_Jzpm_neg}
\end{figure}

Figures~\ref{fig:E_T2_T1planar_Jzpm_pos}  and \ref{fig:E_T2_T1planar_Jzpm_neg} show the  eigenvalue (bands) of the interaction matrix in momentum space, the specific heat, and the spin structure factors  
obtained for the parameters marked by the yellow crosses for the CSL 6 and CSL 7 in Fig.~\ref{fig:phase_diagram_connectivity}, respectively.
As expected, and in agreement with the field analysis performed above, the band structure of the corresponding interaction matrices for these systems host four low-energy flat bands in their spectra (resulting from the 8 degrees of freedom and 4 constraints given by the Gauss's laws), see panel  Figs.~\ref{fig:E_T2_T1planar_Jzpm_pos}(a) and \ref{fig:E_T2_T1planar_Jzpm_neg}(a). 
\dlgnew{The equivalence between the band structure of these two models is a consequence of the duality of the Hamiltonian  [Eq.~\eqref{eq:Hex1}] via Eq.~\eqref{eqn_kramers_duality}. }
Neither model shows a sign of a thermodynamic phase transition, as indicated by the smooth evolution of the specific heat of both systems, see Fig.~\ref{fig:E_T2_T1planar_Jzpm_pos}(b) and Fig.~\ref{fig:E_T2_T1planar_Jzpm_neg}(b). 
The spin structure factors  for both systems are shown in Fig.~\ref{fig:E_T2_T1planar_Jzpm_pos}(c)-(d) and Fig.~\ref{fig:E_T2_T1planar_Jzpm_neg}(c)-(d) for the $(hh\ell)$ and $(hk0)$ reciprocal space planes, respectively, showing excellent agreement with the predictions obtained via large-$\mathcal{N}$ in Fig.~\ref{fig:gauge_tunning}.

\begin{figure}[ht!]
    \centering
    \begin{overpic}[width=1.\columnwidth]{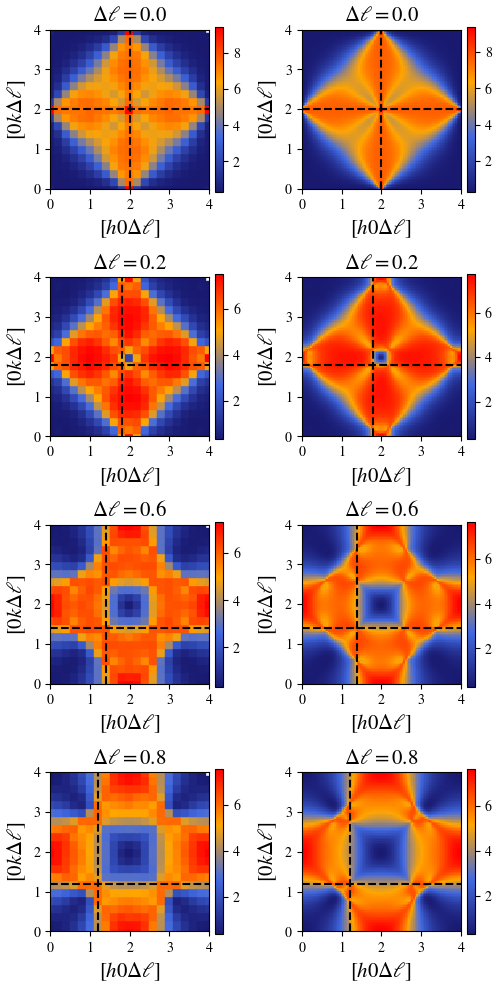}
    \end{overpic}
    \caption{Spin structure factors obtained via classical Monte Carlo simulations at $T=0.03|J_{zz}|$ (left column) and large-$\mathcal{N}$ approximation in the $T\to 0$ limit (right column) for the CSL 6 at various reciprocal plane cuts $(hk\Delta \ell)$, where the value of $\Delta \ell$ is specified at the top of each panel. 
    Here, the black dashed lines are drawn at $h=2-\Delta \ell $ and $k=2-\Delta \ell$. The intersection of these lines marks the $\Delta \ell \times [11\Bar{1}]$ direction which tracks the location of the pinch lines where two-fold pinch points are observed.} \label{fig:E_T2_T1planar_Jzpm_pos_pinch_lines}
\end{figure}

In addition  to the multifold pinch point anisotropies, we note that a distinctive feature in the correlation functions of CSL 7 is the observation of pinch lines, which can be observed in Fig.~\ref{fig:E_T2_T1planar_Jzpm_neg}(c) along the $[111]$ and symmetry-related directions.
In a large-$\mathcal{N}$  theory, these features are produced by a dispersive band which, along the $[111]$ direction, becomes degenerate with the low-energy flat bands, see the first dispersive band marked in red in Fig.~\ref{fig:E_T2_T1planar_Jzpm_pos}(a) and Fig.~\ref{fig:E_T2_T1planar_Jzpm_neg}(a)~\footnote{The reason behind the observation of the pinch-lines in reciprocal space can be tracked to the degeneracy of the low-energy modes in reciprocal space. In the $T\to 0$ limit, the spin correlation functions obtained via SCGA is mainly determined by the contributions of the eigenmodes 
with the smallest eigenvalue.
All others become negligible in the $T\to 0$ limit. In reciprocal space, the pinch-lines are observed for ${\bm q}$-points where a dispersive eigenmodes become degenerate with the low-energy modes. This leads to an enhanced intensity along these directions and therefore the observation of pinch lines.}. Reference~\cite{Benton2016NatComm} noted that for the CSL 7 singular features (i.e.~pinch points) in the reciprocal space correlation function are observed along the location of the pinch line in reciprocal space. In particular, four-fold pinch points can be observed at the intersection of multiple pinch lines. 
Given that the CSL 6 is dual to the CSL 7, pinch lines should naively be observed in the structure factor of CSL 6.
Although it is not immediately clear from the structure factors in Fig.~\ref{fig:E_T2_T1planar_Jzpm_pos}(c) and (d) where these features appear, we may track the location of the pinch lines by studying the spin structure factor in the vicinity of a four-fold pinch point. Figure~\ref{fig:E_T2_T1planar_Jzpm_pos_pinch_lines} illustrates the spin  structure factor for the CSL 6 in the $(hk\Delta \ell)$ plane for increasing values of $\Delta \ell$ obtained via CMC and the large-$\mathcal{N}$ method where the intersection of the black dashed lines tracks the location of one of the four pinch lines intersecting at the $[220]$ point. 
We note that, in both CMC and large-$\mathcal{N}$, as $\Delta \ell$ is increased, the four-fold pinch point (in the $(hk\Delta \ell)$ plane with $\Delta \ell=0$) splits into four two-fold pinch points. 
The location of the two-fold pinch points with respect to the four-fold pinch point is consistent with a vector $\vb*{q}=[\Delta \ell,\Delta \ell,\Delta \ell]$ and symmetry-related directions, corresponding to $\vb*{q}$ vectors where pinch lines are observed.  

\begin{figure}[ht!]
    \centering
    \begin{overpic}[width=\columnwidth]{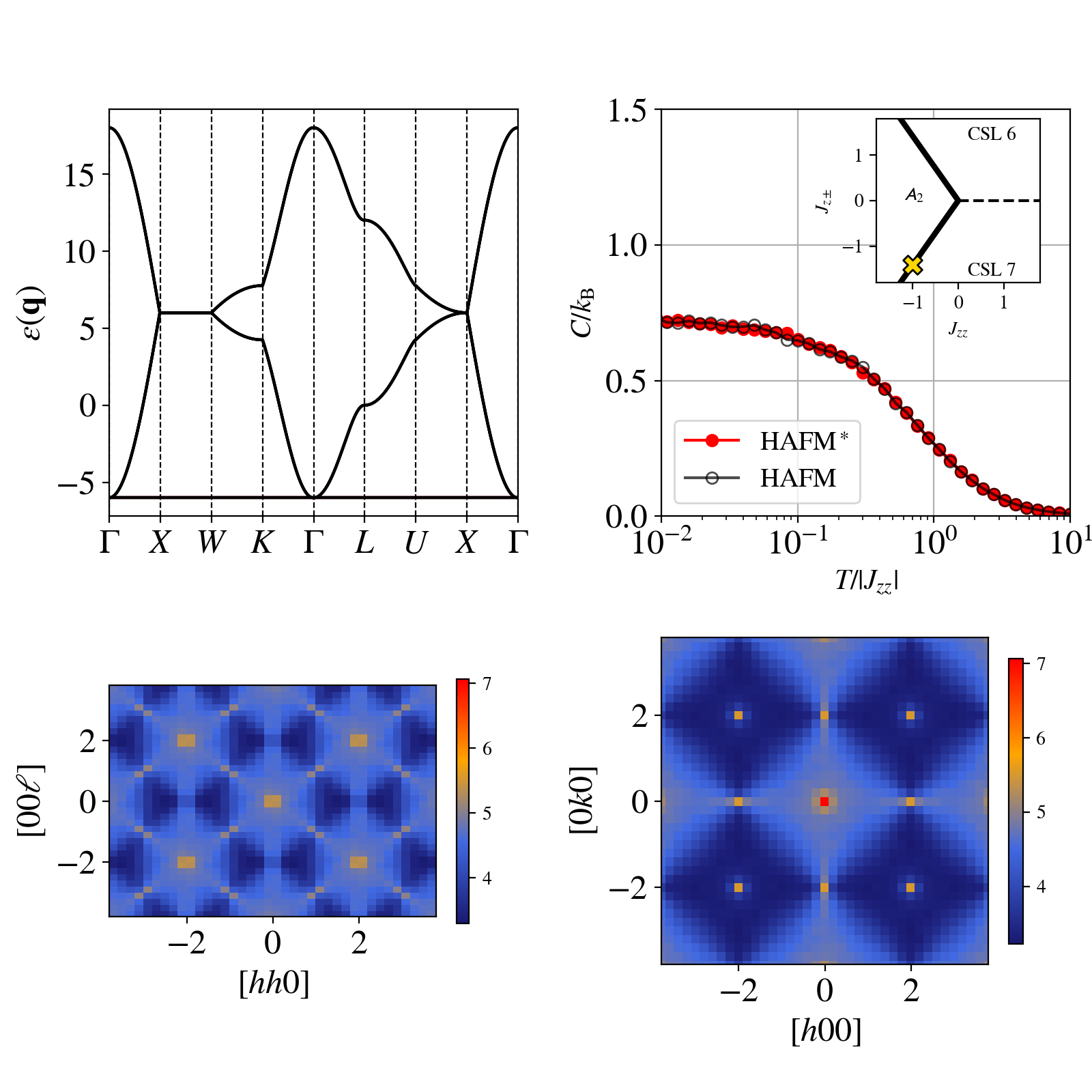}
    \put(0,92) {(a)}  
    \put(50,92) {(b)}  
    \put(0,40) { (c)}  
    \put(50,40) {(d)}  
    \end{overpic}
    \caption{(a) Interaction matrix bands, (b)  specific heat, and structure factors in the $(hh\ell)$ (c) and $(hk0)$ (d) planes of the CSL9, the HAFM* model.
    (a) Diagonalized interaction matrix [Eq.~\eqref{eqn:tetra.hamiltonian}] band in reciprocal space for CSL 9 (HAFM*), using parameters $\{J_{zz}, J_\pm, J_{\pm\pm}, J_{z\pm} \} = \{-1.0, 0.5, 1, -\sqrt{2}\}$. 
    There are six flat bands at the bottom of the spectrum and the dispersive bands are three-fold degenerate. The spectrum is identical to that of CSL 8 (HAFM) due to the duality of Eq.~\eqref{eqn_kramers_duality} between the two models.  (b) Specific heat of the $\rm HAFM^\ast$ and $\rm HAFM$ model, showing no sharp phase transitions from the paramagnetic phase down to the lowest temperature. The inset in this figure shows the precise location of the model in the phase diagram shown in Fig.~\ref{fig:phase_diagram_connectivity}. 
    (c,d) neutron structure factors in the $(hh\ell)$ and $(hk0)$ planes of the model. Although CSL 8 and CSL 9 are dual to each other, their spin structure factor is different. }
    \label{fig:dual_HAFM}
\end{figure}

Lastly, we present in Fig.~\ref{fig:dual_HAFM} the CMC results for CSL 9 corresponding to the dual HAFM model, HAFM*, which is CSL 8.
This model presents the same functional form of the energy bands as the regular HAFM where the six low-energy bands of the interaction matrix are flat; see Fig.~\ref{fig:dual_HAFM}(a). 
Furthermore, and as found for the HAFM~\cite{Reimers1992}, the specific heat of this model presents no sign of symmetry-breaking transition down to the lowest temperatures where it plateaus to a value around the theoretical value $3/4$ ~\cite{Reimers1992,Moessner1998PRB}. 
As in the HAFM, the spin structure factor of the HAFM* exhibits two-fold pinch points, as seen in the $(hk0)$ scattering plane both in CMC, and as predicted by large-$\mathcal{N}$.
We further note that the structure factor of the HAFM$^\ast$ appears as an intensity-inverted version of that of the HAFM. 
In other words, the regions of high intensity in the spin structure factor of one model correspond to regions of low intensity in the dual model. 
This inversion is associated with the precise constraint acting in the ground-state manifold. 
For the HAFM, this constraint translates to a vanishing single-tetrahedron magnetization (which implies vanishing correlations around the $\Gamma$ points).
On the other hand, for the HAFM$^\ast$,  this constraint allows for a ferromagnetic configuration in the ground-state manifold (which results in a finite intensity at the $\Gamma$ points) and forbids the antiferromagnetic configurations with a non-vanishing $\mathsf{T_{1,+}}$ components.

\section{The failed CSL candidates}
\label{sec:Unsuccessfuk_CSL}
In this section, we briefly discuss the failed CSL candidates that sit on the boundaries between three magnetically ordered phases but yet present a symmetry-breaking transition at low temperatures. 
So far we have provided a list of the CSLs stable down to the lowest temperatures and therefore lack any type of conventional symmetry-breaking transition. 
In the large-$\mathcal{N}$ approximation, a CSL is realized whenever the lowest band (or bands) of the interaction matrix is flat, resulting in an extensive degeneracy in the ground states. 
The extensive degeneracy associated with low-energy flat bands in the large-$\mathcal{N}$ approximation, however, does not impede the onset of an order-by-disorder selection of a symmetry-breaking  spin configuration. 
Such entropically driven selection mechanisms are typically described by high-order terms in the free energy which are not captured by the large-$\mathcal{N}$ approximation~\cite{lozanogomez2023arxiv}. 
Consequently, the stability of a CSL must be ascertained using CMC.  

It turns out that, broadly speaking, these models are located at the three edges of the green pyramid of the $\mathsf{A_2}$ phase (all-in-all-out magnetic order) Fig.~\ref{fig:nonkramer.phase}(a) and Fig.~\ref{fig:kramer.phase} for the non-Kramers and Kramers Hamiltonian, respectively. 
These edges correspond to models in parameter space where three $ a_\mathsf{X}$ are minimal, with the constraint that one of these minimal fields must be the $\mathsf{A_2}$ field. 
As a shorthand notation, we refer to the models defined on these edges by the minimal irreps describing their ground states, namely, $\mathsf{A_2}\oplus\mathsf{T_{1,p}}\oplus\mathsf{T_{2}}$ (where the $a_\mathsf{A_2}$, $a_\mathsf{T_{1,p}}$, and $a_\mathsf{T_2}$ are minimal )~\cite{Francini2024nematicR2}, $\mathsf{A_2}\oplus \mathsf{E}\oplus\mathsf{T_2}$ (where the $a_\mathsf{A_2}$, $a_\mathsf{T_{E}}$, and $a_\mathsf{T_2}$ are minimal), and $\mathsf{A_2}\oplus\mathsf{E}\oplus\mathsf{T_{1,p}}$ (where the $a_\mathsf{A_2}$, $a_\mathsf{E}$, and $a_\mathsf{T_{1,p}}$ are minimal) for the non-Kramers case. For the Kramers case, we use the same labeling but replace the $\mathsf{T_{1,p}}$ field by $\mathsf{T_{1,-}}$ field.

An analysis of the large-$\mathcal{N}$ interaction matrix bands for the non-Kramers Hamiltonian models reveals that the $\mathsf{A_2}\oplus\mathsf{T_{1,p}}\oplus\mathsf{T_{2}}$ model possesses four low-energy flat bands in its interaction matrix spectrum, whereas both the $\mathsf{A_2}\oplus \mathsf{E}\oplus\mathsf{T_2}$ and the $\mathsf{A_2}\oplus\mathsf{E}\oplus\mathsf{T_{1,p}}$ models yield two low-energy flat bands. 
On the other hand, for the Kramers models, both the $\mathsf{A_2}\oplus\mathsf{T_{1,-}}\oplus\mathsf{T_{2}}$ model and the $\mathsf{A_2}\oplus \mathsf{E}\oplus\mathsf{T_2}$ yield two low-energy flat bands, while the $\mathsf{A_2}\oplus\mathsf{E}\oplus\mathsf{T_{1,-}}$ has no flat bands in its interaction matrix spectrum. 
As previously mentioned, all of these models (with the exception of the $\mathsf{A_2}\oplus\mathsf{E}\oplus\mathsf{T_{1,-}}$ Kramers model) would have been predicted to be a CSL according to the large-$\mathcal{N}$ analysis. 
However, a detailed analysis of CMC simulation data for these two models reveals that all of them undergo a symmetry-breaking phase transition at low temperatures (not shown in this paper).
We leave the study of details of these failed CSL candidates for future work.

\section{Roadmap of quantum phase diagram}
\label{sec:roadmap}

\dlgnew{
In the preceding sections, we systematically investigated the CSLs that arise within the phase diagram of the classical pyrochlore spin system with anisotropic bilinear nearest-neighbor interactions. 
The parameter regimes in which CSLs appear correspond to regions of maximal frustration where competing interactions suppress conventional magnetic ordering by leaving the energy-minimizing spin configurations underconstrained. 
This frustration is generally expected to persist upon quantization of the model---except, in cases where quantum order-by-disorder intervenes, analogous to the thermal order-by-disorder mechanisms discussed in Sec.~\ref{sec:Unsuccessfuk_CSL}. 
Deep within classically ordered phases, the corresponding quantum models will typically also exhibit symmetry-breaking long-range order, sometimes stabilized by quantum order-by-disorder, closely resembling their classical counterparts.
In contrast, parameter regimes associated with CSLs are much more likely, when carried over to the quantum model, to support quantum spin liquids.
These considerations suggest that the classical results presented in this work may serve as a heuristic baseline for further exploration of the corresponding quantum pyrochlore spin systems. In the remainder of this section, we briefly review what is currently known about the quantum phase diagram, in particular for spin-$1/2$, and highlight the regions that remain largely unexplored, including the CSLs that we have identified above.
}

\dlgnew{
The best known example is quantum spin ice~\cite{Gingras_2014}, a form of quantum spin liquid arising from quantum dynamics applied to the classical spin ice model (CSL 1). A large variety of analytical and quantum numerical techniques have been employed to study its physics~\cite{yan2023experimentally,Gingras2014RPPh,udagawaSpinIce2021,Hermele2004, Huang2018PhysRevLett,shannon2012PRL,leeGenericQuantumSpin2012,Benton2012PhysRevB,Pace2021PhysRevLett,DesrochersPRL,
savaryCoulombicQuantumLiquids2012,Pore2024,Gao2025,Smith_Ce2Zr2O7}.
Another case with a long history of intensive study is the Heisenberg antiferromagnetic model (CSL 8) 
\cite{PhysRevLett.80.2933,PhysRevLett.90.147204,PhysRevB.79.144432,PhysRevX.9.011005,PhysRevX.11.041021,PhysRevLett.126.117204,PhysRevLett.131.096702,pohle2023groundstates12pyrochlore}. 
}

\begin{figure}[th!]
    \centering
    \begin{overpic}[width=\columnwidth]{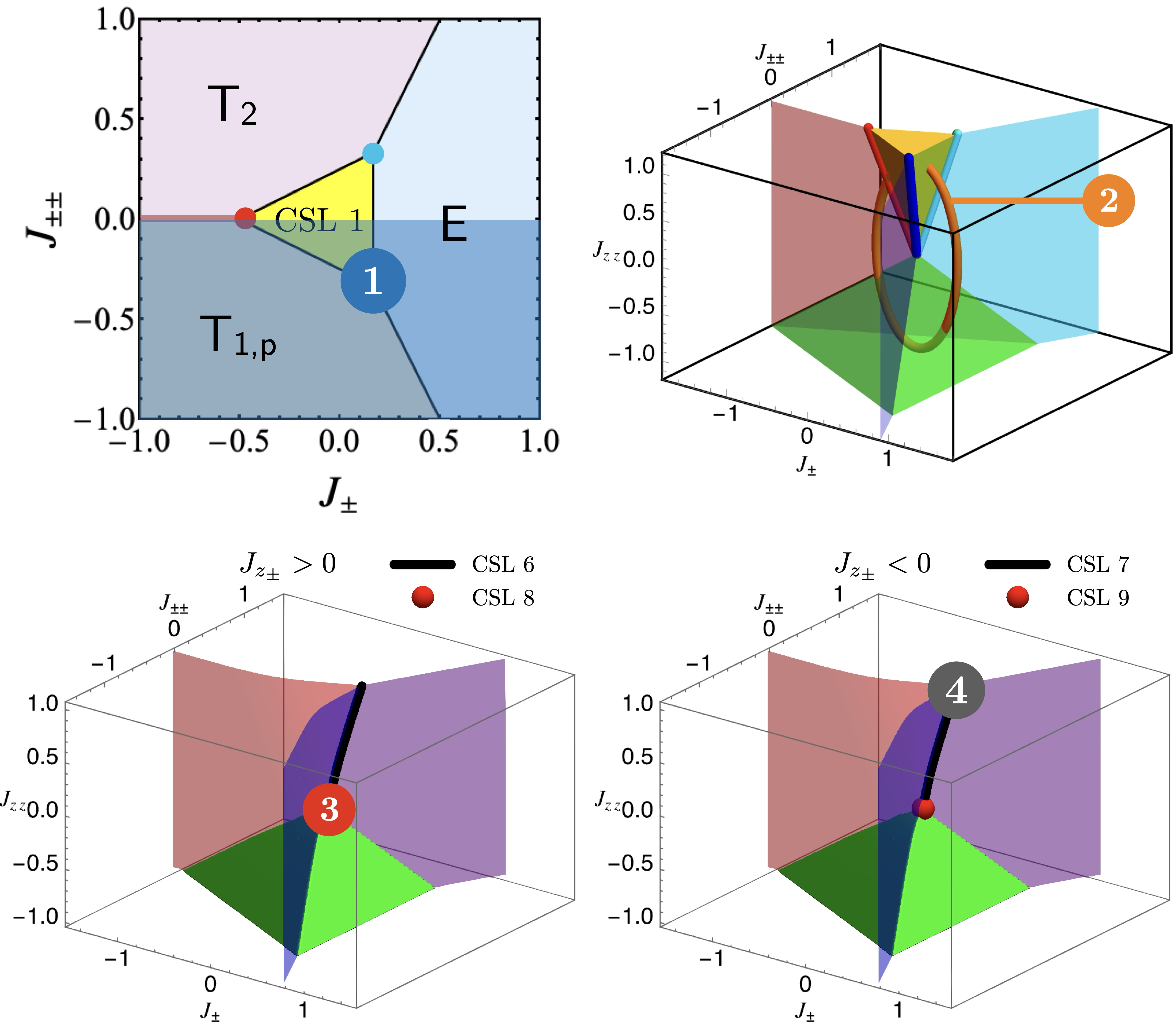}
    \put(0,85) {(a)}  
    \put(50,85) {(b)}  
    \put(0,38) { (c)}   
    \put(50,38) { (d)}   
    \end{overpic}
    \caption{\dlgnew{Classical phase diagram for the non-Kramers Hamiltonian, panels (a) and (b), and Kramers Hamiltonian, panels (c) and (d).
    Here, the colored numerals label in these phase diagrams illustrate the regions in $\{J_{uv}\}$ parameters space for which the $S=1/2$ quantum Hamiltonian has been studied. 
    The numeral labels correspond to the four rows of Table.~\ref{Table_quantum_phase}. 
    In panel (c) and (d), the Kramers Hamiltonian phase diagram is obtained for $J_{z\pm}>0$ and for $J_{z\pm}<0$, respectively, with $|J_{z\pm}|=0.1$. 
    These two panels show identical phase diagrams as a consequence of the duality of Eq.~\eqref{eqn_kramers_duality} for the Kramers Hamiltonian. 
    The CSLs labeled in panel (c) correspond to the CSL 6 and the CSL 8 the HAFM spin liquid, whereas in panel (d) the CSLs are the CSL 7 and the CSL 9. 
    In panel (d), the label `4' indicates a particular realization of CSL 7, whose immediate vicinity was studied in Ref.~\cite{Gresista_QPLSL}.
    }
    } 
    \label{fig:Fig_quantum_all}
\end{figure}

\dlgnew{
Departing from the above two prototypical classical spin liquids and their quantum counterparts, recent efforts have been made to understand other parts of the phase diagram with high interaction frustration.
One of them was targeted at the neighborhood of CSL 4 identified in Ref.~\cite{lozanogomez2023arxiv}. 
A very recent work~\cite{SchadenPhysRevB.111.134442} explored a one-parameter trajectory in the non-Kramers phase diagram, namely
\begin{equation}
    \begin{split}
        J_{zz}&=\cos\theta \\
        J_{\pm}&= -\frac{1}{2}\sin \theta \\ 
        J_{\pm\pm} &=0 \\  
        J_{z\pm}&=0 
    \end{split} 
\label{eqn:quantum_non-Kramers_path}
\end{equation} 
which realizes the CSL 1, 2, and 3 for $\theta\in[-0.1 \pi ,0.25\pi]$, $\theta\in [0.25\pi,0.602\pi]$, and $\theta=0.25\pi$, respectively. 
Furthermore, in the Kramers phase diagram, Ref.~\cite{Noculak_PhysRevB.107.214414} studied the Heisenberg plus Dzyaloshinskii-Moriya Hamiltonian with the parametrization 
\begin{equation}
    \begin{split}
        J_{zz}&=\frac{1}{3}\left(-\cos\theta-4\sin\theta\right) \\
        J_{\pm}&=\frac{1}{6}\left(\cos\theta-2\sin\theta\right) \\
        J_{\pm\pm}&=\frac{1}{3}\left(\cos\theta+\sin\theta\right)\\
        J_{z\pm}&=\frac{1}{3\sqrt{2}}\left(2\cos\theta-\sin\theta\right)  
    \end{split} \label{eqn:quantum_1_para}
\end{equation}
which, in the classical model, realizes the CSL 4 and CSL 8 for $\theta=\pi+\arctan(2)$ and $\theta=0$, respectively. 
We note in passing that for $\tan\theta=2$, the parametrization in Eq.~\eqref{eqn:quantum_1_para} yields $J_{z\pm}=0$, which is enforced for the non-Kramers system.
More recently, a quantum phase diagram for a 2D region of the Kramers Hamiltonian with $J_{z\pm} < 0$ in the vicinity of CSL 7 was reported in Ref.~\cite{Gresista_QPLSL}. 
Since the phase diagram in that work was parameterized in terms of $J_1,J_2,J_3,J_4$ (see Appendix B, Eqs.~\eqref{eqn:j1}-\eqref{eqn:j4}, for the relation between different notation conventions  of the exchange parameters), the full range of parameters explored in Ref.~\cite{Gresista_QPLSL} cannot be entirely displayed in the phase diagram of Fig.~\ref{fig:Fig_quantum_all}(d) (where $J_{z\pm}$ was fixed to $J_{z\pm}=-0.1$; see caption of that figure) as the interaction parameter $J_{z\pm}$ varies as a function of the couplings $J_1,J_2,J_3,J_4$.
These recent works on QSLs are summarized in Table~\ref{Table_quantum_phase}, and their location in the phase classical diagrams indicated in Fig.~\ref{fig:Fig_quantum_all}.
}

\begin{table}[th!]
\def\arraystretch{1.5}%
\caption{\label{Table_quantum_phase}
  \dlgnew{Quantum spin models of the pyrochlore spin-1/2 Hamiltonian numerically investigated in the literature. The different parameter regimes are shown in the pertinent panels of  Fig.~\ref{fig:Fig_quantum_all}.
  }
    }
    \vspace*{5mm}
\begin{tabular}{|c|c|c|}
\hline Label  & Parameter regimes & Reference  \\
\hline \circo[myblue1]
{\small{\textcolor{white}{1}}} &  CSL 4, and shadowed  blue rectangle &  \cite{lozanogomez2023arxiv}\\
\hline \circo[orange]{\small{\textcolor{white}{2}}}  & $(J_z, J_\pm) = (\cos\theta,-\tfrac{1}{2}\sin\theta)$; orange ring & \cite{SchadenPhysRevB.111.134442}\\
\hline \circo[red]{\small{\textcolor{white}{3}}}  & CSL 8, Eq.~\eqref{eqn:quantum_1_para} with $\theta=0$; red dot & \cite{Noculak_PhysRevB.107.214414} \\
\hline \circo[gray]{\small{\textcolor{white}{4}}} &  Vicinity of CSL 7; gray dot & \cite{Gresista_QPLSL} \\
\hline
\end{tabular}
\end{table}

\dlgnew{
To summarize, using a variety of numerical schemes, a number of studies have found evidence for the existence of quantum spin liquid phases that surround locations or regions where the classical spin liquids  discussed earlier in this paper are realized. 
Apart from the extensively studied quantum variants of classical models located near CSL 1 (classical spin ice) and CSL 8 (Heisenberg antiferromagnetic model), very recent results are providing some insight on the possibility of quantum spin liquids in the vicinity of CSL 4~\cite{lozanogomez2023arxiv} and CSL 7~\cite{Gresista_QPLSL}.
By the duality of the Kramers and non-Kramers Hamiltonian in Eq.~\eqref{eqn_kramers_duality} and Eq.~\eqref{eqn_non_kramer_duality}, respectively, CSLs 4, 7 and 8 listed in Table ~\ref{Table_quantum_phase}  are dual to CSLs 5, 6, and 9, respectively. 
We thus expect the thermodynamic properties and the low-energy effective theories of these last three CSLs  to be identical to those of CSLs 4, 7 and 8, respectively.
From this brief survey, we note, however, that various regions in parameter space where a CSL is realized have not been systematically studied in the corresponding quantum model. 
Studying these unexplored regions yields the potential of realizing novel quantum spin liquid as well as non-conventional magnetic orders which are not observed in the classical model~\cite{Gresista_QPLSL}. 
}

\dlgnew{
\textbf{Underexplored quantum region 1:} 
The extent and precise nature of the disordered state observed in the quantum $S=1/2$ model for the $\{J_{uv}\}$ parameters corresponding to the CSL 2
[see phase diagram Fig.~\ref{fig:nonkramer.phase}(a)] and its immediate vicinity in the non-Kramers phase diagram has not been fully elucidated.
Reference~\cite{lozanogomez2023arxiv} found no conventional magnetic order in the vicinity of the CSL 2 and in the proximity of CSL 3~\footnote{Ref.~\cite{lozanogomez2023arxiv}, using energy scale set by $J_{zz}=3$, identified a region of no conventional magnetic order for which $J_{\pm\pm}/J_{zz}\in [-0.5,0.0]$ in the vicinity of the line of $\{J_{uv}\}$ parameters defining  CSL 2, which would also encompass the region 
$J_{\pm\pm}/J_{zz}\in [0.0,0.5]$ by invoking the duality~\eqref{eqn_non_kramer_duality}.} for a finite range of the $J_{\pm}$ and $J_{\pm\pm}$ parameters. 
In addition, Ref.~\cite{SchadenPhysRevB.111.134442} reported an absence of magnetic order for the $\{J_{uv}\}$ parameters defining  CSL 2 [i.e.,  where $J_{\pm\pm}=0$ and \emph{all} $J_{zz}$ and $J_{\pm}$ values on the classical phase boundary, i.e. red plane in Fig.~\ref{fig:Fig_quantum_all}(b)]. 
Putting the results of Ref.~\cite{lozanogomez2023arxiv} and Ref.~\cite{SchadenPhysRevB.111.134442} together, it is not clear whether the finite 
range of $J_{\pm\pm}\neq0$ for which no order is seen in Ref.~\cite{lozanogomez2023arxiv} (for fixed $J_{zz}$) extends over the complete classical phase boundary of CSL 2 (of varying $J_{zz}$, see phase diagram Fig.~\ref{fig:nonkramer.phase}(a)) or whether a quantum spin liquid state with $|J_{\pm\pm}|\neq 0$ is realized only in the vicinity of the $\{J_{uv}\}$ parameters of CSL 3. 
While it seems plausible that the region near CSL 2, but within the CSL 1 (classical spin ice) is a $\pi$-flux quantum spin ice~\cite{SchadenPhysRevB.111.134442,Desrochers-symmetry_fract,Desrochers-exclusive_bosons}, the nature of the QSL, or QSLs, in the vicinity of the surface defined by CSL 2 in the phase diagram of Fig.~\ref{fig:nonkramer.phase}(a) is rather unknown at the present time.
}

\dlgnew{
\textbf{Underexplored quantum region 2:}  
Another region of interest in the  $\{J_{uv}\}$
parameter space that has not yet been systematically explored in the quantum model is the one connecting CSL 8 and CSL 6 and, by the $J_{z\pm} \rightarrow -J_{z\pm}$ duality, CSL 9 and CSL 7, respectively [see Fig.~\ref{fig:Fig_quantum_all}(c) and (d)].
In other words, the question here is what might the quantum analogue of the classical phase diagram shown in Fig.~\ref{fig:phase_diagram_connectivity} be.
As stated above, CSL 6 and CSL 9 have, to the best of our knowledge, not been studied before and prior to the present work (see  Table~\ref{Table_all_csls}). 
We first note that a QSL ``similar'' to CSL 7 was found in the pseudofermion functional renormalization-group (PFFRG) calculations of Ref.~\cite{Gresista_QPLSL}.
In that work, the anisotropic features observed in the correlation functions of the identified QSL, i.e. pinch-lines as well as twofold and fourfold pinch points, were found to resemble those realized for CSL 7.
The observation and resemblance of these features suggest that this QSL is a pinch-line rank-2 quantum spin liquid, i.e. the quantum analogue of the classical pinch-line spin liquid~\cite{Benton2016NatComm}. 
As discussed in Ref.~\cite{Gresista_QPLSL}, the pinch lines, twofold and fourfold pinch point features observed in the correlation functions of the quantum model descend from those observed in the classical model, which we discussed in Sec.~\ref{sec:Connectibity}, as shown in the second and fourth row of Fig~\ref{fig:gauge_tunning}, and in Sec.~\ref{sec:numerics}, Fig.~\ref{fig:E_T2_T1planar_Jzpm_pos_pinch_lines}.
On the other hand, Refs.~\cite{PhysRevX.9.011005,Noculak_PhysRevB.107.214414} showed, again using PFFRG, that the structure factors of the QSL observed for the $\{J_{uv}\}$ parameters corresponding to the CSL 8 (the Heisenberg antiferromagnet) exhibit only twofold pinch points, \dlgnew{indicating} 
that the underlying theory for this QSL is described by a rank-1 field -- as naively expected~\cite{Benton2012PhysRevB}.
}

\dlgnew{Referring to the classical phase diagram illustrated in Fig.~\ref{fig:Fig_quantum_all}(c) and (d), the two QSLs discussed above, that is the quantum pinch-line spin liquid and the quantum Heisenberg antiferromagnet spin liquid, as well as their respective dual partners, may be continuously connected in parameter space by following the path indicated by the black line connecting CSLs 6 and 8 (for $J_{z\pm}>0$) or CSLs 7 and 9 (for $J_{z\pm}<0$).
These two paths can also be viewed as those indicated in  Fig.~\ref{fig:phase_diagram_connectivity}, but without crossing the CSL 5.
}

\dlgnew{To summarize, it would appear worthwhile to study  the low-temperature properties of these four presumed QSLs that emerge from CSL 6 (and its dual, CSL 7) and CSL 8 (and its dual, CSL 9) for example, using pseudo Marojana fermion RG~\cite{Schaden-XXZ}.
Such an approach would allow to explore the $T>0$ properties of these models
In particular, a detailed exploration of the anisotropic reciprocal space structures of their spin-spin correlations to provide valuable insights into their  effective low-energy quantum gauge theory, and how the latter relates or not
at $T>0$ with the classical gauge theories described in the present work.
}

\section[Proxy]
{Proximity to a classical spin liquid and materials outlook}
\label{sec:proximity_to_CSL}

\dlgnew{
In this section, we first discuss how the CSLs occupying a zero-measure region of the parameter space at zero temperature control a finite parameter volume at finite temperature, making them relevant to actual materials. 
}

\dlgnew{In the previous sections, we identified all CSLs of the most general bilinear nearest-neighbor spin Hamiltonian on the pyrochlore lattice that remain stable down to  $T=0$.
As illustrated in Table~\ref{Table_all_csls}, apart from CSL 1 (classical spin ice) all other $T=0$ CSLs are restricted to a fined-tuned sub-dimensional manifold of the full four-dimensional $\{J_{uv}\}$ parameter space. 
However, previous works~\cite{Yan-2017,Taillefumier2017PhysRevX} have shown that systems with couplings $\{J_{uv}\}$ close to those of a particular CSL often exhibit signatures above their transition temperature to long-range order -- most notably finite-temperature spin structure factors -- that resemble those of the nearby CSL.
Having noted above that several CSLs can ``collide'' or ``intersect'' in the $\{J_{uv}\}$ parameter space (see Fig.~\ref{fig:phase_diagram_connectivity}), and building on the observations of Refs.~\cite{Yan-2017,Taillefumier2017PhysRevX}, we illustrate in 
subsection~\ref{sec:proxy-liquid-model}  below how proximity to \emph{two} ``nearby'' CSLs influences the physics of a system in its cooperative paramagnetic regime above the critical temperature where it develops long-range order.
In the following subsection~\ref{sec:proxy-liquid-materials}, we provide a survey of some rare-earth pyrochlore materials that either naturally reside in the proximity of a CSL, or may be experimentally tuned to lie close to one thanks to departure from perfect stoichiometry.
}

\subsection{An example: all-in--all-out long-range order with adjacent CSL 8 and nearby CSL 6}
\label{sec:proxy-liquid-model}

\begin{figure}[ht!]
    \centering
    \begin{overpic}[width=\columnwidth]{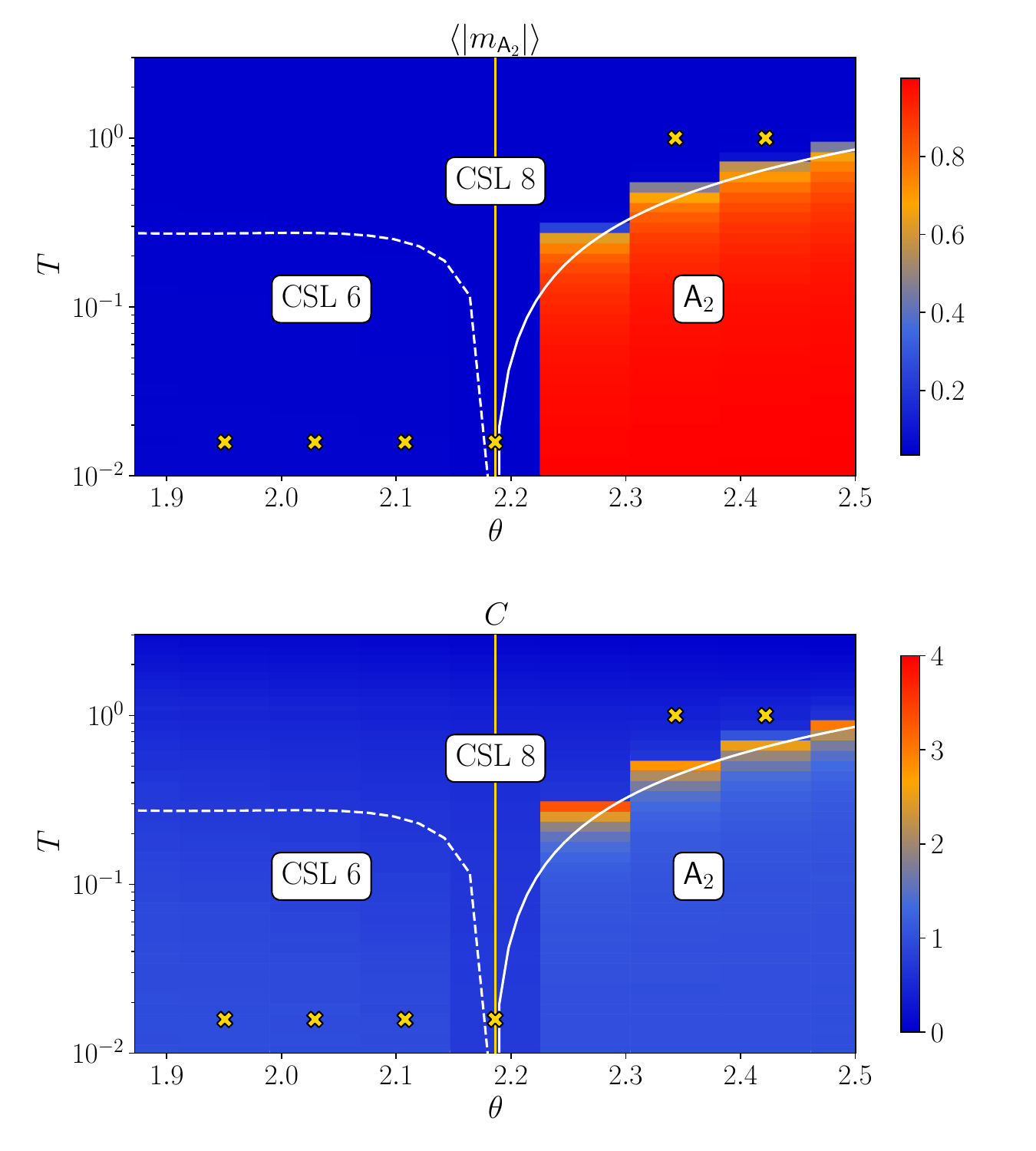}
    \put(7,100) {(a)}  
    \put(7,50) {(b)}  
    \end{overpic}
    \caption{\dlgnew{(a) $\mathsf{A_2}$ order parameter, and (b) specific heat, obtained via CMC for a family of models parametrized by  $(J_{zz},J_{z\pm})=(\cos\theta,\sin\theta)$, where the remaining $J_{\pm}$ and $J_{\pm\pm}$ couplings are given by Eq.~\eqref{eq:CL6_constraint}.
    The white dashed and solid lines in these figures are guides to the eye for the crossover between CSL 6 and CSL 8, and the phase transition between the CSL 8 and the all-in-all-out (AIAO, $\mathsf{A}_2$) phase, respectively. 
    The vertical gold colored line marks the angle $\theta$ corresponding to the $\{J_{uv}\}$ parameters for the CSL 8.
    In these plots, the temperature is measured in units of $\sqrt{J_{zz}^2+J_{z\pm}^2}$ 
    with $k_{\textrm B}\equiv 1$. 
    The gold crosses mark the $(\theta,T)$ values for which the spin structure factors in Fig.~\ref{fig:phase_diagram_structure} were computed. }
     }
    \label{fig:phase_diagram_A2_C}
\end{figure}

\dlgnew{To illustrate the effects of the proximity of two CSLs on  the development of a long-range-ordered phase upon cooling from the paramagnetic phase, we consider, as an example, the phase diagram shown in Fig.~\ref{fig:phase_diagram_connectivity}.
We introduce the parametrization $(J_{zz},J_{z\pm})=(\cos\theta,\sin\theta)$ with $\theta$ varying in the vicinity of the value where the CSL 8 is stabilized, i.e. $\theta_{\mathrm{CSL\ 8}}\simeq 2.19$, while the remaining $J_{\pm}$ and $J_{\pm\pm}$ interaction parameters are given by Eq.~\eqref{eq:CL6_constraint}. 
Figure~\ref{fig:phase_diagram_A2_C} illustrates the temperature evolution of the thermally averaged $\mathsf{A}_2$ order parameter, $\langle \vert m_{\mathsf{A}_2}\vert \rangle$,  and specific heat, $C$, obtained via CMC in panels (a) and (b), respectively.
In this figure, the dashed white line serves as a guide to the eye to mark the CSL 6--CSL 8 crossover, obtained by tracking the inflection point of $\langle m_\mathsf{A_2}^2\rangle $ evolution~\footnote{One could also identify an additional crossover line between the CLS 8 and the paramagnetic phase obtained via mean-field theory~\cite{Enjalran2004TbTiO} given as the $\rm{max}_{m,\bm q}[\varepsilon_m(\bm q)]/3$, where $\varepsilon_m(\bm q)$ are the eigenvalues of the interaction matrix $\bm{J}_{\mu\nu}^{\alpha,\gamma}(\bm q)$. 
However, this crossover is predominantly above the highest temperature considered in Fig.~\ref{fig:phase_diagram_A2_C} and is therefore omitted}. On the other hand, the solid white line labels the symmetry-breaking transition between the CSL 8 and the $\mathsf{A}_2$ long-range ordered phase. 
For more details on the definitions of these lines, we refer the reader to Appendix~\ref{appedix:Thermodynamics_temp}.
As shown in panels (a) and (b) of Fig.~\ref{fig:phase_diagram_A2_C}, as $\theta$ exceeds $\theta_{\mathrm{CSL\ 8}}$, the system enters the $\mathsf{A}_2$  phase, characterized by a non-zero $\mathsf{A}_2$ order parameter below a  critical temperature $T_c(\theta)$ where the specific heat displays a sharp peak [indicated by the color bar of Fig.~\ref{fig:phase_diagram_A2_C}(b)].
}

\begin{figure}[t!]
    \centering
    \begin{overpic}[width=\columnwidth]{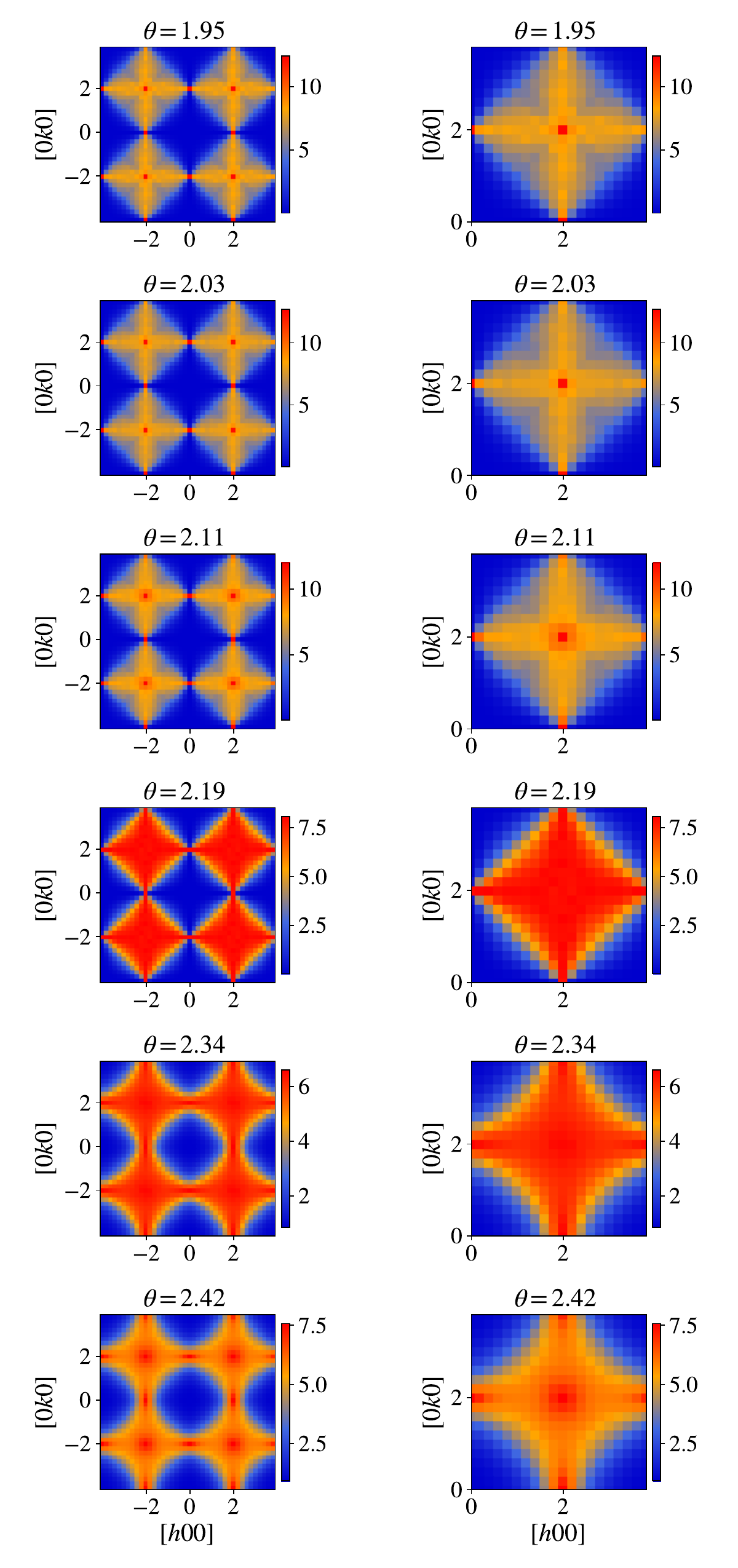}
    \put(2,98) {(a)}  
    \put(25,98) {(b)}  
    \put(2,82) {(c)}  
    \put(25,83) {(d)}  
    \put(2,66) {(e)}  
    \put(25,66) {(f)}  
    \put(2,50) {(g)}  
    \put(25,50) {(h)}  
    \put(2,34) {(i)} 
    \put(25,34) {(j)} 
    \put(2,18) {(k)} 
    \put(25,18) {(l)} 
    \put(15,95){\fontsize{7}{16}\textcolor{white}{$\swarrow$}}
    \put(13,93){\fontsize{7}{16}\textcolor{white}{$\nearrow$}}
    \put(15,93){\fontsize{7}{16}\textcolor{white}{$\nwarrow$}}
    \put(13,95){\fontsize{7}{16}\textcolor{white}{$\searrow$}}
    \put(9.4,91.3){\fontsize{7}{16}\textcolor{white}{$\leftarrow$}}
    \put(7.6,91.3){\fontsize{7}{16}\textcolor{white}{$\rightarrow$}}
    \put(9.4,42.8){\fontsize{7}{16}\textcolor{white}{$\leftarrow$}}
    \put(7.6,42.8){\fontsize{7}{16}\textcolor{white}{$\rightarrow$}}
    \put(39,95.5){\textcolor{white}{$\swarrow$}}
    \put(31,95.5){\textcolor{white}{$\searrow$}}
    \put(31,87.5){\textcolor{white}{$\nearrow$}}
    \put(39,87.5){\textcolor{white}{$\nwarrow$}}
    \end{overpic}
\caption{\dlgnew{Spin structure factors in the $(hk0)$ plane from CMC for two ranges of reciprocal vectors for the same $\theta$ parametrization as in Fig.~\ref{fig:phase_diagram_A2_C} and at temperatures $T=0.015$ [panels (a-h)] and $T=1$ [panels (i-l)]. 
Each row corresponds to a particular $\theta$  (given above each panel) corresponding to the gold crosses in Fig.~\ref{fig:phase_diagram_A2_C}(a,b).
In panels (a) and (g), the four (inward-pointing) and two (sideways-pointing) white arrows indicate the twofold and fourfold pinch points, respectively.
Panel (b) is a zoom on the fourfold pinch point at $220$
observed in panel (a).}
}
    \label{fig:phase_diagram_structure}
\end{figure}

\dlgnew{
The effect of tuning the parameter $\theta$ on the low-temperature physics of the system can be further exposed by studying the evolution of the spin structure factor, ${\mathcal{S}}(\vb*{q})$, as a function of this parameter. 
Figure~\ref{fig:phase_diagram_structure} illustrates this structure factor for six sets of $\{J_{uv}\}$ parameters at temperatures and angles corresponding to those labeled by golden crosses in Fig.~\ref{fig:phase_diagram_A2_C}. 
In this figure, each row corresponds to a set of correlation functions sampled via CMC in the $(hk0)$ plane at an angle $\theta$ and temperature $T$ corresponding to the gold crosses in Fig.~\ref{fig:phase_diagram_A2_C}.
For panels (a-f) the system is at a finite temperature within CSL 6, where twofold and fourfold pinch point features in ${\mathcal{S}}(\vb*{q})$ are observed, as indicated by the white arrows in panels (a) and (b) of Fig.~\ref{fig:phase_diagram_structure}, respectively. 
The presence  of fourfold pinch points is a consequence of the effective low-temperature U(1) rank-2 theory describing the spin liquid phase. 
We note that, as $\theta$ increases and the $\{J_{uv}\}$ parameters approach those defining CSL 8, the energy cost associated with the $\mathsf{A_2}$ field decreases, leading to a (temperature-driven) proliferation of the (progressively decreasing energy) gauge charges associated with this field and, consequently, the broadening of the fourfold pinch points, see Fig.~\ref{fig:phase_diagram_structure}(b), (d), and (f). 
When the system enters the CSL 8 phase, the $m_\mathsf{A_2}$ field becomes part of the ground-state manifold.
In the CSL 8, only twofold pinch points remain, and the effective low-temperature theory is now described by three copies of a rank-1 field, where each field fulfills a U(1) Gauss' law, see Fig.~\ref{fig:phase_diagram_structure}(g-h)%
}

\dlgnew{
For 
the two parameter sets which eventually result in long-range all-in/all-out 
($\mathsf{A_2}$) order, broad pinch point features at temperatures above the critical transition are observed in ${\mathcal{S}}(\vb*{q})$, 
see Fig.~\ref{fig:phase_diagram_structure}(i-l). 
The broad pinch points are associated with the closest CSL, i.e. CSL 8, where the broadness of these reflect a thermal population of the gauge charges associated to the CSL.
Indeed, the spin structure factors obtained above the $\mathsf{A_2}$ ordered phase do not show the modulation around the $\langle 220\rangle$ points associated with the fourfold pinch point observed in the CSL 6 phase, compare Fig.~\ref{fig:phase_diagram_structure}(j) and (l) with Fig.~\ref{fig:phase_diagram_structure}(b), (d), and (f). 
As the system approaches the critical temperature $T_c(\theta)$ from the paramagnetic phase, the pinch point features become sharper (signaling the progressive depopulation of the gauge charges associated with the CSL 8), however, the overall intensity of the correlation function starts to cluster around the points in reciprocal space where sharp magnetic Bragg peaks associated with the $\mathsf{A}_2$ order (i.e. the $\langle 200\rangle$ and $\langle 220\rangle$ points) start to progressively develop, see Fig.~\ref{fig:phase_diagram_structure}(j) and (l). 
We note that the correlation functions shown in  Fig.~\ref{fig:phase_diagram_structure}(i-l) do not possess features that can be associated with the fourfold pinch points or pinch lines, implying that the physics in the cooperative paramagnetic regime are mainly determined by the closest CSL, which in this case is CSL 8, and not CSL 6.
The evolution of the anisotropic features observed in ${\mathcal{S}}(\vb*{q})$ as a function of $T$ and $\theta$ therefore indicates that a spin liquid regime prevails above the transition temperature in ordered phases, placing the emergent gauge field description and its associated gauge charges as a natural framework to describe phenomena observed in these temperature regimes.
}

\dlgnew{
Although we have considered in Fig.~\ref{fig:phase_diagram_A2_C} and \ref{fig:phase_diagram_structure} a specific parametrization of the $\{J_{uv}\}$ couplings to illustrate the manifestation of anisotropic features associated to a CSL at a temperature above an ordered phase, similar results generally apply to  other CSLs.
This suggests that, even though the CSLs we identified are constrained to low-dimensional manifolds in the four-dimensional $\{J_{uv}\}$ parameter space, any model that falls in the vicinity of a CSL will display features associated with the competition characteristic of the proximate CSL at intermediate and high temperatures. 
As such, our study is not merely on fine-tuned parameters at the zero-temperature phase boundary --- it is also relevant to the physics in a finite volume of parameter space near the phase boundary at finite temperature which is strongly influenced by the nearby zero-temperature CSLs.
}

\subsection{Rare-earth pyrochlore materials in close proximity to a CSL}
\label{sec:proxy-liquid-materials}

\dlgnew{We now ask whether there exist materials with magnetic rare-earth ions on the pyrochlore lattice that lie close to one of the CSLs identified in Sections~\ref{sec:CSL_non_kramers} and~\ref{sec:CSL_kramers}. 
As illustrated in Subsection~\ref{sec:proxy-liquid-model}, a material described by Eq.~\eqref{eq:Hex1} may, even without being precisely tuned to one of these CSLs, exhibit spin correlations over a finite temperature window that are inherited from that nearby CSL. 
In this sense, although such a material may ultimately avoid the CSL and instead develop long-range order at low temperatures, an understanding of the emergent gauge fields of the neighboring CSL, particularly their associated temperature-dependent structure factor, remains essential for rationalizing the material's properties in the collective paramagnetic regime above the ordering temperature.
On the other hand, a material whose exchange couplings $\{J_{uv}\}$ place it close to a CSL, yet fails to develop long-range order down to the lowest temperatures while exhibiting a structure factor that deviates significantly from the predictions of the corresponding gauge theory, would point to the presence of additional physics. 
Such effects could arise, for example, from extra interactions (e.g., beyond nearest-neighbor) or from quantum corrections to the classical gauge theory~\cite{lozanogomez2023arxiv}. 
Observations of this kind would therefore motivate extensions of the theoretical framework developed above.
With this perspective in mind, we now turn to a discussion of several insulating rare-earth pyrochlore materials that appear to lie in close proximity to some of the CSLs considered above.
}

\subsubsection{Tb$_2$Ti$_2$O$_7$}
\label{sec:proxy-liquid-materials-TTO}

\dlgnew{
 Along with the Ho$_2$Ti$_2$O$_7$ and Dy$_2$Ti$_2$O$_7$ spin ice materials~\cite{Harris1997PhysRevLett,RamirezDTO,Bramwell-Science,Bramwell-Ho2Ti2O7}, Tb$_2$Ti$_2$O$_7$ was one of the first magnetic rare-earth pyrochlore oxide materials found to not develop long-range order down to at least $T\sim 50$ mK~\cite{Gardner-TbTiO,Rau2019ARCMP}.
Later, in an important work, Taniguchi {\it et al.}~\cite{Taniguchi-2013} discovered an extreme sensitivity of the thermodynamic behavior of Tb$_{2+x}$Ti$_{2-x}$O$_{7+y}$ to its precise stoichiometry.
Specifically, it was found that a very slight variation of $x$ around the stoichiometric value $x=0$ tunes the system from a spin-liquid state for $x \lesssim x_c$ to a partially magnetically ordered state for $x \gtrsim x_c$, with $x_c \sim -0.005$. 
For $x=0.005$, heat capacity measurements reveal a sharp peak at $T_c \approx 0.5\,$K, indicating a phase transition into a state with a small antiferromagnetic ordered moment.
Such extreme sensitivity to sample composition suggests that a hidden predominant order, possibly accompanied by a weak antiferromagnetic moment, is realized  for $x=0.005>x_c$~\cite{Taniguchi-2013}.
Subsequent work by the same group~\cite{Takatsu_TbTiO} suggested that Tb$_{2+x}$Ti$_{2-x}$O$_{7+y}$ lies near a quadrupolar phase (the proposed hidden order in Ref.~\cite{Taniguchi-2013}) and a QSL, possibly a quantum spin ice phase, with small stoichiometric variations driving the transition between them.
}

\dlgnew{
Within the nearest-neighbor model of Eq.~\eqref{eq:Hex1}, with $J_{z\pm}=0$), non-Kramers ions on the pyrochlore lattice admit three quadrupolar long-range ordered phases~\cite{leeGenericQuantumSpin2012} that correspond to the phases labeled $\sf T_{1,p}$, $\sf T_{2}$ and $\sf E$ in the phase diagram of Fig.~\ref{fig:Fig_quantum_all}(a).
Therein, two triple points are realized; one where $\sf T_{1,p}$, $\sf E$ and the spin ice (CSL 1) meet, corresponding to CSL 4, and one where $\sf T_{2}$, $\sf E$ and CSL 1 meet, corresponding to CSL 5.
Of note, for the quantum $S=1/2$ version of the model, recent numerical work using PFFRG found an absence of conventional magnetic order at $T=0$ in an extended $\{ J_\pm/J_{zz}, J_{\pm\pm}/J_{zz}<0 \}$ region that roughly tracks the $\sf T_{1,p}-\sf E$ boundary and includes the triple point where CSL 4 occurs. } 
 \dlgnew{
By the duality Eq.~\eqref{eqn_non_kramer_duality}, this corridor without magnetic order would be replicated for $J_{\pm\pm}/J_{zz}>0$ at the $\sf T_{2} - \sf E$ phase boundary, including  CSL 5,  where Tb$_{2+x}$Ti$_{2-x}$O$_{7+y}$ has been proposed to reside for $x\lesssim x_c$~\cite{Takatsu_TbTiO}.
On the other hand, a more recent inelastic neutron scattering work argues, on the basis of an analysis that ignores the quadrupolar interactions between the  Tb$^{3+}$ ions but incorporates the effects of the low-lying crystal field of Tb$^{3+}$ (at an energy $E\sim 1.5$ meV~\cite{Molavian2007,McClarty-TTO}), that Tb$_2$Ti$_2$O$_7$ is located at the boundary between several magnetic phases, such as a splayed ferromagnet and antiferromagnetic ones~\cite{Roll_TbTiO}.
These magnetic \dlgnew{(i.e. conventional dipolar-ordered) phases} are not realized for the non-Kramers version of Eq.~\eqref{eq:Hex1} with $J_{z\pm}=0$.
They would, however, somewhat correspond to the phases found for Kramers ions in Fig.~\ref{fig:kramer.phase} for $J_{\pm\pm}$ negative (splayed ferromagnet) and positive (antiferromagnet) -- such behavior arising due to the incorporation of the excited crystal field states~\cite{McClarty-TTO} in the model considered in
Ref.~\cite{Roll_TbTiO}.
}

\dlgnew{While stoichiometric Tb$_2$Ti$_2$O$_7$ may be close to a CSL, it is the small change in stoichiometry in Tb$_{2+x}$Ti$_{2-x}$O$_{7+y}$ that has allowed this proximity to be exposed by driving a transition from a long-range ordered state to a spin liquid~\cite{Taniguchi-2013}.
How the change of stoichiometry causes the ground state transition from a spin liquid to an ordered phase in Tb$_{2+x}$Ti$_{2-x}$O$_{7+y}$ as $x$ increases is not understood, but there are likely multiple microscopic mechanisms operating concurrently.}

\dlgnew{One possible contributing factor is an effective negative chemical pressure, as suggested
\dlgnew{by the observation~\cite{Taniguchi-2013} that the cubic unit-cell size $a$ increases as 
$x$ increases from $x=-0.015$ to $x=0.01$.}
In contrast, the stoichiometric sister compounds Tb$_2$Sn$_2$O$_7$~\cite{Tb2Sn2O7-1,Tb2Sn2O7-2} and Tb$_2$Ge$_2$O$_7$~\cite{Tb2Ge2O7-1,Tb2Ge2O7-2}, which respectively
have a larger and smaller $a$ value than Tb$_2$Ti$_2$O$_7$, order both into a $\Gamma_9$ splayed ferromagnetic (SFM) phase
\footnote{Note that a non-Kramers ion system strictly described by Hamiltonian \eqref{eq:Hex1} with $J_{z\pm}=0$ cannot support such a SFM phase. This follows because the $\mathsf{T_{1,p}}$ and $\mathsf{T_{1,i}}$ irreps do not mix, corresponding to quadrupolar and magnetic (dipolar) order parameters, respectively. Consideration of the excited crystal electric field levels is therefore necessary to generate the effective couplings required to stabilize such a phase~\cite{Molavian2007,McClarty-TSO,Tb2Sn2O7-2,Roll_TbTiO}.}.
Moreover, and in paradoxical contrast to the $x$-dependence of $a$ in Tb$_{2+x}$Ti$_{2-x}$O$_{7+y}$~\cite{Taniguchi-2013} with $\partial a/\partial x<0$, earlier neutron diffraction measurements had found that applying high (positive) pressure ($P\gtrsim 1$ GPa, which decreases $a$) using sapphire anvil cells \emph{induces} long-range antiferromagnetic order~\cite{Mirebeau-Nature,Pressure_wrong_way}.
How a hypothetically ideal (positive or negative) uniform hydrostatic pressure~\cite{Pressure_wrong_way} may change the ground state of Tb$_2$Ti$_2$O$_7$ currently lacks a microscopic explanation. 
Other mechanisms that are presumably operating in the ground state transition of Tb$_{2+x}$Ti$_{2-x}$O$_{7+y}$~\cite{Taniguchi-2013,Takatsu_TbTiO} include:
 (1) a modification of the Tb$^{3+}$$-$Tb$^{3+}$ interactions (both their average and their random distribution); (2) the lowering of the local point-group symmetry away from $D_{3d}$ allowing for additional $\{J_{uv}\}$ couplings in Eq.~\eqref{eq:Hex1}; and, perhaps most importantly, (3) the degeneracy-lifting of the ground doublet of non-Kramers Tb$^{3+}$, an effect  minimally described by adding an effective (random, site-dependent) transverse field in Eq.~\eqref{eq:Hex1}~\cite{Benton_instab,Pr2Zr2O7_disorder}.
Which of these effects is predominantly responsible for the ground state change in Tb$_{2+x}$Ti$_{2-x}$O$_{7+y}$ is an open and interesting question.
}

\dlgnew{
Evidence has thus progressively accumulated over the past three decades that Tb$_2$Ti$_2$O$_7$ resides near a CSL, possibly CSL~5~\cite{Takatsu_TbTiO}, which may be promoted to a quantum spin liquid by quantum fluctuations~\cite{lozanogomez2023arxiv}, with sample stoichiometry potentially driving a ground-state transition between a quadrupolar phase and the CSL. 
While appealing, this interpretation faces several challenges. 
First, the lowest excited crystal-field doublet of Tb$^{3+}$ in Tb$_2$Ti$_2$O$_7$ lies only about 1.5~meV above the ground state. 
This relatively small energy separation renders a description of the material in terms of an effective spin-1/2 model---such as Eq.~\eqref{eq:Hex1}, augmented by long-range magnetic dipole--dipole interactions and obtained by projecting all multipolar Tb--Tb interactions onto the ground doublet while neglecting virtual crystal-field fluctuations (VCFF)~\cite{Molavian2007,McClarty-TTO,Rau-Petit,Rau2019ARCMP,Roll_TbTiO}---potentially questionable.
Second, neutron scattering experiments on various Tb$_2$Ti$_2$O$_7$ samples at temperatures below $\sim 100$ mK reveal intense scattering at wave vector ${\bm q}=\left(\tfrac{1}{2}, \tfrac{1}{2}, \tfrac{1}{2}\right)$~\cite{Fritsch,Kadowaki-halfhalfhalf}.
Such sharp features are not accounted for by the correlations predicted for CSL 5 (see Appendix ~\ref{appendix:DQQ_ast_neutron}).  Accounting for the ${\bm q}=\left ( \tfrac{1}{2}, \tfrac{1}{2},\tfrac{1}{2}\right )$ correlations has proven challenging, requiring unrealistic very long-range interactions between the $S^z$ components of the pseudospin-1/2~\cite{Kadowaki-halfhalfhalf}.
However, and interestingly, Ref.~\cite{Kadowaki-3body} has argued that the three-spin interaction generated by VCFF~\cite{Rau2019ARCMP} and ignored in Eq.~\eqref{eq:Hex1} (with $J_{z\pm}=0$) could provide an explanation for those correlations.
}

\dlgnew{
After three decades~\cite{Gardner-TbTiO} of investigation, it is now abundantly clear that Tb$_2$Ti$_2$O$_7$, along with its finely tuned off-stoichiometric variant Tb$_{2+x}$Ti$_{2-x}$O$_{7+y}$, is a highly intriguing compound that harbors strongly competing magnetic and quadrupolar phases and, quite possibly, a quantum-disordered phase---either an intrinsic CSL, such as CSL 5, or one arising from a disorder-induced splitting of the ground-state doublet. 
With an understanding of the gauge theory describing CSL 5, a strong candidate for an intrinsic CSL realizable in Tb$_{2+x}$Ti$_{2-x}$O$_{7+y}$, we now have a valuable conceptual framework to guide the interpretation of future experimental studies and also build upon.
In particular, determining to what extent the QSL observed in this compound relates or not to, CSL 5 will help pave the way toward a definitive understanding of this long-standing perplexing material.
}

\subsubsection{Yb$_2$Ti$_2$O$_7$}

\dlgnew{
Heat-capacity measurements on Yb$_2$Ti$_2$O$_7$, revealing a broad anomaly near $T \sim 2$ K followed by a sharp peak at $T_c \sim 214$ mK signaling a phase transition, possibly to long-range magnetic order, were first reported nearly sixty years ago by Bl\"ote  \emph{et al.}~\cite{Blote}.
The first indication that the low-temperature state of Yb$_2$Ti$_2$O$_7$ may be unconventional was provided by Hodges \emph{et al.}~\cite{Hodges}, who confirmed the heat-capacity anomaly, but found no evidence for long-range order from neutron diffraction and muon spin relaxation measurements.
This work helped catalyze extensive experimental and theoretical investigations of this compound, along with its sister materials Yb$_2$$X_2$O$_7$ ($X=\mathrm{Sn}, \mathrm{Ge}, \mathrm{Pt}$), spanning the past twenty-five years (see Refs.~\cite{Hallas-AnnRevCMP,Rau2019ARCMP} for reviews).
To this day, the central question regarding Yb$_2$Ti$_2$O$_7$ remains the one posed in Ref.~\cite{Hodges}: what is the nature of the state below $T \sim 0.2$ K?
}

\dlgnew{With Yb$^{3+}$ being a Kramers ion with a large energy separation of $\sim 77$ meV ($\sim 900$ K) to its lowest excited crystal field level~\cite{Gaudet-YbTiO-CEF} and with its small magnetic moment of $\sim 2$ $\mu_{\textrm{B}}$, and thus negligible magnetic dipole--dipole interactions, Yb$_2$Ti$_2$O$_7$ is expected to be well described by Eq.~\eqref{eq:Hex1}. 
Indeed, a significant effort has been devoted for the past fifteen years to precisely determine the set of  $\{J_{uv}\}$ parameters
for this compound~\cite{ThompsonPRL2011,RossPRX2011,Robert2015,Thompson_PRL2016,Scheie_YbTiO}.
The most prevalent approach for determining those couplings has been to fit the magnon spectrum obtained from inelastic neutron scattering (INS) measurements on single crystals subject to a large polarizing magnetic field~\cite{RossPRX2011,Thompson_PRL2016,Scheie_YbTiO}.
The currently accepted refined exchange parameters place Yb$_2$Ti$_2$O$_7$ very close to the classical phase boundary between a (splayed) ferromagnetic phase and an antiferromagnetic one --- the  $\mathsf{T_{1,-}}$ and ${\mathsf{E}}$ phases, respectively, as illustrated in Fig.~\ref{fig:kramer.phase}(b). }

\dlgnew{
Such INS experiments have revealed an intriguing evolution of the magnon excitations as the magnetic field is reduced.
Rather than the sharp modes typically expected in a well-ordered magnet, the magnons broaden significantly at lower fields, accompanied by the emergence of relatively flat low-energy bands. 
This incompletely understood behavior is thought to be linked form of ``multiphase competition''~\cite{Jaubert2015PhysRevLett}---such as the material's proximity to the $\mathsf{T_{1,-}}$ and $\mathsf{E}$ phases~\cite{Scheie_YbTiO,Robert2015,Scheie_dynamics_YbTiO,zhang2025intrinsicquantumdisorderyb2ti2o7,
Thompson_PRL2016}---the precise mechanism of which is unclear. 
Moreover, in a manner reminiscent of the situation discussed above for Tb$_{2+x}$Ti$_{2-x}$O$_{7+y}$, such proximity to a phase boundary is also believed to underlie the pronounced sensitivity of the heat capacity of Yb$_2$Ti$_2$O$_7$ to its precise stoichiometry, which itself depends sensitively on the details of sample synthesis~\cite{Hallas-AnnRevCMP,Arpino,
Ross-YbTiO-stuffing,Mostaed-YbTiO}.
However, in contrast to Tb$^{3+}$, Yb$^{3+}$ is a Kramers ion and, consequently, local disorder arising from off-stoichiometry does not split the ground-state doublet. 
That said, such disorder is expected to (1) generate additional $\{J_{uv}\}$ couplings through the local lowering of $D_{3d}$ symmetry and (2) introduce randomness into these 
$\{J_{uv}\}$.
}

\dlgnew{
As in Tb$_2$Ti$_2$O$_7$, one may also ask about the role of chemical or applied pressure. 
Beginning with Yb$_2$Sn$_2$O$_7$, which orders into a splayed ferromagnetic state, substitution of Sn by Ti
(i.e. Yb$_2$(Ti$_{1-x}$Sn$_x$)$_2$O$_7$) reduces the cubic lattice parameter $a$ and drives the system into an \emph{antiferromagnetic} phase over a range of compositions~\cite{Zhang_local_distortion_PRB_2024}.
Upon approaching $x \sim 0$, corresponding to Yb$_2$Ti$_2$O$_7$, the material exhibits a splayed ferromagnetic ground state in some samples~\cite{Zhang_local_distortion_PRB_2024}, depending sensitively on synthesis conditions as discussed above. 
Conversely, starting from Yb$_2$Ti$_2$O$_7$ and substituting Ti by Ge, (i.e. Yb$_2$(Ti$_{1-y}$Ge$_y$)$_2$O$_7$) further \emph{decreases} the lattice parameter as $y$ increases from 0. 
This drives the system back into an antiferromagnetic state, even for small $y$, that persists up to $y = 1$~\cite{Zhang_local_distortion_PRB_2024}, as realized in Yb$_2$Ge$_2$O$_7$, which orders in the $E$ phase below a N\'eel temperature of 0.57 K~\cite{Sarkis_Yb2Ge2O7,Hallas-YbGeO}. 
While Yb$_2$Ge$_2$O$_7$ with its smallest $a$ value orders in the $E$ phase, neutron diffraction studies on Yb$_2$Ti$_2$O$_7$ 
under hydrostatic pressure finds that a pressure larger than $\sim 1$ kbar causes the systems to order in a splayed ferromagnet (e.g. $\mathsf{T_{1,-}}$) at a critical temperature 
$\sim 0.25$K~\cite{Kermarrec-pressure-YbTiO}
}

\dlgnew{
Thus, similarly to Tb$_2$Ti$_2$O$_7$, the exotic behavior of
Yb$_2$Ti$_2$O$_7$ appears strongly related to its proximity to (at least) two competing phases, namely $\mathsf{T_{1,-}}$ and $\mathsf{E}$. According to the discussion in Section \ref{sec:CSL_kramers}, when the $\mathsf{T_{1,-}}$ --- $\mathsf{E}$ boundary collides with the $\mathsf{T_{2}}$ phase it admits a CSL, either CSL 6 or CSL 7 depending on the sign of $J_{z\pm}$ 
Assuming that $J_{z\pm}<0$ for 
Yb$_2$Ti$_2$O$_7$~\cite{Thompson_PRL2016}, the relevant CSL would thus be CSL 7.
However, estimates of the $\{J_{uv}\}$ parameters for Yb$_2$Ti$_2$O$_7$~\cite{Scheie_YbTiO,Thompson_PRL2016,Robert2015}  all put the material extremely close along the $\mathsf{T_{1,-}}$ and $\mathsf{E}$ phase boundary, but a fair distance from the location of CSL 7 in the phase diagram~\cite{Gresista_QPLSL,zhang2025intrinsicquantumdisorderyb2ti2o7}.
On the other hand, for the quantum $S = 1/2$ version of Eq.~\eqref{eq:Hex1}, appropriate to Yb$_2$Ti$_2$O$_7$, both numerical PFFRG~\cite{Gresista_QPLSL} and exact diagonalization calculations~\cite{zhang2025intrinsicquantumdisorderyb2ti2o7} identify, despite quantitative differences, \dlgnew{extended corridors along the $\mathsf{T_{1,-}}$---$\mathsf{E}$ and $\mathsf{T_{1,-}}$---$\mathsf{T_2}$ phase boundaries} in which conventional long-range magnetic order is absent and finding, instead, an unidentified ``non-conventional'' emergent quantum phase separating the $\mathsf{T_{1,-}}$ splayed ferromagnetic and $\mathsf{E}$ antiferromagnetic phases, where Yb$_2$Ti$_2$O$_7$ is believed to reside~\cite{Jaubert2015PhysRevLett,zhang2025intrinsicquantumdisorderyb2ti2o7}.
Given that the position of Yb$_2$Ti$_2$O$_7$ in the phase diagram appears rather  removed from CSL 7, one may ask whether any unconventional quantum state realized at nonzero temperature could nonetheless retain remnants of the SV-R2U1$^*$ nature of that CSL described (see Sec.~\ref{sec:CSL_kramers} and Table~\ref{Table_all_csls}) or is a distinct intrinsically quantum phase as recently suggested~\cite{zhang2025intrinsicquantumdisorderyb2ti2o7,Gresista_QPLSL}.
In light of the complex behavior exhibited by Yb$_2$Ti$_2$O$_7$ as a function of stoichiometry, whether through slight off-stoichiometry, substitution of Ti by Sn or Ge, or the application of hydrostatic pressure, it is natural to ask whether such perturbations might tune the system closer to this CSL, along with its exotic gauge-theory structure, unconventional low-energy excitations, and rich set of emergent charge-conservation laws. 
More work, experimental, theoretical and numerical is unquestionably needed to address these interesting and challenging questions.
}

\subsubsection{Er$_2$X$_2$O$_7$ (X=Ti, Ge, Sn, Pt)}

\dlgnew{
Er$_2$X$_2$O$_7$ (X = Ti, Ge, Sn, Pt) is the last family of materials we discuss in this section. 
These compounds have arguably received less attention than Tb$_2$Ti$_2$O$_7$ and Yb$_2$Ti$_2$O$_7$. 
In part, this may be because the first systematically studied member of the series, Er$_2$Ti$_2$O$_7$, exhibits a sharp phase transition at $T_{\textrm{N}} \approx 1.2$ K~\cite{Blote,Champion-ETO}, with the ordered state correctly identified early on as the antiferromagnetic ($\mathsf{E}$) phase~\cite{Champion-ETO}.
At first glance, this behavior appeared relatively conventional and thus perhaps less likely to warrant extensive further investigation.
This perception is, however, misleading. 
Champion \emph{et al.}~\cite{Champion-ETO} already suggested that Er$_2$Ti$_2$O$_7$, often described as an $XY$ antiferromagnet, similarly to the Yb pyrochlores discussed above~\cite{Hallas-AnnRevCMP}, could realize the intriguing phenomenon of quantum order by disorder, stabilizing one of the two basis states, $\psi_2$ or $\psi_3$, of the $\mathsf{E}$ irrep (see Fig.~\ref{fig:irreps}). 
Subsequent studies indeed provided compelling evidence that both quantum and thermal order-by-disorder mechanisms cooperate in selecting the $\psi_2$ state in Er$_2$Ti$_2$O$_7$~\cite{Savary2012,Ross2014_gap,ZhitomirskyPRL2012,Oitmaa-Er2Ti2O7}.~\footnote{It remains unresolved to what extent VCFF effects~\cite{Molavian2007} contribute to the selection of $\psi_2$ in this compound~\cite{McClarty-Curnoe,Petit_ErTiO,Rau-Petit}.}.
More recently, Er$_2$Ge$_2$O$_7$ was likewise found to order in the antiferromagnetic $\mathsf{E}$ phase~\cite{Dun-Er2Ge2O7}.
The N\'eel temperatures of Er$_2$Ti$_2$O$_7$ and Er$_2$Ge$_2$O$_7$, 1.2 K and 1.4 K, respectively, are the highest among the $XY$ pyrochlores~\cite{Hallas-AnnRevCMP}. 
Notably, these two compounds are also unique within $XY$ pyrochlores in that they do not exhibit the broad high-temperature anomaly in the heat capacity at $T \in [2,5]$ K, a feature commonly interpreted as evidence of proximity to a competing magnetic phase~\cite{Hallas-AnnRevCMP}.
This is consistent with the view that they lie deep within the $\mathsf{E}$ phase~\cite{Wong2013}.
}

\dlgnew{
In contrast to Er$_2$Ti$_2$O$_7$ and Er$_2$Ge$_2$O$_7$, Er$_2$Sn$_2$O$_7$ and Er$_2$Pt$_2$O$_7$ order in the 
$\mathsf{T_2}$ (so-called Palmer-Chalker~\cite{Palmer-Chalker}) phase, with N\'eel temperatures of 0.11 K~\cite{Petit-ErSnO} and 0.38 K~\cite{Hallas2017_PhaseComp}, respectively. 
Both compounds exhibit a broad heat-capacity anomaly in the range $T \in [2,5]$ K~\cite{Hallas-AnnRevCMP}, presumably signaling proximity to either the $\mathsf{T_{1,-}}$ or $\mathsf{E}$ phase.
While it is unambiguous that Er$_2$Sn$_2$O$_7$ orders in the $\mathsf{T_2}$ phase~\cite{Petit-ErSnO}, a mean-field analysis of neutron diffraction and magnetization data places this compound in a region of exchange-parameter space that straddles the $\mathsf{T_2}$---$\mathsf{E}$ phase boundary~\cite{Guitteny-ErSnO-straddles}.
In that context, one notes that Ref.~\cite{Shirai-ErTiSnO} has proposed that substituting Ti by Sn in the Er$_2$Ti$_{2-x}$Sn$_x$O$_7$ may drive a quantum phase transition separating the $\mathsf{E}$ phase found for $x\lesssim 1.7$ and the $\mathsf{T_{2}}$ in  Er$_2$Sn$_2$O$_7$.
Interestingly, the aforementioned exact  diagonalization~\cite{zhang2025intrinsicquantumdisorderyb2ti2o7} and  PFFRG~\cite{Gresista_QPLSL} 
calculations had, in addition to a nonconventional quantum phase \dlgnew{separating the $\mathsf{T_{1,-}}$ and $\mathsf{E}$ phases}, also found evidence for a range of exchange parameters separating the $\mathsf{T_2}$ and $\mathsf{E}$  phases where the tendency towards magnetic order may be quite reduced. 
Moreover, the PFFRG calculations of Ref.~\cite{Gresista_QPLSL} found a quantum triple point where the $\mathsf{T_2}$, $\mathsf{E}$ and the nonconventional quantum phase meet.
One may then ask whether the finite temperature properties and magnon excitations of Er$_2$Sn$_2$O$_7$ and Er$_2$Pt$_2$O$7$, or Er$_2$Ti$_{2-x}$Sn$_x$O$_7$ as $x$ is varied, 
display signatures of this nonconventional quantum phase and perhaps, as for the Yb$_2$Ti$_2$O$_7$ compound, 
quantum-renormalized relics of the CSL 7 which is located at the classical triple point where  $\mathsf{T_2}$ (Palmer-Chalker), $\mathsf{E}$ and  $\mathsf{T_{1,-}}$ (splayed ferromagnetic) phases meet, see Fig.~\ref{fig:kramer.phase}(b).
}
 
\dlgnew{The experimental investigation of the above questions are likely to prove challenging and necessitate high-quality single crystals of Er$_2$Sn$_2$O$_7$, possibly studied under high-pressure~\cite{Kermarrec-pressure-YbTiO}.
However, as such crystals are now available~\cite{Yahne2021}, one may perhaps proceed by investigating the magnon excitation spectrum of this material in a magnetic field, similar to the approach taken in Ref.~\cite{Thompson_PRL2016} for Yb$_2$Ti$_2$O$_7$.
}

\section{Application to other corner-sharing frustrated-unit lattices}
\label{sec:Application_other_lattices}

\begin{figure}[th!]
    \centering
    \includegraphics[width=\columnwidth]{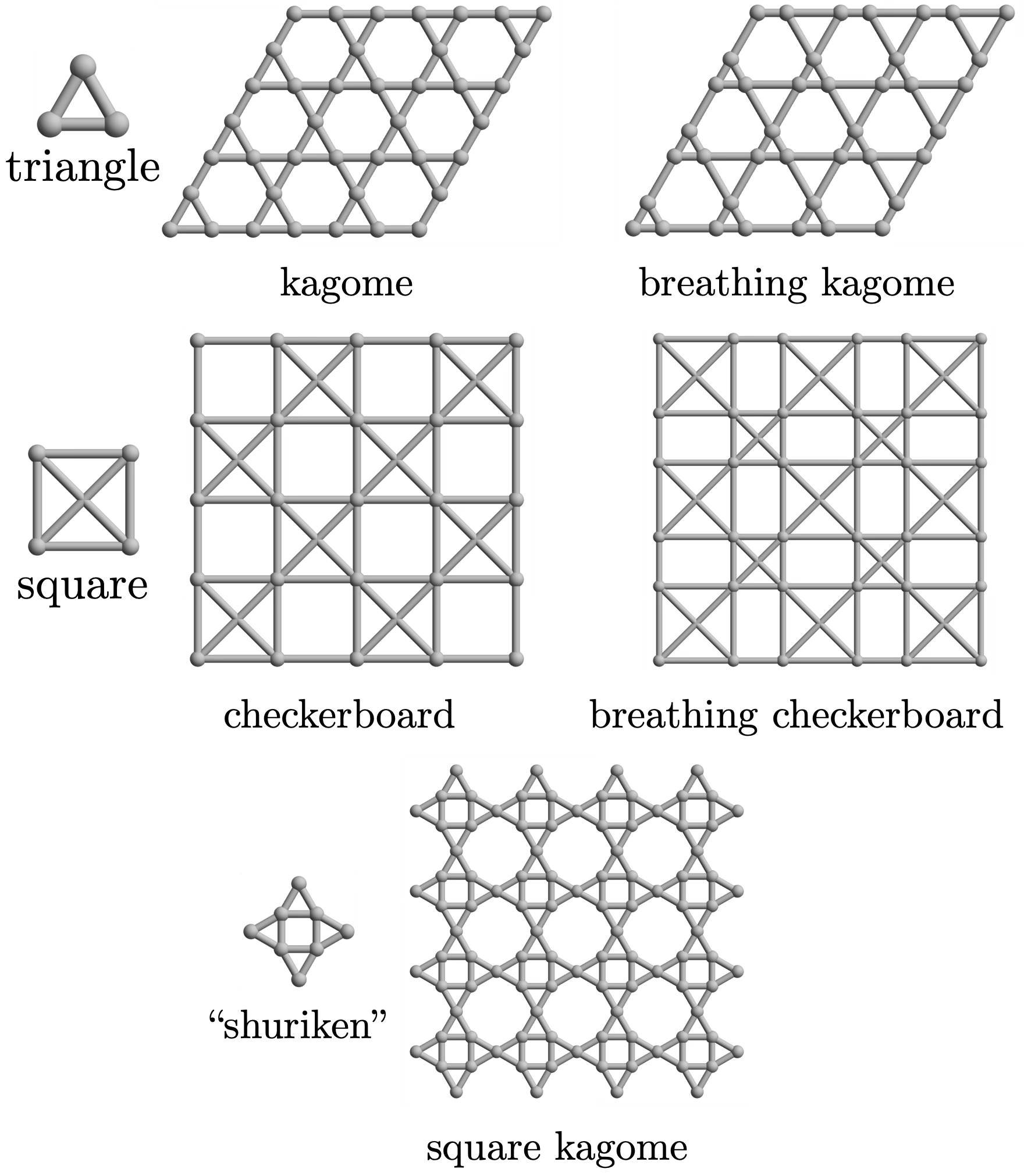} 
    \caption{\dlgnew{Examples of bipartite corner-sharing frustrated-unit lattices in two dimensions, with the basic frustrated unit shown too. }
    }
    \label{fig:Fig_2D_laticces}
\end{figure}

\begin{figure}[th!]
    \centering
    \includegraphics[width=\columnwidth]{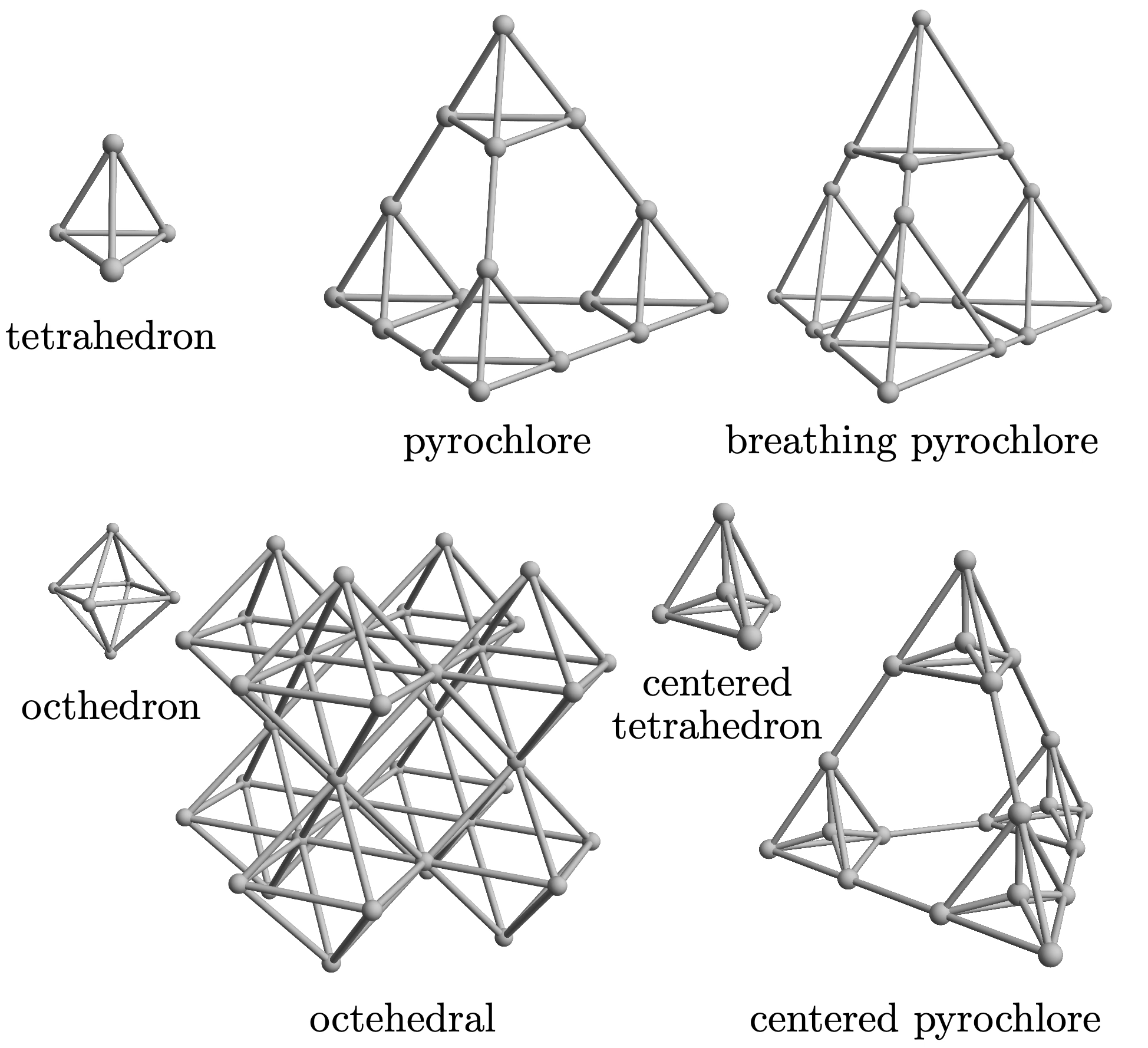} 
    \caption{\dlgnew{Examples of bipartite corner-sharing frustrated-unit lattices in three dimensions, with the elementary frustrated unit shown too. }
    }
    \label{fig:Fig_3D_laticces}
\end{figure}

\refthreenew{The same construction extends to a broad, well-defined class of corner-sharing frustrated-unit lattices for which the centers of the frustrated units form a bipartite parent lattice and the relevant soft sector occurs at $\vb*{q}=\vb*{0}$. This class includes the kagom\'e, checkerboard, and their breathing variants in two dimensions [Fig.~\ref{fig:Fig_2D_laticces}], as well as the pyrochlore, centered-pyrochlore, and octochlore lattices in three dimensions [Fig.~\ref{fig:Fig_3D_laticces}]. For a specified Hamiltonian in this class, decomposition of the single-unit Hamiltonian into its complete set of irreps, followed by enumeration of all single-irrep regions, all realizable phase-boundary surfaces, and their lower-dimensional intersections, exhausts the candidate CSL manifolds. The compatibility relations between neighboring units are then used to derive their long-wavelength gauge constraints. The application to the kagom\'e lattice below demonstrates this procedure explicitly.

We do not claim that this construction exhausts CSLs in Hamiltonians outside this defined class, such as models without the required cluster decomposition, with soft modes away from $\vb*{q}=\vb*{0}$, or with longer-range or multispin interactions. Such systems require separate extensions of the framework.}

Many of these lattices have been studied both theoretically and experimentally. Examples include the kagom\'e~\cite{RevModPhys.88.041002,Han2012,Yan2011,PhysRevB.77.224413,Essafi_kagome_2017,Essafi2017_natcomm}, breathing kagom\'e~\cite{PhysRevB.107.174406,Hirschberger2019}, checkerboard~\cite{PhysRev.162.162,PhysRevLett.18.692,PhysRevB.85.205122}, square kagom\'e~\cite{PhysRevB.107.245115,Gonzalez2025,PhysRevResearch.5.043204,Fujihala2020,PhysRevB.94.014429,Yogendra2024PhysRevB}, breathing pyrochlore~\cite{Yan2020PRL_Rank2,Dissanayake2022,PhysRevB.107.L140408,PhysRevMaterials.7.104404,Mazzotti_2025}, octochlore~\cite{PhysRevLett.127.107202}, and centered pyrochlore~\cite{PhysRevResearch.5.L022018} lattices, illustrating the wide range of possible applications of our generalized theoretical framework.

\subsection{General Recipe}
\label{subsec:general_recipe}
\dlgnew{
Following the approach used above for the pyrochlore lattice, we now recapitulate the general procedure for identifying the set of CSLs stabilized down to zero temperature in a given corner-sharing lattice. This procedure can be divided into five stages, which we describe below.}

Stage $\#1$: \textit{Perform an irrep decomposition.}\\
\dlgnew{As a starting point, we rewrite the most general bilinear nearest-neighbor Hamiltonian $H$ in these lattices in the form
\begin{eqnarray}
    H=\sum_{\mathsf{u}}H_\mathsf{u} ,
\end{eqnarray}
where $H_\mathsf{u}$ is the single frustrated-unit Hamiltonian~\footnote{For example, for the kagom\'e lattice, this would correspond to the Hamiltonian of single triangle, whereas for the octochlore lattice, it would be the Hamiltonian for a single octahedron.}. 
This rewriting can be performed since the lattices considered are corner-sharing with bilinear spin interactions confined to a single frustrated unit $\mathsf{u}$. 
Using $H_\mathsf{u}$, we identify the irreps $\mathsf{X}^{(\mathsf{u})}$ and their corresponding energies $a_\mathsf{X}^{(\mathsf{u})}$ of a single frustrated unit $u$, such that the single frustrated-unit Hamiltonian takes the form
\begin{eqnarray}
    H_\mathsf{u}=\sum_\mathsf{X} a_\mathsf{X}^{(\mathsf{u})} |\vb*{m}_\mathsf{X}^{(\mathsf{u})}|^2. 
\end{eqnarray}
Here, $\vb*{m}_\mathsf{X}^{\mathsf{u}}$ correspond to the possible conventional magnetic orders observed over the full corner-sharing lattice~\cite{Yan-2017}, while the $a_\mathsf{X}^{(\mathsf{u})}$ are linear functions of the $\{J_{uv}\}$ parameters~\footnote{
\dlgnew{Here, we assume the existence of a diagonal irrep basis yielding this form, which may require an additional diagonalization step, such as that performed in Sec.~\ref{subsec:single.tetra.ham}, to decouple interacting copies of a given irrep.}
}
}

Stage $\#2$: \textit{Identify sets of interaction parameters.}\\
\dlgnew{Using the irrep energies $a_\mathsf{X}^{(\mathsf{u})}$, we define the minimal energy of a frustrated-unit $\mathcal{E}_0\equiv\mathrm{min}_{\mathsf{X}}(a_\mathsf{X}^{(\mathsf{u})})$ and identify sets of the $\{J_{uv}\}$ parameters where multiple irreps become minimal~\footnote{It is also possible that having a single minimal irrep results in the stabilization of a CSL down to the lowest temperature. For example, this is the case of the CSL 1 in Table~\ref{Table_all_csls}.}, i.e. $\mathcal{E}_0=a_{\mathsf{X}_1}=\cdots=a_{\mathsf{X}_m}$. 
The energy degeneracy between multiple irreps implies the identification of a phase boundary in the $T=0$ phase diagram between the ordered phases corresponding to the degenerate irreps.
As such, the identified points are expected to display a strong competition (i.e., frustration) between the competing magnetic orders.}\\

Stage $\# 3$: \textit{Identify the set of parameters with flat bands.}\\
\dlgnew{
Having identified a set of $\{J_{uv}\}$ parameters where multiple irreps achieve  minimal $a_\mathsf{X}^{(\mathsf{u})}$, we proceed to obtain the band structure in reciprocal space of the full Hamiltonian, i.e.
\begin{eqnarray}
H=\frac{1}{2}\sum_{\vb*{q}}\sum_{n=1}^{\mathcal{N}\times N_\mu} \varepsilon_n(\vb*{q})|\vb*{S}_n(\vb*{q})|^2,
\end{eqnarray}
where $\mathcal{N}$ corresponds to the number of spin components, and $N_\mu$ to the number of sublattices, and $\vb*{S}_n(\vb*{q})$ is a $\mathcal{N}\times N_\mu$ dimensional eigenvector associated to the band energy $\varepsilon_n(\vb*{q})$. 
In this stage, we only keep those $\{J_{uv}\}$ parameter sets obtained in Stage $\# 2$ that possess a non-zero number of low-energy flat bands in the spectrum, i.e. $\varepsilon_n(\vb*{q})=\mathcal{E}_0$.
As previously mentioned, according to an SCGA analysis, all the sets of $\{J_{uv}\}$ parameters harboring at least one low-energy flat band could become CSLs. 
}\\

Stage $\#4 $: \textit{Perform a CMC analysis}\\
\dlgnew{Next, a CMC analysis of the parameter sets identified in Stage $\# 3$ would allow to verify that no symmetry-breaking transition takes place down to the $T\to 0 $ limit. For this analysis, there are cases where it is crucial to implement some type of cluster-like update, i.e. beyond single spin-flip, to permit a more accurate sampling of the ground-state manifold~\cite{lozanogomez2023arxiv}.
It is worth noting that the observation of a symmetry-breaking transition for these parameter sets would imply a thermal order-by-disorder origin as the internal energy associated with the distinct magnetic orders is equivalent~\footnote{This is guaranteed as the irrep energy is equivalent for the competing magnetic orders.} and therefore only the entropy term in the free energy function is capable of causing a selection between the competing magnetic orders. }\\

Stage $\#5$: \textit{Derive the long-wavelength theory.}
\dlgnew{After having identified the set of $\{J_{uv}\}$ parameters which, according to the CMC study, realize a CSL down to the lowest temperatures, we may next derive the constraints on the ground-state manifold by identifying the set of irreps which become thermally depopulated in the $T\to 0$ limit.
To this end, one would use the irrep modes $\vb*{m}_\mathsf{X}^{(u)}$ and construct an expansion as described in Subsection~\ref{subsec:inter} based on the inter-unit constraints. 
Interestingly, for \emph{any} corner-sharing lattices whose interactions can be described by a Hamiltonian on a single frustrated-unit, the constraints on the ground-state manifold will take the form of a generalized Gauss' law as those presented in Table~\ref{Table_all_csls}. This is guaranteed for these type of lattices as a band-touching point between the low-energy flat bands and the excited bands takes place at the $\Gamma$ point whenever the number of flat bands is less than the total number of degrees of freedom of the degenerate irreps~\footnote{The only possible alternative is that the number of flat bands is equivalent to the total number of degrees of freedom of the degenerate irreps. In this case, the identified CSL is a ``fragile topological'' spin liquid~\cite{Yan2024a,Yan2024b}.}. 
Moreover, one can already identify the type of effective long-wavelength theory describing the low-temperature physics by studying the dispersion away from the band touching point~\cite{Pretko-2018}. 
}

\subsection{Application to Kagom\'e Lattice}

\dlgnew{To demonstrate concretely the implementation of our general framework for identifying  CSLs in corner-shared frustrated units, we next present the field-theory derivation for the kagom\'e lattice. 
The kagom\'e unit cell contains three spins, which we take to be those on a triangle defined by the basis vectors}
\begin{eqnarray}
    \vb*{r}_0&=&(0,0),\\
    \vb*{r}_1&=&\frac{a_d}{2}(1,0),\\
    \vb*{r}_2&=&\frac{a_d}{4}\left(1,-\sqrt{3}\right),
\end{eqnarray}
\dlgnew{where $a_d$ is the distance between the centers of two neighboring triangles. For this lattice, the symmetry group associated to the corner-sharing unit, i.e. a single triangle, is given by the group $C_{3v}$. 
A symmetry analysis of the possible interaction of this lattice identifies the most general bilinear Hamiltonian in the global basis parametrized by the couplings $\{J_x,J_y,J_z,D_z\}$, i.e. an XYZ model with an additional DM interaction where the DM vectors are defined along the global $z$ direction~\cite{Essafi_kagome_2017}, namely }
\begin{equation}
    \begin{split}
    \mathcal{H}=&\sum_{\langle ij\rangle} J_x S_i^xS_j^x+J_y S_i^yS_j^y+J_z S_i^zS_j^z\\
    &+ D_z\sum_{\langle ij\rangle}(S_i^xS_j^y-S_i^yS_j^x).        
    \end{split}
\end{equation}

\dlgnew{
Using the most general nearest-neighbor bilinear Hamiltonian for the kagom\'e lattice, we give a general overview of how the stages listed in Subsection~\ref{subsec:general_recipe} are implemented. 
In particular, we focus on three CSLs identified in previous works~\cite{Essafi2017_natcomm,Essafi_kagome_2017}, and derive their yet-unidentified effective low-energy theories.}

\dlgnew{
To this end, we follow Stages $\#1$, $\# 2$, $\# 3$ and $\#5$, which do not require significant numerical calculations.} \\

Stage $\# 1$: \textit{Perform an irrep decomposition}\\ 
\dlgnew{
An irreducible representation analysis of this group identifies six distinct irreps~\cite{Essafi_kagome_2017}, namely the $\{\mathsf{A}_1,\mathsf{A}_{2z},\mathsf{A}_{2\perp},\mathsf{E}_{\mathrm{FM}},\mathsf{E}_{\mathrm{AFM}},\mathsf{E}_{z}\} $, leading to the single triangle Hamiltonian of the form}
\begin{eqnarray}
    H_\triangle&=&\frac{1}{2}\Big( a_{\mathsf{\mathsf{A}_1}}m^2_{\mathsf{A}_1} +a_{\mathsf{\mathsf{A}_{2z}}}m^2_{\mathsf{A}_{2z}} +a_{\mathsf{A_{2\perp}}}m^2_{\mathsf{A}_{2\perp}}\nonumber\\
   &&+  a_{\mathsf{E}_\mathrm{FM}} \vb*{m}^2_{\mathsf{E}_\mathrm{FM}} 
      +a_{\mathsf{E}_\mathrm{AFM}}\vb*{m}^2_{\mathsf{E}_\mathrm{AFM}} +a_{\mathsf{E_{z}}}\vb*{m}^2_{\mathsf{E}_{z}}\nonumber\\
   && +  a_{\mathrm{mix}} \vb*{m}_{\mathsf{E}_\mathrm{FM}} \cdot \vb*{m}_{\mathsf{E}_\mathrm{AFM}}   
    \Big),\label{eq:kagome_irreps}
\end{eqnarray}
\dlgnew{where $ \vb*{m}_{I}$ are the local order parameters for the irreps and $a_{I}$ are defined in Table~\ref{Table_irrep_kagome} in terms of the interaction parameters $\left\{J_x,J_y,J_z,D_z\right\}$. 
In this table, the term $a_{\mathrm{mix}}$ allows for mixing of two of the $\mathsf{E}$ irreps.  }\\
\begin{table}[th!]
\def\arraystretch{1.5}%
\caption{\label{Table_irrep_kagome}
    Coefficients of the irreducible representations in Eq.~\eqref{eq:kagome_irreps} in terms of exchange parameters $\left\{J_x,J_y,J_z,D_z\right\}$ for the kagom\'e lattice.
    }
    \vspace*{5mm}
\begin{tabular}{|c|c|}
\hline Coefficient & \begin{tabular}{c} 
Definition in terms of exchange \\
parameters $\left\{J_x,J_y,J_z,D_z\right\}$
\end{tabular} \\
\hline   $a_{\mathsf{A}_1}$ & $\frac{1}{2}(J_x-3J_y)-\sqrt{3}D_z$ \\
\hline$ a_{\mathsf{A}_{2z}}$  & $2J_z$ \\
\hline$ a_{\mathsf{A}_{2\perp}}$  & $\frac{1}{2} (  - 3 J_x+J_y) -\sqrt{3} D_z $ \\
\hline$  a_{\mathsf{E}_{\mathrm{FM}}}$ & $J_x+J_y$ \\
\hline$ a_{\mathsf{E}_{\mathrm{AFM}}}$ & $-\frac{1}{2}(J_x+J_y) + \sqrt{3}D_z$ \\
\hline$ a_{\mathsf{E}_{z}}$ & $-J_z$ \\
\hline$a_{\mathrm{mix}}$ & $J_x-J_y$ \\
\hline
\end{tabular}
\end{table}

\begin{table}[th!]
\def\arraystretch{1.5}%
\caption{\label{Table_irrep_m_kagome} Order parameters associated with the irreducible representations for the kagom\'e lattice. 
The spin components are defined in the global basis.
}
    \vspace*{5mm}
\begin{tabular}{|c|c|}
\hline Irrep &   
Definition  \\
\hline   $m_{\mathsf{A}_1}$ & $\frac{1}{3}\left(\frac{1}{2} S_0^x+\frac{\sqrt{3}}{2} S_0^y+\frac{1}{2} S_1^x-\frac{\sqrt{3}}{2} S_1^y-S_2^x\right)$ \\
\hline$ m_{\mathsf{A}_{2z}}$  & $\frac{1}{3}\left(S_0^z+S_1^z+S_2^z\right)$ \\
\hline$ m_{\mathsf{A}_{2\perp}}$  & $\frac{1}{3}\left(-\frac{\sqrt{3}}{2} S_0^x+\frac{1}{2} S_0^y+\frac{\sqrt{3}}{2} S_1^x+\frac{1}{2} S_1^y-S_2^y\right)$ \\
\hline$  \vb*{m}_{\mathsf{E}_{\mathrm{FM}}}$ & $\frac{1}{3}\begin{pmatrix}
{S_0^x+S_1^x+S_2^x}\\
{S_0^y+S_1^y+S_2^y}
\end{pmatrix}$\\
\hline $\vb*{m}_{\mathsf{E}_{\mathrm{AFM}}}$ & $\frac{1}{3}\begin{pmatrix}
{\frac{1}{2} S_0^x-\frac{\sqrt{3}}{2} S_0^y+\frac{1}{2} S_1^x+\frac{\sqrt{3}}{2} S_1^y-S_2^x}\\
{-\frac{\sqrt{3}}{2} S_0^x-\frac{1}{2} S_0^y+\frac{\sqrt{3}}{2} S_1^x-\frac{1}{2} S_1^y+S_2^y}
\end{pmatrix}
$ \\
\hline $\vb*{m}_{\mathsf{E}_{z}}$ & $\frac{1}{3}\begin{pmatrix}
    {\sqrt{\frac{3}{2}}\left(S_0^z-S_1^z\right)}\\
    {\frac{1}{\sqrt{2}}\left(S_0^z+S_1^z-2 S_2^z\right)}
\end{pmatrix}$ \\ 
\hline
\end{tabular}
\end{table}

Stage $\# 2$: \textit{Identify sets of interaction Parameters}\\  
\dlgnew{Using the irrep coefficients provided in Table~\ref{Table_irrep_kagome} we may now find sets of the interactions parameters $\left\{J_x,J_y,J_z,D_z\right\}$ for which multiple irreps become degenerate. 
Since  we are here only providing a general overview of the application of this framework, we solely consider here the simpler case with $J_x=J_y=J_{xy}$,  where 
the $a_{\mathrm{mix}}$ coefficient vanishes,  
} 
\begin{eqnarray}
a_{\mathsf{A}_1} &=& -J_{xy}-\sqrt{3}D_z,  \\
a_{\mathsf{A}_{2z}} &=& 2J_z, \\
 a_{\mathsf{A}_{2\perp}} &=& -J_{xy} -\sqrt{3} D_z,  \\
a_{\mathsf{E}_{\mathrm{FM}}} &=& 2J_{xy}, \\
a_{\mathsf{E}_{\mathrm{AFM}}} &=&-J_{xy} + \sqrt{3}D_z, \\
a_{\mathsf{E}_{z}} &=&-J_z, 
\end{eqnarray}
\dlgnew{where we note from the $a_{\mathsf{A}_1}$ and $a_{\mathsf{A}_{2\perp}}$ expression above that the $\mathsf{A}_1$ and $\mathsf{A}_{2\perp}$ irreps are degenerate. 
We now determine the points where there exist at least three irreps that are degenerate with $\mathsf{E}_{z}$ being one of these, where CMC simulations have been previously performed and reported by Ref.~\cite{Essafi_kagome_2017,Essafi2017_natcomm}.
This leads to the consequential identification of three sets of parameters, which we henceforth refer to as triple points. 
The first triple point corresponds to the case when the $\mathsf{E}_{z}$, $\mathsf{E}_{\rm{FM}}$, and $\mathsf{E}_{\mathrm{AFM}}$ irreps (the $\mathsf{E}_{z}\oplus\mathsf{E}_{\mathrm{FM}}\oplus\mathsf{E}_{\mathrm{AFM}}$ point)  are degenerate, which happens when
}
\begin{eqnarray}
    J_{z}&>&0,\\
    J_{xy}&=&-\frac{J_z}{2},\\
    D_{z}&=&-\frac{\sqrt{3}J_{z}}{2}.
\end{eqnarray}
\dlgnew{The second triple point corresponds to the case when the $\mathsf{A}_{1}$, $\mathsf{A}_{2\perp}$, $\mathsf{E}_{z}$, and $\mathsf{E}_{\mathrm{FM}}$ (the $\mathsf{A}_{1}\oplus\mathsf{A}_{2\perp}\oplus\mathsf{E}_{z}\oplus \mathsf{E}_{\mathrm{FM}}$ point) irreps are degenerate. This degeneracy if obtained for the set}
\begin{eqnarray}
     J_{z}&>&0,\\
    J_{xy}&=&-\frac{J_z}{2},\\
    D_{z}&=&\frac{\sqrt{3}J_{z}}{2}.
\end{eqnarray}
\dlgnew{The third and final triple point is the one for which the $\mathsf{A}_{1}$, $\mathsf{A}_{2\perp}$, $\mathsf{E}_{z}$, and $\mathsf{E}_{\mathrm{AFM}}$ (the $\mathsf{A}_{1}\oplus\mathsf{A}_{2\perp}\oplus\mathsf{E}_{z}\oplus \mathsf{E}_{\mathrm{AFM}}$ point) irreps are degenerate, corresponding to the parametrization}
\begin{eqnarray}
    J_{z}&>&0,\\
    J_{xy}&=&J_z,\\
    D_{z}&=&0.
\end{eqnarray}
It is worth noting that this last triple point corresponds to the antiferromagnetic Heisenberg model. 
\dlgnew{Furthermore, and of particular interest, we note that all of these high-degeneracy points just identified have been studied in previous works~\cite{Essafi_kagome_2017,Essafi2017_natcomm} where a CSL is stabilized down to the lowest temperature at each one of these points.}\\

Stage $\# 3$: \textit{Identify the set of parameters with flat bands}\\
\dlgnew{Taking as a starting point the above three triple points identified for the kagom\'e lattice, one would next study the band structure of the Hamiltonian for these parameters sets.
All of the triple points identified have three low-energy flat bands in the matrix spectra and are therefore predicted to realize a CSL according to a SCGA analysis.}\\

Stage $\# 4:$ \textit{Perform a CMC analysis}\\
\dlgnew{Since the three triple points identified in Stage $\#$ 2 have already been studied in Refs.~\cite{Essafi_kagome_2017,Essafi2017_natcomm}, which confirmed that these points do realize a CSL phase to the lowest temperatures, we do not carry the CMC analysis and continue with Stage $\#$ 5 to derive the low-energy theories of these CSLs.} \\

Stage $\# 5$: \textit{Derive the long-wavelength theory}\\
\dlgnew{
As a final stage to exemplify the application of the framework to find all CSLs relevant to the kagom\'e lattice, we provide the constraints on the ground-state manifold of the CSLs identified in Stage $\# 3$. 
For this, we consider the CSLs which take place at the $\mathsf{E}_{z}\oplus\mathsf{E}_{\mathrm{FM}}\oplus\mathsf{E}_{\mathrm{AFM}}$, 
$\mathsf{A}_1\oplus \mathsf{A}_{2\perp}\oplus \mathsf{E}_{z}\oplus\mathsf{E}_{\mathrm{FM}}$, and $\mathsf{A}_1\oplus \mathsf{A}_{2\perp}\oplus \mathsf{E}_{z}\oplus\mathsf{E}_{\mathrm{AFM}}$, with their corresponding ground-state manifold constraints provided by the coarse-grained theory for the kagom\'e lattice.
The general inter-triangle constraints, equivalent to the inter-tetrahedra constraints in Eq.~\eqref{eq:nonK.constraintA2}-\eqref{eq:nonK.constraint+}, yield 
}
\begin{widetext}
\begin{align}
m_{\mathsf{A}_1} = 0  \implies & m_{\mathsf{A}_1} = 
\frac{a_d}{2}\left(-\partial_x m_{\mathsf{E}_\text{FM}}^y+\partial_x m_{\mathsf{E}_\text{AFM}}^y+\partial_y m_{\mathsf{E}_\text{FM}}^x-\partial_y m_{\mathsf{E}_\text{AFM}}^x\right)=0 \\
m_{\mathsf{A}_{2\perp}} = 0 \implies &   m_{\mathsf{A}_{2\perp}} = \frac{a_d}{2} \nabla \cdot\left({\vb*{m}}_{\mathsf{E}_\text{FM}}+{\vb*{m}}_{\mathsf{E}_\text{AFM}}\right)=0 \\
m_{\mathsf{A}_{2z}} = 0 
\implies &  m_{\mathsf{A}_{2z}} =  \frac{a_d}{\sqrt{2}}\left( \partial_x m_{\mathsf{E}_z}^x+   \partial_y m_{\mathsf{E}_z}^y\right)=0 \\
\vb*{m}_{\mathsf{E}_\text{FM}}= 0  \implies &   \vb*{m}_{\mathsf{E}_\text{FM}}= \frac{a_d}{2}\binom{\partial_x m_{\mathsf{A}_{2\perp}}+\partial_y m_{\mathsf{A}_1}+\partial_x m_{\mathsf{E}_\text{AFM}}^y+\partial_y m_{\mathsf{E}_\text{AFM}}^x}{-\partial_x m_{\mathsf{A}_1}+\partial_y m_{\mathsf{A}_{2\perp}}+\partial_x m_{\mathsf{E}_\text{AFM}}^x-\partial_y m_{\mathsf{E}_\text{AFM}}^y}=0 \\
\vb*{m}_{\mathsf{E}_\text{AFM}} = 0 \implies &\vb*{m}_{\mathsf{E}_\text{AFM}} =   \frac{a_d}{2}\binom{\partial_x m_{\mathsf{A}_{2\perp}}-\partial_y m_{\mathsf{A}_1}+\partial_x m_{\mathsf{E}_\text{FM}}^y+\partial_y m_{\mathsf{E}_\text{FM}}^x}{\partial_x m_{\mathsf{A}_1}+\partial_y m_{\mathsf{A}_{2\perp}}+\partial_x m_{\mathsf{E}_\text{FM}}^x-\partial_y m_{\mathsf{E}_\text{FM}}^y}=0 \\
\vb*{m}_{\mathsf{E}_z} = 0 \implies & \vb*{m}_{\mathsf{E}_z} =  \frac{a_d}{\sqrt{2}} (
\partial_x m_{\mathsf{A}_{2z}}, \partial_y m_{\mathsf{A}_{2z}}
) =0   ,
\end{align} 
\end{widetext}
\dlgnew{
where the above constraints are imposed depending on which irreps are not in the ground-state manifold and are therefore expected to be thermally depopulated in the $T\to 0$ limit. 
Starting with the $\mathsf{E}_{z}\oplus\mathsf{E}_{\mathrm{FM}}\oplus\mathsf{E}_{\mathrm{AFM}}$ point, the three $\mathsf{A}$ irreps become thermally depopulated. These constraints yield the following effective Gauss's laws}
\begin{eqnarray}
 m_{\mathsf{A}_{2z}}=0&\implies&\vb*{\nabla} \cdot \vb*{E}^{(1)}=0,\label{eq:A2z_constraitn}\\
    m_{\mathsf{A}_1}=0&\implies& \vb*{\nabla} \cdot \vb*{E}^{(2)}=0,\label{eq:A1_constraitn}\\
    m_{\mathsf{A}_{2\perp}}=0&\implies& \vb*{\nabla} \cdot \vb*{E}^{(3)}=0,\label{eq:A2p_constraitn}    
\end{eqnarray}
\dlgnew{where the electric fields $\vb*{E}^{(1,2,3)}$ are defined as}
\begin{eqnarray}
\vb*{E}^{(1)}&=&  \begin{pmatrix}
        m_{\mathsf{E}_{z}}^x\\
        m_{\mathsf{E}_{z}}^y
    \end{pmatrix}\\
   \vb*{E}^{(2)}&=& \begin{pmatrix}
        m_{\mathsf{E}_{\mathrm{AFM}}^y}-m_{\mathsf{E}_{\mathrm{FM}}^y}\\
        m_{\mathsf{E}_{\mathrm{FM}}^x}-m_{\mathsf{E}_{\mathrm{AFM}}^x}
    \end{pmatrix},\\
    \vb*{E}^{(3)}&=&  \begin{pmatrix}
        m_{\mathsf{E}_{\mathrm{AFM}}^x}+m_{\mathsf{E}_{\mathrm{FM}}^x}\\
        m_{\mathsf{E}_{\mathrm{AFM}}^y}+m_{\mathsf{E}_{\mathrm{FM}}^y}
    \end{pmatrix}.
\end{eqnarray}
\dlgnew{For the $\mathsf{A}_1\oplus \mathsf{A}_{2\perp}\oplus \mathsf{E}_{z}\oplus\mathsf{E}_{\mathrm{FM}}$ point, the corresponding ground-state constraint yield } 
\begin{eqnarray}
     m_{\mathsf{A}_{2z}}=0&\implies&\vb*{\nabla} \cdot \vb*{E}^{(1)}=0,\\
     \vb*{m}_{\mathsf{E}_{\mathrm{AFM}}}=0&\implies&\left
     \{\begin{matrix}
         \vb*{\nabla} \cdot \vb*{E}^{(4)}=0\\
          \vb*{\nabla} \cdot \vb*{E}^{(5)}=0 ,\\
     \end{matrix} \right . 
\end{eqnarray}
\dlgnew{where the electric fields $\vb*{E}^{(1)}$ is defined in Eq.~\eqref{eq:A2z_constraitn}, and the remaining fields are defined as}
\begin{eqnarray}
   \vb*{E}^{(4)}&=& \begin{pmatrix}
        m_{\mathsf{A}_{2\perp}}+m_{\mathsf{E}_{\mathrm{FM}}^y}\\
        m_{\mathsf{E}_{\mathrm{FM}}^x}-m_{\mathsf{A}_1}
    \end{pmatrix},\\
    \vb*{E}^{(5)}&=&  \begin{pmatrix}
        m_{\mathsf{A}_{1}}+m_{\mathsf{E}_{\mathrm{FM}}^x}\\
        m_{\mathsf{A}_{2\perp}}-m_{\mathsf{E}_{\mathrm{FM}}^y}
    \end{pmatrix}.
\end{eqnarray}
\dlgnew{
Lastly, for the $\mathsf{A}_1\oplus \mathsf{A}_{2\perp}\oplus \mathsf{E}_{z}\oplus\mathsf{E}_{\mathrm{AFM}}$  point, the ground-state manifold constraints yield}

\begin{eqnarray}
     m_{\mathsf{A}_{2z}}=0&\implies&\vb*{\nabla} \cdot \vb*{E}^{(1)}=0,\\
     \vb*{m}_{\mathsf{E}_{\mathrm{FM}}}=0&\implies&\left\{\begin{matrix}
         \vb*{\nabla} \cdot \vb*{E}^{(6)}=0\\
          \vb*{\nabla} \cdot \vb*{E}^{(7)}=0 ,\\
     \end{matrix}\right. 
\end{eqnarray}
\dlgnew{where the electric fields $\vb*{E}^{(6,7)}$ are defined as}
\begin{eqnarray}
   \vb*{E}^{(6)}&=& \begin{pmatrix}
        m_{\mathsf{A}_{2\perp}}+m_{\mathsf{E}_{\mathrm{AFM}}^y}\\
        m_{\mathsf{E}_{\mathrm{AFM}}^x}+m_{\mathsf{A}_1}
    \end{pmatrix},\\
    \vb*{E}^{(7)}&=&  \begin{pmatrix}
        -m_{\mathsf{A}_{1}}+m_{\mathsf{E}_{\mathrm{AFM}}^x}\\
        m_{\mathsf{A}_{2\perp}}-m_{\mathsf{E}_{\mathrm{FM}}^y}
    \end{pmatrix}.
\end{eqnarray}

\dlgnew{We note that the constraints found for all the identified CSLs imply that the ground-state manifolds of these CSLs are \dlgnew{all} described by three copies of rank-1 U(1) theories -- these results being consistent with \dlgnew{arguments put forward in} Ref.~\cite{Essafi_kagome_2017}.
Having considered the example of the kagom\'e lattice, we have shown that our framework is readily applicable to other corner-shared frustrated-unit type of lattices whose parent lattice is bipartite (e.g. those illustrated in Figs.~\ref{fig:Fig_2D_laticces} and  \ref{fig:Fig_3D_laticces}), and allowing to identify all CSLs for such models.}

\section{Discussion and Conclusion}
\label{sec:discussion}

\refthreenew{ 
In this work, we have developed a comprehensive framework to identify all CSLs of the nearest-neighbor pyrochlore spin model. Using three independent methods---the coarse-grained field theory, the large-$\mathcal{N}$ theory, and CMC---we cross-checked all parameter-space manifolds on which ordered phases are degenerate and exhaustively identified the CSL phases.

The resulting nine CSLs in Table~\ref{Table_all_csls} therefore exhaust the CSL phases of this Hamiltonian, with no additional CSL remaining elsewhere in its parameter space. This complete map [Figs.~\ref{fig:nonkramer.phase}, \ref{fig:kramer.phase}, and~\ref{fig:phase_diagram_connectivity}] locates each CSL within the surrounding ordered-phase landscape and provides a basis for future theoretical studies and materials searches.

A single microscopic representation also derives the long-wavelength gauge fields and Gauss's laws of all nine CSLs without separate case-by-case constructions. This identifies two previously unstudied phases, CSLs~6 and~9, and yields new mixed scalar- and vector-charge constraints with associated infinite families of multipole conservation laws. Mapping the full parameter space further reveals the connectivity and evolution of the gauge constraints and the exact dualities between CSLs~6 and~7 and between CSLs~8 and~9 [Eq.~\eqref{eqn_kramers_duality}], while their microscopic gauge-field embeddings and physical spin correlations distinguish the dual partners. These classification, gauge-theory, and connectivity results were not established previously, and the construction provides a systematic recipe for analogous atlases on the broader lattice class described in Sec.~\ref{sec:Application_other_lattices}.}

From the work presented here, we now have in hand a comprehensive overview of all the CSLs on the pyrochlore lattice and how these transform into each other as a consequence of modifying the lowest degenerate irreps by varying the anisotropic bilinear spin-spin couplings.
Such insight is crucial for understanding the exotic physics of anisotropic spin models on the pyrochlore lattice, especially for materials that happen to be finding themselves near the phase boundaries of pseudospin long-range order~\cite{RossPRX2011,savaryCoulombicQuantumLiquids2012,Guitteny2013PhysRevB,Jaubert2015PhysRevLett,Yan-2017,Hallas-AnnRevCMP}.
\dlgnew{Furthermore, we have discussed in Sec.~\ref{sec:roadmap} how the atlas of classical spin liquids in the pyrochlore lattice provides \dlgnew{key insights} 
to guide the ultimate construction of a similar atlas for the quantum model, and highlighted various regions of interest for future work. 
Additionally, we discussed in Sec.~\ref{sec:proximity_to_CSL} how phenomena typically associated with a CSL phase can be observed at finite temperatures in systems whose interaction parameters $\{J_{uv}\}$ lie in proximity of a CSL but ultimately order into a symmetry-broken phase.
Our atlas of CSLs on the pyrochlore lattice with nearest-neighbor couplings is a crucial step toward a complete understanding of one of the most representative frustrated magnetic systems.
}

\dlgnew{Lastly, we have illustrated how the same framework used in the identification of the atlas of CSLs on the pyrochlore lattice can be used to identify the CSLs in other corner-sharing lattices \dlgnew{with bi-partite parent lattices} in two and three dimensions.
In particular, we provided a concrete example by applying this framework to the kagom\'e lattice, and identify the general set of constraints through which the effective low-energy theories for the CSLs in this lattice can be derived.}

\begin{acknowledgments}
The authors acknowledge useful discussions with Ludovic Jaubert, Nic Shannon, Roderich Moessner, Yasir Iqbal, Rajiv Singh, Jaan Oiitma  Johannes Reuther, Andriy Nevidomskyy, Kristian Chung, and Matthias Vojta. 
D.L.-G. acknowledge financial support from the DFG through the W\"urzburg-Dresden Cluster of Excellence on Complexity,
Topology and Dynamics in Quantum Matter -- \textit{ctd.qmat} (EXC 2147, project-id 390858490) and
through SFB 1143 (project-id 247310070). D.L.-G. is supported by the Hallwachs-R\"ontgen Postdoc Program of \textit{ctd.qmat}.
H.Y. acknowledges the 2024 Toyota Riken Scholar Program from the Toyota Physical 
and Chemical Research Institute, and the  Grant-in-Aid for Research Activity Start-up from Japan Society
for the Promotion of Science (Grant No. 24K22856).
The work at the University of Waterloo was supported by the NSERC of Canada and the Canada Research Chair (Tier I, M.J.P.G.) program. 
\end{acknowledgments}

\appendix

\section{Local coordinates}
\label{appendix:local_coordinates}
In this appendix, we provide the convention that we follow for the local $x$ and local $z$ direction of the $A$ tetrahedra shown in Fig.~\ref{fig:pyrochlore_lattice_unit_cell}.  
\begin{eqnarray}
    \vb*{x}_0&=\frac{1}{\sqrt{6}}(-2,1,1),\\
    \vb*{x}_1&=\frac{1}{\sqrt{6}}(2,-1,1), \\ 
     \vb*{x}_2&=\frac{1}{\sqrt{6}}(2,1,-1), \\ 
     \vb*{x}_3&=\frac{1}{\sqrt{6}}(-2,-1,-1);
\end{eqnarray}
\begin{eqnarray}
    \vb*{y}_0&=\frac{1}{\sqrt{2}}(0,-1,1),\\
    \vb*{y}_1&=\frac{1}{\sqrt{2}}(0,1,1), \\
     \vb*{y}_2&=\frac{1}{\sqrt{2}}(0, -1,-1), \\  
     \vb*{y}_3&=\frac{1}{\sqrt{2}}(0,1,-1),
\end{eqnarray}
and
\begin{eqnarray}
    \vb*{z}_0&=\frac{1}{\sqrt{3}}(1,1,1),\\
    \vb*{z}_1&=\frac{1}{\sqrt{3}}(-1,-1,1), \\ 
     \vb*{z}_2&=\frac{1}{\sqrt{3}}(-1,1,-1), \\  
      \vb*{z}_3&=\frac{1}{\sqrt{3}}(1,-1,-1).
\end{eqnarray}

\section{Hamiltonian in the ``spin vector basis'' and in the global basis}
\label{appendix:global_basis_hamiltonian}
In this appendix, we provide an alternative basis for the general bilinear nearest-neighbor Hamiltonian in the pyrochlore lattice in Eq.~\eqref{eq:Hex1}~\cite{ThompsonPRL2011}.
Doing so will further emphasize how the form of the spin Hamiltonian appears significantly different when considering the dual forms discussed above.
The Hamiltonian can be parametrized in what we
refer to as a ``spin vector interaction'' basis 
(SVI) basis with Heisenberg $J$, local Ising $J_{\rm Ising}$, pseudo-dipole $J_{\rm PD}$, and Dzyaloshinskii-Moriya (DM) $J_{\rm DM}$ interactions~\cite{McClarty-Curnoe,ThompsonPRL2011}.
In this SVI basis, the  Hamiltonian in Eq~\eqref{eq:Hex1} reads
\begin{eqnarray}
\mathcal{H}&=&\mathcal{H}_{\rm Heis}+\mathcal{H}_{\rm Ising}+\mathcal{H}_{\rm PD}+\mathcal{H}_{\rm DM}\label{eq:H_general}\\
\mathcal{H}_{\rm Heis}&=&J\sum_{\langle ij \rangle}\vb*{S}_i \cdot \vb*{S}_j \\
\mathcal{H}_{zz}&=& J_{\rm Ising} \sum_{\langle ij \rangle}\left(\vb*{S}_i\cdot \vb*{z}_i\right) \left(\vb*{S}_j\cdot \vb*{z}_j\right)\\
\mathcal{H}_{\rm PD}&=&J_{\rm PD}\sum_{\langle ij \rangle}\left(\vb*{S}_i\cdot \vb*{r}_{ij}\right) \left(\vb*{r}_{ij}\cdot  \vb*{S}_j \right)\\
\mathcal{H}_{\rm DM}&=&J_{\rm DM}\sum_{\langle ij \rangle}\vb*{d}_{ij}\cdot (\vb*{S}_i \times \vb*{S}_j),
\end{eqnarray}
where the spins $\vb*{S}_i$ are here defined in the global Cartesian basis, $\vb*{z}_i$ is the local-$z$ (cubic $[111]$) direction at the $i$th lattice site, $\vb*{r}_{ij}$ is the separation vector between lattices sites $i$ and $j$, and $\vb*{d}_{ij}$ labels the direct DM vectors
\begin{eqnarray}
\vb*{d}_{03}&=&(-1,1,0),\quad \vb*{d}_{02}=(1,0,-1), \\
 \vb*{d}_{01}&=&(0,-1,1),\quad \vb*{d}_{32}=(0,1,1),\\
\vb*{d}_{31}&=&(-1,0,-1),\quad  \vb*{d}_{21}=(1,1,0).
\label{eq:dm_vectors2}
\end{eqnarray}
as defined in Ref.~\cite{Noculak_PhysRevB.107.214414}. 
The relation between the interaction parameters basis $\{J_{zz}, J_{\pm},J_{\pm\pm},J_{z\pm}\}$ and the above $\{J,J_{\rm Ising},  J_{\rm PD},J_{\rm DM}\}$ basis yields
\begin{eqnarray}
J&=&2(J_{\pm} + J_{\pm\pm}),\\
J_{\rm Ising}&=&-10 J_{\pm} - 2 J_{\pm\pm} + 4 \sqrt{2} J_{z\pm} + J_{zz},\\
J_{\rm PD}&=&-4 (2 J_{\pm} + J_{\pm\pm} - \sqrt{2} J_{z\pm}),\\
J_{\rm DM}&=&-4 J_{\pm} + \sqrt{2} J_{z\pm}.
\end{eqnarray} 
Consider now the duality between CSL 8 and CSL 9 that is generated upon the change of sign of $J_{z\pm}$.  
In SVI basis,  the former is simply an antiferromagnetic Heisenberg model, namely $\{J,J_{\rm Ising},  J_{\rm PD},J_{\rm DM}\}=\{\frac{3}{\sqrt{2}}J_{z\pm},0,0,0\}$, while the CSL 9 is a model where all parameters in the SVI basis are non-vanishing, i.e. $\{J,J_{\rm Ising},  J_{\rm PD},J_{\rm DM}\}=\{-\frac{3}{\sqrt{2}}J_{z\pm},8\sqrt{2}J_{z\pm},8\sqrt{2}J_{z\pm},2\sqrt{2}J_{z\pm}\}$.
We further note that, in addition to the local basis, i.e. $\{J_{zz}, J_{\pm}, J_{\pm\pm}, J_{z\pm}\}$, and the SVI basis, $\{J, J_{\rm Ising}, J_{\rm PD}, J_{\rm DM}\}$, 
a global parameter basis, namely $\{J_1,J_2,J_3,J_4\}$ is also extensively used in the literature. In the global basis, the spin components are expressed in global Cartesian coordinates, and the spin Hamiltonian takes the form 
\begin{eqnarray}
    H_{\sf ex}=\sum_{\langle i j \rangle} \vb*{S}_i\cdot \mathcal{J}_{ij} \cdot\vb*{S}_j,
\end{eqnarray}
where the exchange matrix $\mathcal{J}_{ij}$ couples the spins at the $i$ and $j$ sites. For instance, for sublattices $0$ and $1$, this exchange matrix takes the form
\begin{eqnarray}
    \mathcal{J}_{01}= \begin{pmatrix}
        J_2 & J_4 & J_4 \\
        -J_4 & J_1 & J_3 \\
        -J_4 & J_3 & J_1
    \end{pmatrix},
\end{eqnarray}
while all other exchange matrices can be obtained by applying the transformation of the single tetrahedron group $T_d$. For the explicit form of these exchange matrices we refer the reader to Ref.~\cite{Yan-2017}.
The relationship between the local and the global basis is given by the transformation
\begin{eqnarray} 
\label{eqn:j1}
    J_1&=& \frac{1}{3}\left(-J_{zz}+4J_{\pm}+2J_{\pm\pm}+2\sqrt{2}J_{z\pm}\right),\\
    J_2&=& \frac{1}{3}\left(J_{zz}-4J_{\pm}+4J_{\pm\pm}+4\sqrt{2}J_{z\pm}\right),\\
    J_3&=& \frac{1}{3}\left(-J_{zz}-2J_{\pm}-4J_{\pm\pm}+2\sqrt{2}J_{z\pm}\right),\\
    \label{eqn:j4}
    J_4&=& \frac{1}{3}\left(-J_{zz}-2J_{\pm}+2J_{\pm\pm}-\sqrt{2}J_{z\pm}\right).
\end{eqnarray}
In the global basis, the parametrization of the CSL 8 and 9 is given by $\{J_1,J_2,J_3,J_4\}=\{\frac{3J_{z\pm}}{\sqrt{2}},\frac{3J_{z\pm}}{\sqrt{2}},0,0\}$ and $\{J_1,J_2,J_3,J_4\}=\{-\frac{J_{z\pm}}{3\sqrt{2}},\frac{7J_{z\pm}}{3\sqrt{2}},\frac{4\sqrt{2}J_{z\pm}}{3},-\frac{2\sqrt{2}J_{z\pm}}{3}\}$, respectively.

\section{Irreducible representations and their exchange coefficients in the global basis}

In the main text, we discussed the degeneracy of different magnetic orders using the irrep description with the spin degrees of freedom expressed in the local basis. 
In this appendix, we provide the irrep modes $\{\vb*{m}_{\mathsf{X}}\}$ in the global Cartesian basis as well as their exchange coefficients, $a_\mathsf{X}$, in the global and in SVI basis above. 
Table~\ref{Table_irrep_global} gives the irrep modes in terms of the spin DOFs defined in the global basis while 
 Table~\ref{Table_irrep_para_global} gives the exchange coefficients, $a_\mathsf{X}$, in terms of the global and SVI basis.

\begin{table*}

\caption{\label{Table_irrep_global}
    Relation between irreducible representations of the point group and the spins in the global basis.
    } 
\begin{tabular}{ | c | c |}
\hline 
\multirow{2}{*}{}
      Local & 
   Definition in terms  of spin components
\\
    order parameter& 
 \\
field
&
\\
\hline
\multirow{1}{*}{}
   $m_{\sf A_2}$ & 
   $\frac{1}{\sqrt{3}} 
     \left(S_0^x+S_0^y+S_0^z+S_1^x-S_1^y-S_1^z-S_2^x+S_2^y-S_2^z-S_3^x-S_3^y+S_3^z
     \right)$ 
\\   
\hline
\multirow{1}{*}{} 
   ${\bf m}_{\sf E}$ & 
   $\begin{pmatrix}
         \frac{1}{\sqrt{6}} \left( -2 S_0^x + S_0^y + S_0^z - 2 S_1^x - S_1^y-S_1^z+2 S_2^x + S_2^y-
              S_2^z +2 S_3^x-S_3^y +S_3^z \right) \\
         \frac{1}{\sqrt{2}} \left( -S_0^y+S_0^z+S_1^y-S_1^z-S_2^y-S_2^z+S_3^y+S_3^z \right)
     \end{pmatrix}$ \\
\hline
\multirow{1}{*}{}
   ${\bf m}_{\sf T_2} $ & 
   $\begin{pmatrix}
        \frac{1}{\sqrt{2}} 
        \left(
         -S_0^y+S_0^z+S_1^y-S_1^z+S_2^y+S_2^z-S_3^y-S_3^z
        \right) 
        \\
        \frac{1}{\sqrt{2}} 
        \left(
        S_0^x-S_0^z-S_1^x-S_1^z-S_2^x+S_2^z+S_3^x+S_3^z
        \right) \\
        \frac{1}{\sqrt{2}}
        \left(
        -S_0^x+S_0^y+S_1^x+S_1^y-S_2^x-S_2^y+S_3^x-S_3^y
        \right)
      \end{pmatrix} $  
\\ 
\hline
\multirow{1}{*}{}
   ${\bf m}_{\sf T_{1,i}}$  & 
   $\begin{pmatrix}
        \frac{1}{\sqrt{3}} (S_0^x+S_0^y+S_0^z
        +S_1^x-S_1^y-S_1^z
        +S_2^x-S_2^y+S_2^z
        +S_3^x+S_3^y-S_3^z) \\
        \frac{1}{\sqrt{3}} (S_0^x+S_0^y+S_0^z
        -S_1^x+S_1^y+S_1^z
        -S_2^x+S_2^y-S_2^z
        +S_3^x+S_3^y-S_3^z) \\
        \frac{1}{\sqrt{3}} (S_0^x+S_0^y+S_0^z
        -S_1^x+S_1^y+S_1^z
        +S_2^x-S_2^y+S_2^z
        -S_3^x-S_3^y+S_3^z)
   \end{pmatrix} $ 
    \\
\hline
\multirow{1}{*}{}
   ${\bf m}_{\sf T_{1,p}}$  & 
   $\begin{pmatrix}
        \frac{1}{\sqrt{6}} (-2 S_0^x+S_0^y+S_0^z-2S_1^x-S_1^y-S_1^z-2S_2^x -S_2^y+S_2^z-2 S_3^x+S_3^y-S_3^z)  \\
        \frac{1}{\sqrt{6}} (S_0^x-2S_0^y+S_0^z-S_1^x-2S_1^y+S_1^z-S_2^x-2S_2^y-S_2^z+S_3^x-2S_3^y-S_3^z)   \\
        \frac{1}{\sqrt{6}} (S_0^x+S_0^y-2S_0^z-S_1^x+S_1^y-2S_1^z+S_2^x-S_2^y-2S_2^z-S_3^x-S_3^y -2 S_3^z)  
   \end{pmatrix}$ 
\\
\hline
\end{tabular}
\end{table*}

\begin{table*}[th!]
\def\arraystretch{1.5}%
\caption{\label{Table_irrep_para_global}
    Coefficients of the irreducible representations used in Eq.~\eqref{eqn:single.ion} and Eq.~\eqref{eq:H_general}
    }
\begin{tabular}{|c|c|c|}
\hline Coefficient & \begin{tabular}{c} 
Definition in terms of exchange \\
parameters $\left\{J_1, J_2, J_3, J_4, D\right\}$
\end{tabular} & \begin{tabular}{c} 
Definition in terms of exchange \\
parameters $\{J,J_{\rm Ising},  J_{\rm PD},J_{\rm DM},D\}$
\end{tabular} \\
\hline$a_{\mathsf{A}_2}$ & $-2 J_1+J_2-2\left(J_3+2 J_4\right)+D$ & $-J -4J_{\rm DM}+3J_{\rm Ising}+D$ \\
\hline$a_{\mathsf{E}}$ & $-2 J_1+J_2+J_3+2 J_4$ & $\frac{1}{2}\left(-2J + 4J_{\rm DM} - J_{\rm PD}\right)$ \\
\hline$a_{\mathsf{T}_2}$ & $-J_2+J_3-2 J_4$ & $\frac{1}{2}\left(-2J-4J_{\rm DM}+J_{\rm PD}\right)$ \\
\hline$a_{\mathsf{T}_1 \text { ice }}$ & $\frac{1}{3}\left(2 J_1-J_2+2 J_3+4 J_4\right)+D$ & $\frac{1}{3}\left(J+4J_{\rm DM}-3J_{\rm Ising}+2J_{\rm PD}\right)+D$ \\
\hline$a_{\mathsf{T}_1 \text { planar }}$ & $\frac{1}{3}\left(4 J_1+J_2-5 J_3+2 J_4\right)$ & $\frac{1}{6}\left(10J+4J_{\rm DM}-J_{\rm PD}\right)$ \\
\hline$a_{\mathsf{T}_1 \text { mixing }}$ & $-\frac{4 \sqrt{2}}{3}\left(J_1+J_2+J_3-J_4\right)$ & $-\frac{4\sqrt{2}}{3} \left(2J-J_{\rm DM}+J_{\rm PD}\right)$ \\
\hline
\end{tabular}
\end{table*}

\section{Correlation functions}
\label{appendix:correlation_functions}

In this appendix, we provide the expressions for the correlation functions presented in the main text. 
In reciprocal space, and for a generic non-Bravais lattice with a sublattice structure, the general correlation between spins is given by the expression 
\begin{align}
      \mathcal{S}^{\alpha\gamma}_{\mu\nu}=\langle S_{\mu}^\alpha (\vb*{q}) S_{\nu}^\gamma(-\vb*{q})\rangle ,
\end{align}
where $\mu,\nu$ label the sublattices and $\alpha,\beta$ label the spin components.  
In particular, we compute the additive contribution of the spin-component diagonal elements of the general correlation function
\begin{eqnarray}
\mathcal{S}(\vb*{q})&=&\sum_{\alpha}\sum_{\mu,\nu}\mathcal{S}^{\alpha\alpha}_{\mu\nu}\nonumber\\
&=&\sum_{\alpha}\sum_{\mu,\nu}\langle S_{\mu}^\alpha (\vb*{q}) S_{\nu}^\alpha(-\vb*{q})\rangle,
\end{eqnarray}
which, for short, we refer simply to as the spin structure factor. In addition to the spin structure factor, we have also computed the experimentally measurable unpolarized neutron structure factor, given by the expression
\begin{align}
 \mathcal{S}_\perp(\vb*{q}) &=&\sum_{\alpha,\beta}\sum_{\mu,\nu}\left(\delta_{\alpha,\beta} -\hat{\vb*{q}}^\alpha \hat{\vb*{q}}^\beta\right)\langle m_{\mu}^\alpha (\vb*{q}) m_{\nu}^\beta(-\vb*{q})\rangle,\label{eq:unpolarized}
\end{align}
which measures the correlation between the magnetic moments, $\vb*{m}_{i\mu}$, of the systems considered where the sub-index $i$ labels the primitive FCC vectors $\vb*{R}_i$ and $\mu$ denotes the sublattice sites. The relationship between the magnetic moments and the spin degrees of freedom is provided by an anisotropic $g$-tensor
\begin{eqnarray}
    m_{i\mu}^\alpha=\sum_{\beta } g_\mu^{\alpha\beta}S_{i\mu}^\beta,
\end{eqnarray}
where $g_\mu^{\alpha\beta}$ are the components of  the $g$-tensor associated to the $\mu$-sublattice. 

In addition to the unpolarized neutron structure factor, we have also considered the polarized neutron structure factor defined in terms of the incident neutrons' polarization 
$\hat{z}_{\textrm N}$~\cite{Chung_flatband}.
This polarization analysis separates the neutron structure factor in Eq.~\eqref{eq:unpolarized} into two channels, the non-spin-flip (NSF) channel which studies the correlations that are parallel to the incident polarization 
$\hat{z}_{\textrm N}$
\begin{eqnarray}
 \mathcal{S}_\perp^{\mathrm{NSF}}(\vb*{q})&=&\sum_{\alpha,\beta}\sum_{\mu,\nu}\left(\hat{z}_{\textrm N}
^\alpha \hat{z}_{\textrm N}
^\beta\right)\langle m_\mu^\alpha m_\nu^\beta \rangle,\label{eq:polarized_NSF}
\end{eqnarray}
and the spin-flip channel which is the complement of the NSF channel, defined as 
\begin{align}
 \mathcal{S}_\perp^{\mathrm{SF}}(\vb*{q})=\mathcal{S}_\perp (\vb*{q})-\mathcal{S}_\perp^{\mathrm{NSF}}(\vb*{q}).\label{eq:polarized_SF}
\end{align}

We note that the introduction of an anisotropic $g$-tensor in Eqs.~\eqref{eq:unpolarized}-\eqref{eq:polarized_NSF} effectively ``distorts'' in momentum (${\vb*q}$) space the correlation functions we have presented~\cite{Castelnovo2019rods,Kadowaki_2015,lozanogomez2023arxiv} which were considered to discernibly expose the underlying gauge theory. 
For this reason, when reporting in the main text neutron structure factors (e.g. see Fig.~\ref{fig:gauge_tunning}), we chose an isotropic $g$-tensor (i.e. $g_\mu^{\alpha\beta}=g\delta_{\alpha\beta}$). 
To reiterate, in doing so, we make no assumptions on the nature (Kramers or non-Kramers) of the spin degrees of freedom in order to more clearly expose the underlying gauge theory of the CSLs considered.

\section{Determination of the phase transitions and crossovers between distinct CSLs}
\label{appedix:Thermodynamics_temp}
\begin{figure}[ht!]
    \centering
    \begin{overpic}[width=\columnwidth]{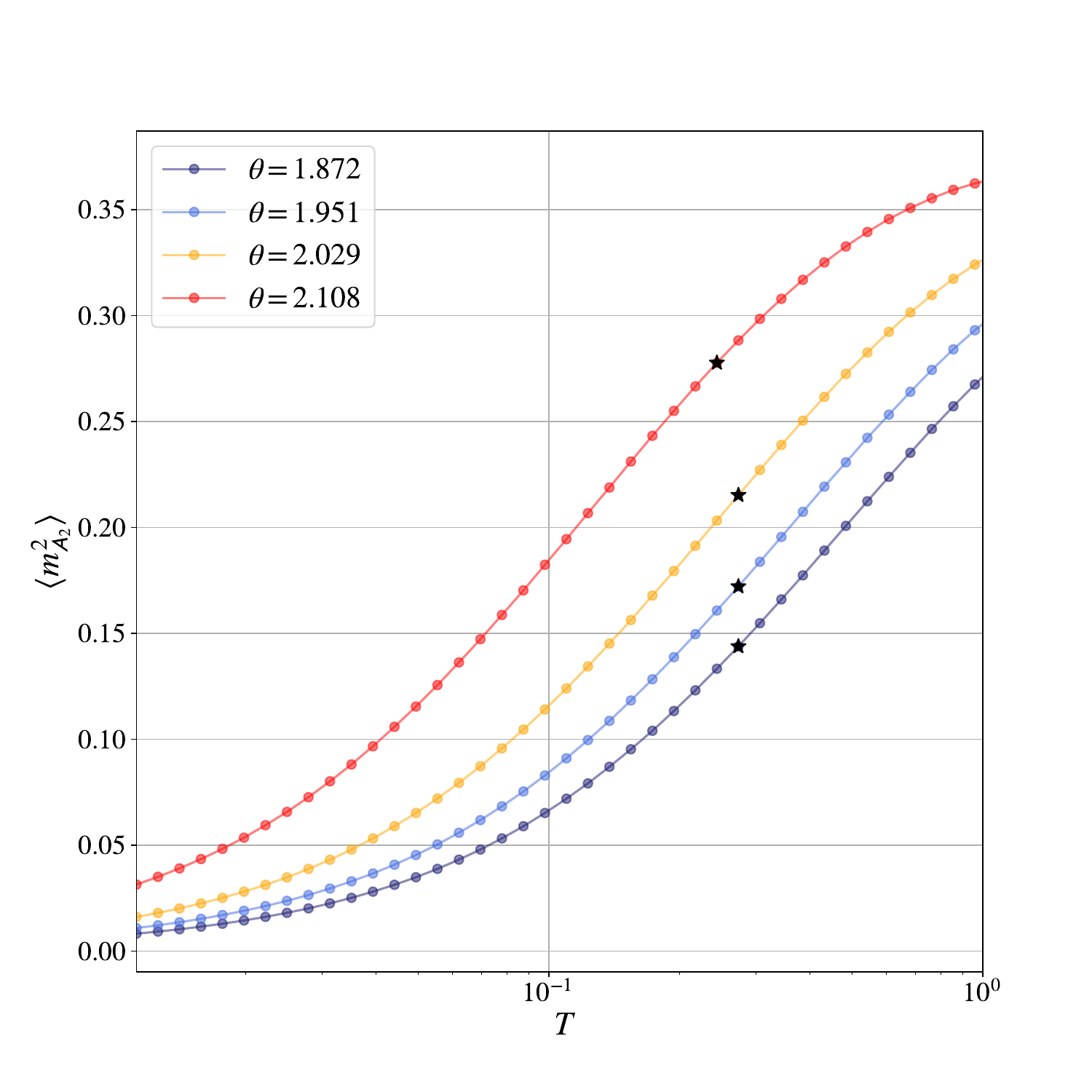}
    \end{overpic}
    \caption{Temperature evolution of the $\langle m_{\mathsf{A}_2}^2\rangle$ parameter. In this figure, each curve corresponds to a parametrization discussed in Sec.~\ref{sec:proximity_to_CSL}, i.e. $(J_{zz},J_{z\pm})=(\cos\theta,\sin\theta)$, with the remaining parameters given by Eq.~\eqref{eq:CL6_constraint}. Here the black stars label the inflection point of the curves and estimate the crossover between the two CSLs in Fig.~\ref{fig:phase_diagram_A2_C}(a) and (b).}
    \label{fig:MA2}
\end{figure}

\dlgnew{
In Sec.~\ref{sec:proximity_to_CSL} we illustrate a phase diagram in Fig.~\ref{fig:phase_diagram_A2_C} (a) and (b) where we identify three distinct phases; two CSL phases and one $\vb*{q}=0$ symmetry-breaking phase.
The boundary between the CSL 8 and the symmetry-breaking $\mathsf{A}_2$ phase, marked by a white line in Fig.~\ref{fig:phase_diagram_A2_C} (a) and (b), is identified by tracking the peak of the specific heat 
\begin{eqnarray}
    C=\frac{\langle \mathcal{H}^2\rangle-\langle \mathcal{H}\rangle^2}{Nk_\mathrm{B}T^2},
\end{eqnarray}
where $\langle \mathcal{\cdots }\rangle$ signifies a thermodynamic average
$N$ is the number of spins in the system, and $k_{\mathrm{B}}$ is the Boltzmann constant which is set to $1$. 
In contrast to the symmetry-breaking transition, the measured specific heat does not show any distinctive feature separating the two CSLs. 
To estimate the crossover between these two CSLs, we follow a similar approach to the one taken in Ref.~\cite{Taillefumier2017PhysRevX} where we measure the thermodynamic expectation of the square of the $m_{\mathsf{A}_2}$ parameter, namely $\langle m_{\mathsf{A}_2}^2 \rangle$, and identity the point of inflection as shown in Fig.~\ref{fig:MA2}. 
We use this quantity to signal the crossover between CSL 8 and CSL 6.
\dlgnew{The motivation for doing so is that the $m_{\mathsf{A_2}}$ field corresponds to the scalar charge excitation associated to the CSL 6, which becomes thermally depopulated as the system crosses over into the CSL 6 regime.}
We measure the square of the $\mathsf{A}_2$ parameter so that we have a measure of how many charges (positive and negative) are still present in the system. }

 \section{Comparison between coarse-grained theory and large-$\mathcal{N}$ theory}
 \label{app.theory.largeN}

\begin{figure}%
    \centering
    \begin{overpic}[width=\columnwidth]{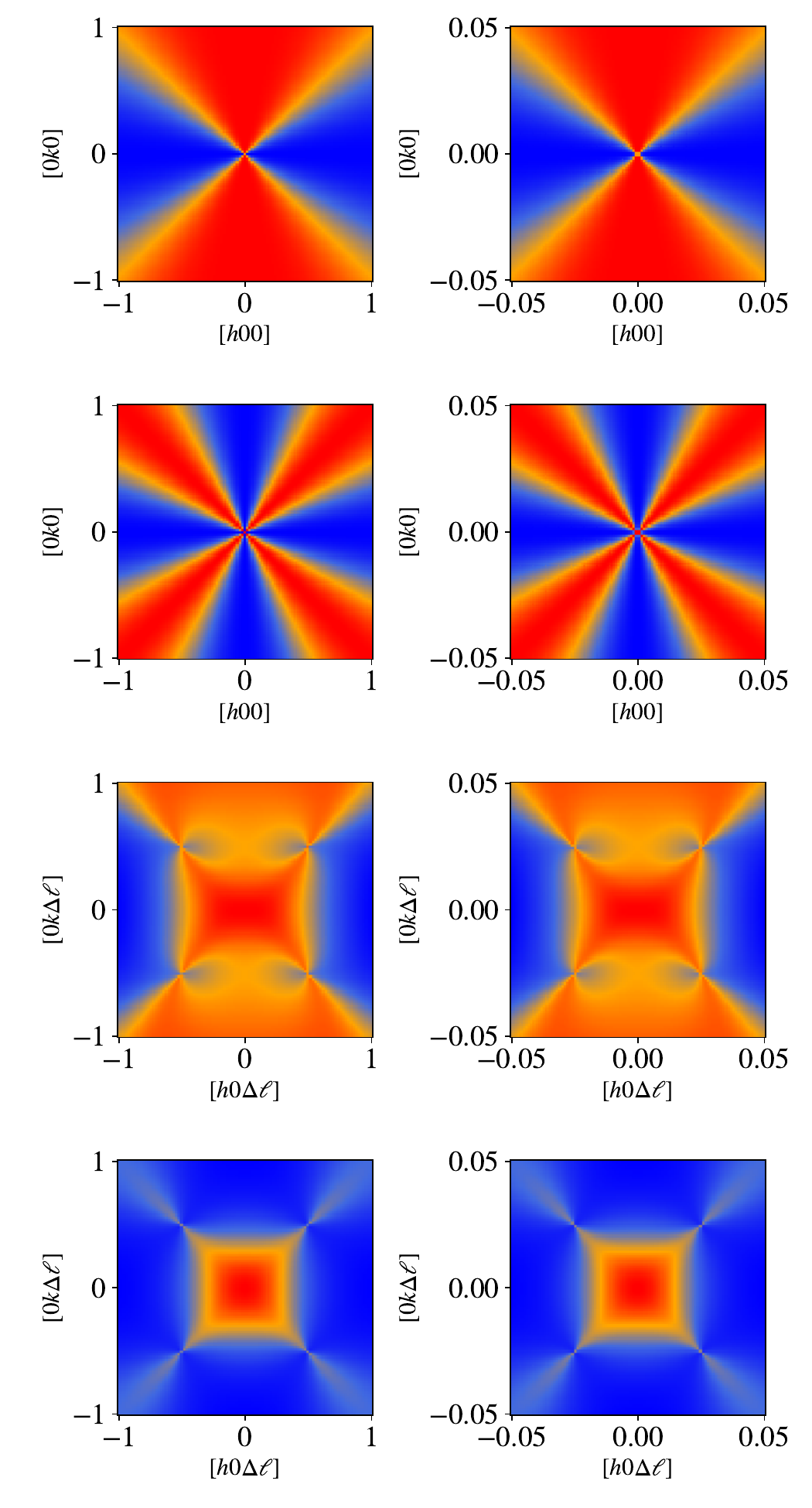}
    \put(5,100) {(a)}  
    \put(30,100) {(b)}  
    \put(5,75) {(c)}  
    \put(30,75) {(d)}  
    \put(5,50) {(e)}  
    \put(30,50) {(f)}  
    \put(5,25) {(i)} 
    \put(30,25) {(j)} 
    \end{overpic}
\caption{\dlgnew{Irrep correlation functions $\langle m_X^\alpha(\vb*{q})m_X^\alpha(-\vb*{q})\rangle $ in the $[hk\Delta \ell]$ plane obtained from the coarse-grained theory (left column) and computed from large-$\mathcal{N}$ (right column) for the CSL 6 using the same parameters as those considered in Fig.\ref{fig:E_T2_T1planar_Jzpm_pos_pinch_lines}. Here panels (a), (b), (e) and (f) illustrate the $\langle m_{\mathsf{E}}^1(\vb*{q})m_{\mathsf{E}}^1(-\vb*{q})\rangle$, whereas panels (c), (d), (g) and (h) illustrate the $\langle m_{\mathsf{T}_2}^z(\vb*{q})m_{\mathsf{T}_2}^z(-\vb*{q})\rangle$. For panels (e-j) we considered $\Delta\ell=0.1$ in the coarse-grained theory and $\Delta\ell=0.025$ in the large-$\mathcal{N}$ calculation.}
}
    \label{fig:coarse-grain_large_N}
\end{figure}

 \dlgnew{
 In this section, we provide a comparison between the coarse-grained field theory deducted from the inter-tetrahedra constraint and the large-$\mathcal{N}$ SCGA results (the latter has been cross-checked with CMC extensively in the main text). 
 We consider a concrete example of  the CSL 6. 
 Its low-energy coarse-grained theory identifies the rank-2 field $ \vb*{E}^{\mathsf{SV}}$, defined as 
 \begin{equation}
 \begin{split}
      \vb*{E}^{\mathsf{SV}} =& \begin{pmatrix}
	2 \mEScalar^1 & \sqrt{3} m_{T_2}^z    &    - \sqrt{3} m_{T_2}^y   \\
	- \sqrt{3} m_{T_2}^z  & -\mEScalar^1 + \sqrt{3} \mEScalar^2 &  \sqrt{3} m_{T_2}^x \\
	 \sqrt{3} m_{T_2}^y  &-\sqrt{3} m_{T_2}^x& -\mEScalar^1 - \sqrt{3} \mEScalar^2
\end{pmatrix}\\
& - 
3\sin\phi
\begin{pmatrix}
	0 & m_{T_{1,-}}^z &     m_{T_{1,-}}^y \\
	 m_{T_{1,-}}^z &0&   m_{T_{1,-}}^x \\
	 m_{T_{1,-}}^y &   m_{T_{1,-}}^x & 0
\end{pmatrix} . 
\end{split}
\end{equation}   
which fulfills two constraints 
\begin{align}
    m_{\mathsf{A_2}} = 0  \implies & |\epsilon^{ijk}| \partial_i E_{jk}^{\mathsf{SV}}  = 0,   \\
	\vb*{m}_{\mathsf{T_{1,+}}} = 0  \implies & \partial_i E_{ij}^{\mathsf{SV}} =0.  
\end{align}
We can thus write down the low-energy effective Hamiltonian as 
\[
\mathcal{H}_\text{CSL6}
 = \int \mathrm{d}v \  (|\epsilon^{ijk}| \partial_i E_{jk}^{\mathsf{SV}})^2 + \partial_i E_{ij}^{\mathsf{SV}}\partial_k E_{kj}^{\mathsf{SV}} 
\]
which, in Fourier space, can be expressed as 
\[
\mathcal{H}_\text{CSL6}
 = \int  \mathrm{d}^3q \  E_{I}^{\mathsf{SV}}H_{IJ}(\vb*{q})E_{J }^{\mathsf{SV}}
\]
Here, $E_{I}^{\mathsf{SV}}$ is \dlgnew{introduced to organize the degree of freedoms in $E_{ij}^{\mathsf{SV}}$ in a vector form.}
$H_{IJ}(\vb*{q})$ is the interaction matrix in the Fourier transformed momentum space. 
It has four flat bands at the bottom, because there are 8 degrees of freedom in $E_{ij}^{\mathsf{SV}}$ and 4 scalar constraints. 
The equal time correlation function 
$\langle E_{ij}^{\mathsf{SV}}(\vb*{q})E_{kl}^{\mathsf{SV}}(-\vb*{q})\rangle $ and $\langle m_X^\alpha(\vb*{q})m_X^\alpha(-\vb*{q})\rangle $ can then be extracted from the eigenvectors of the four bottom flat bands.

In parallel, one can compute the same correlation functions from large-$\mathcal{N}$ 
which, as shown in Fig.~\ref{fig:E_T2_T1planar_Jzpm_pos_pinch_lines}, 
is found to be in good agreement with CMC simulations.
In Fig.~\ref{fig:coarse-grain_large_N}, we illustrate the comparison between the irrep correlation functions $\langle m_{\mathsf{E}}^1(\vb*{q})m_{\mathsf{E}}^1(-\vb*{q})\rangle$ and $\langle m_{\mathsf{T}_2}^z(\vb*{q})m_{\mathsf{T}_2}^z(-\vb*{q})\rangle$ obtained via the coarse-grained theory and large-$\mathcal{N}$ in the $T\to 0$ limit.
We observe an excellent agreement between the two methods. 
This provides numerical proof of the validity of the coarse-grained theory framework we have developed.}

\newpage

\section{Dependence of the neutron structure factors of the CSL 5 as a function of the $g$-factor. }
\label{appendix:DQQ_ast_neutron}

\begin{figure}[ht!]
    \centering
    \begin{overpic}[width=\columnwidth]{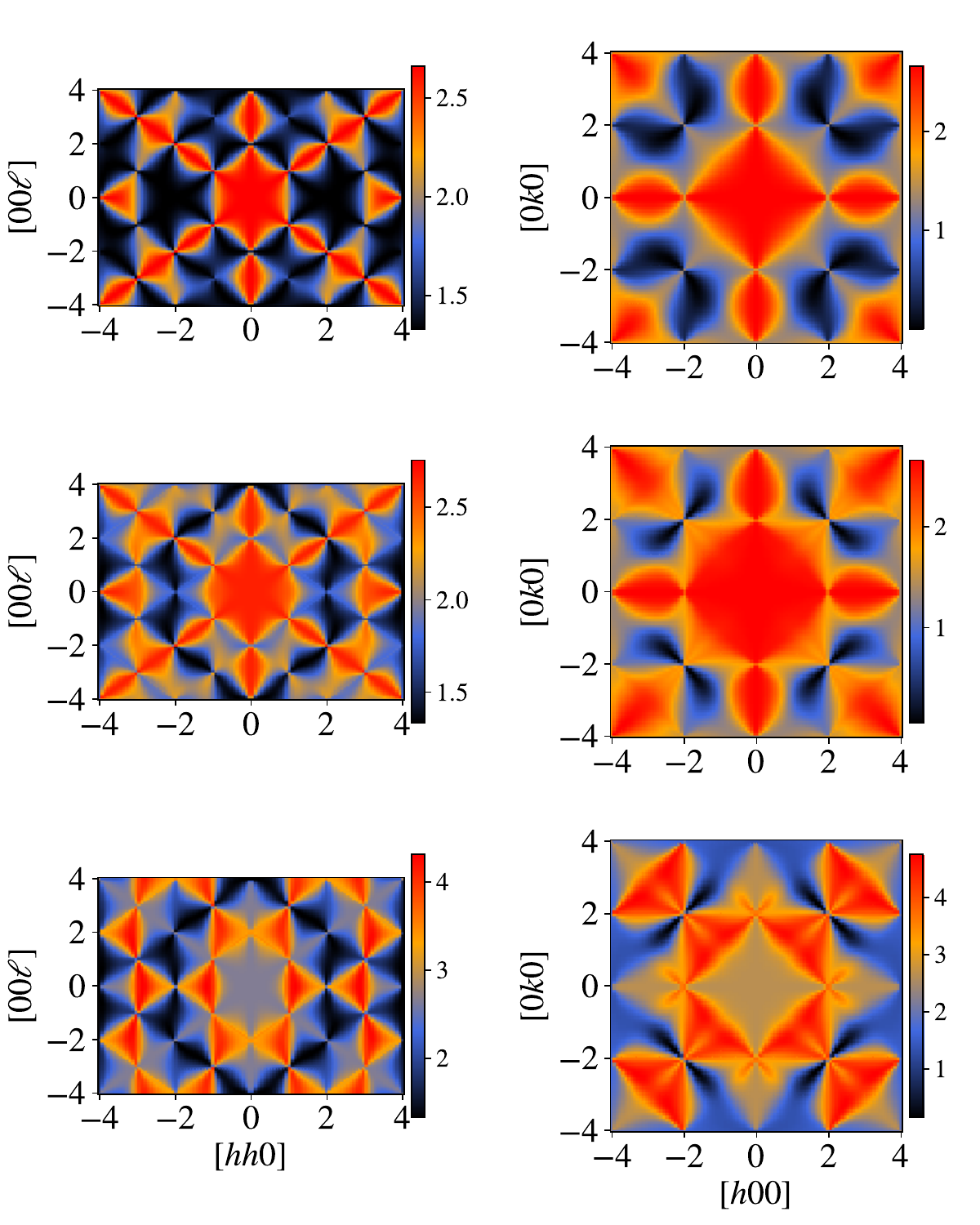}
    \put(2,100) {(a)}  
    \put(15,94) {$g_\perp=0$}  
    \put(43,100) {(b)}  
    \put(56,97) {$g_\perp=0$}

    \put(2,67) {(c)} 
    \put(14,62) {$g_\perp=0.5g_{zz}$}  
    \put(43,67) {(d)}  
    \put(55,65) {$g_\perp=0.5g_{zz}$}  
    
    \put(2,33) {(e)} 
    \put(15,30) {$g_\perp=g_{zz}$}  
    \put(43,33) {(f)} 
    \put(56,33) {$g_\perp=g_{zz}$}  

    \end{overpic}
\caption{\dlgnew{Neutron structure factors obtained via large-$\mathcal{N}$ for a set of parameters describing CSL 5 at low temperatures for distinct values of the $g$-factor as indicated by the title of each panel. Panels (a), (c), and (e) illustrate the neutron structure factor on the $[hh\ell]$ scattering plane, while panels (b), (d), and (f) illustrate the same correlation function in the $[hk0]$ scattering plane.}}
    \label{fig:DQQ*ast_gxy}
\end{figure}

\dlgnew{
In this section, we \dlgnew{report the large-$\mathcal{N}$ calculations of } the unpolarized neutron structure factors obtained for the CSL 5 for various choices of the magnetic $g$-tensor. CSL 5 is realized in a plane in parameter space $\{J_{uv}\}$ defined by the constraint $J_{z\pm}=0$. This plane is most naturally realized for the non-Kramers pyrochlores, where the local-$z$ of the magnetic moments transforms as a dipole, whereas the remaining transverse degree of freedoms transform as quadrupoles. 
As these degrees of freedom are respectively time odd and time even, they do not couple at the bilinear level, hence requiring $J_{z\pm}=0$.
Moreover, in terms of the local $g$-tensor, defined as }
\begin{eqnarray}
    g_{\rm Local}=\begin{pmatrix}
        {g}_\perp & 0 & 0\\
        0 & {g}_\perp & 0 \\
        0 & 0& {g}_{zz}
    \end{pmatrix},
\end{eqnarray}
\dlgnew{non-Kramers ions would normally be characterized by a vanishing $g_{\perp}$ reflecting the nature of an isolated non-Kramers crystal field ground doublet.
\dlgnew{However, it is of interest to explore how virtual crystal-field fluctuations in Tb$^{3+}$ may engender an effective transverse magnetic response with $g_\perp \neq 0$~\cite{Ghosh_2003}, thereby modifying the neutron-scattering form factor that would arise if $g_\perp$ were set to zero.}
\dlgnew{Figure~\ref{fig:DQQ*ast_gxy} illustrates the dependence of the unpolarized neutron structure factor on $g_\perp$, computed using the large-$\mathcal{N}$ method.}
}

\dlgnew{
For $g_\perp=0$, the neutron scattering structure factor shows the typical pattern of classical Ising spin ice since the transverse spin correlator that would exist at the CSL 5 (DQQ$^*$ point)~\cite{lozanogomez2023arxiv} is extinguised by $g_\perp =0$. As $g_\perp$ increases, the pattern gets progressively modified (e.g. $g_\perp =0.5 g_{zz}$) exposing additional pinch point features originating from the correlations between the $xy$ spin degrees of freedom (see for example the twofold pinch points in the $[hh\ell]$ plane at $[220]$.). These additional features become more prominent as the value of $g_\perp$ is tuned to that of $g_{zz}$ (e.g. $g_\perp =g_{zz}$), i.e. where the local $g$-tensor becomes isotropic.
\dlgnew{
Most importantly, and the main point of performing the calculations reported in this appendix, unlike what has been observed experimentally in Tb$_2$Ti$_2$O$_7$~\cite{Fritsch,
Kadowaki-halfhalfhalf}, no evidence of correlations at ${\bm q}=(\frac{1}{2} \frac{1}{2} \frac{1}{2})$ are observed in the calculations (Fig.~\ref{fig:DQQ*ast_gxy}).
}
}

\nocite{uniaxial_component,Mirebeau_uni1,Mirebeau_uni2,Mirebeau_uni3}
\bibliography{atlas_ref}

\end{document}